\documentclass[11pt]{article}

\usepackage{newtxtext,newtxmath}

\usepackage{graphicx}

\usepackage[letterpaper,margin=1in]{geometry}

\renewenvironment{abstract}
	{\quotation}
	{\endquotation}

\date{}

\makeatletter
\renewcommand{\fnum@figure}{\textbf{Figure \thefigure}}
\renewcommand{\fnum@table}{\textbf{Table \thetable}}
\makeatother

\usepackage{scicite}

\usepackage{url}

\usepackage{hyperref}
\usepackage{amsmath}
\usepackage{xcolor}
\usepackage{xspace}
\usepackage{bm}

\definecolor{darkblue}{RGB}{12,94,176}
\hypersetup{
    pdfnewwindow=true,     
    colorlinks=true,       
    linkcolor=darkblue,    
    citecolor=darkblue,    
    filecolor=darkblue,    
    urlcolor=blue          
}

\graphicspath{{./}{figures/}}

\newcommand{\kms}{\ensuremath{\,\rm{km}\,\rm{s}^{-1}}\xspace}
\newcommand{\Rsun}{\ensuremath{\,\rm{R_\odot}}\xspace}
\newcommand{\Msun}{\ensuremath{\,\rm{M}_{\odot}}\xspace}

\newcommand{\Myr}{\ensuremath{\,\rm{Myr}}\xspace}

\def\scititle{
    Evidence of polar and ultralow supernova kicks\\from the orbits of Be X-ray binaries 
}
\title{\bfseries \boldmath \scititle}

\author
{Ruggero Valli$^{1\ast}$, Selma E. de Mink$^{1\ast}$, Stephen Justham$^{1}$, \and
Thomas Callister$^{2}$, Cole Johnston$^{1,3,4}$, Daniel Kresse$^{1,5}$, Norbert Langer$^{6,7}$,\and
Amanda C. Rubio$^{1,8}$, Alejandro Vigna-G{\'o}mez$^{1}$, Chen Wang$^{1,9,10}$\and 
\small{$^{1}$Max-Planck-Institut f{\"u}r Astrophysik, Karl-Schwarzschild-Straße 1, 85748 Garching, Germany.}\and
\small{$^{2}$Kavli Institute for Cosmological Physics, The University of Chicago, Chicago, IL 60637, USA.}\and
\small{$^{3}$Department of Physics, University of Surrey, Guildford, GU27XH Surrey, UK.}\and
\small{$^{4}$Institute of Astronomy, KU Leuven, Celestijnenlaan 200D, 3001 Leuven, Belgium.}\and
\small{$^{5}$Technical University of Munich, TUM School of Natural Sciences, Physics Department,}\and\small{James-Franck-Stra{\ss}e 1, 85748 Garching, Germany.}\and
\small{$^{6}$Argelander-Institut f{\"u}r Astronomie, Universit{\"a}t Bonn, Auf dem H{\"u}gel 71, D-53121 Bonn, Germany.}\and
\small{$^{7}$Max-Planck-Institut f{\"u}r Radioastronomie, Auf dem H{\"u}gel 69, D-53121 Bonn, Germany.}\and
\small{$^{8}$Instituto de Astronomia, Geof{\'i}sica e Ci{\^e}ncias Atmosf{\'e}ricas, Universidade de S{\~a}o Paulo,}\and\small{Rua do Mat{\~a}o 1226, 05508-900 S{\~a}o Paulo, Brazil.}\and
\small{$^{9}$ Department of Astronomy, Nanjing University, Nanjing 210023, People’s Republic of China}\and
\small{$^{10}$ Key Laboratory of Modern Astronomy and Astrophysics (Nanjing University), Ministry of Education,}\and\small{Nanjing 210023, People’s Republic of China}\and
\small{$^\ast$ Corresponding authors: ruvalli@mpa-garching.mpg.de, sedemink@mpa-garching.mpg.de}
}

\begin{document} 

\maketitle

\begin{abstract} \bfseries \boldmath
Supernovae, the explosive deaths of massive stars, create heavy elements and form black holes and neutron stars. These compact objects often receive a velocity at formation, a ``kick" whose physical origin remains debated. We investigate kicks in Be X-ray binaries, containing a neutron star and a rapidly spinning companion. We identify two distinct populations: one with kicks below $10\,\kms$, much lower than theoretical predictions, and another with kicks around $100\,\kms$, that shows evidence for being aligned within 5 degrees of the progenitor’s rotation axis. The distribution of progenitor masses for the two populations have medians around $2.3\,\Msun$ and $4.9\,\Msun$, corresponding to stars with birth masses of about $10\,\Msun$ and $15\,\Msun$.  The second component matches the low-velocity mode observed in isolated pulsars. Combined with the known high-velocity component, which dominates isolated pulsars, this suggests three distinct kick modes. These results reveal previously unrecognized diversity in neutron-star formation.
\end{abstract}


\newpage
\noindent
Neutron stars are the densest known objects in the Universe. They form when the core of an aging massive star collapses under its own weight, reaching temperatures and densities where atoms break apart and protons capture electrons to form neutrons. The release of gravitational energy can power supernova explosions, which drive gas flows in galaxies and enrich the Universe with heavy elements\cite{woosley_evolution_2002}.
Despite their fundamental importance, the death of stars as supernovae and the accompanying birth of neutron stars are still poorly understood. Progress is hampered by uncertainties in the final stages of the lives of massive stars, the complex interaction of matter, radiation, and neutrinos, and the exotic properties of extremely dense matter \cite{burrows_core-collapse_2021, janka_long-term_2025}.  

Observations of Be X-ray binaries, systems consisting of a neutron star orbiting a rapidly-rotating stellar companion, provide unique opportunities to study the nature of supernova explosions and neutron star formation \cite{pfahl_new_2002, fortin_constraints_2022}.  This is possible because (i) their well-understood evolutionary history places strong constraints on the properties and orbit \emph{before} the explosion (\ref{sec:Be X-ray binaries}), while (ii) the pulsed emission of the neutron stars allows for very precise measurements of the present-day orbit (\ref{sec:observational properties}), which constrain the system's state \emph{after} the supernova explosion.  We exploit these features to infer fundamental properties of the explosion: the mass ejected during the supernova and the velocity imparted to the newborn neutron star due to asymmetries in the explosion, known as the supernova kick. 

\subsection*{Observational sample}
We consider all available Be X-ray binaries with high-quality measurements, starting from the catalog in \cite{fortin_catalogue_2023}, including all systems with accurately measured orbital periods and eccentricities and which satisfy the following additional selection criteria (\ref{sec:selection criteria}):
(i) The system contains a neutron star.
(ii) The companion is a classical Be star, indicative of rapid rotation and the presence of an outflowing disk. This supports the notion that the two stars initially present in the system interacted before the supernova occurred. The interaction stripped the envelope from the progenitor of the neutron star, while spinning up the present-day Be star through accretion. 
It also indicates that tides did not significantly modify the orbit, at least, they did not cause the Be star to spin down or prevent the disk from forming.  This supports our assumption that the present-day properties closely resemble the post-explosion properties.  
(iii) We limit our analysis to Galactic systems. Orbital solutions are more sparsely available for extragalactic systems and their different chemical composition affects the progenitor evolution, introducing additional variability in the sample properties.

The resulting sample contains 23 systems (table~\ref{tab:galacticBeX}). Their orbital periods and eccentricities  are shown in Fig.~\ref{fig:main_plot}. The data shows two well-separated groups \cite{pfahl_new_2002} easily recognized by the eye and recovered with clustering algorithms (\ref{sec:two groups in the sample}).
The main selection effects are well understood: the neutron star must pass close enough to accrete and be bright in X-rays. However, a too-close passage would not provide space for the Be star disk to develop, or would spin down the Be star \cite{liu_population_2024}. This constrains the sample to a limited range in periastron separations and explains why the diagram's upper left and lower right sides are empty but does not explain the appearance of the two groups.
The sample is not homogeneous and affected by biases, however, we find no plausible selection effect capable of producing the observed separation in two groups (\ref{sec:selection biases}).
Therefore, we proceed and interpret the sample assuming that the cause for the appearance of the two groups is of a physical nature.

\subsection*{Modeling and inference}
We use the present-day orbits of Be X-ray binaries to infer the properties of the supernovae that formed the neutron stars in these systems.
Since the explosion happened much faster than the orbital timescale, it can be modeled as a near-instantaneous event, whose impact depends on (i) the amount of mass lost $\Delta M$ and (ii) the magnitude and direction of the birth kick of the neutron star $\bm{v}_{\rm kick}$ \cite{kalogera_orbital_1996}.  
Our goal is to constrain the distributions of these two fundamental explosion properties. We achieve this by modeling the pre-explosion binary population, computing the effect of a supernova on each system, and comparing the predicted post-explosion properties with the observed sample. This allows us to derive posteriors for kick and mass loss distributions.

The properties of the pre-explosion binaries are not well known. Model predictions are highly uncertain and observational constraints are extremely scarce \cite{drout_observed_2023}. 
Therefore, we infer the properties of the pre-explosion population alongside the properties of the supernovae (\ref{sec:methods_model}), and verify that our results are not sensitive to our assumed functional form for the distribution of the pre-explosion donor mass and orbital period (\ref{sec:effect_of_pre-SN_distributions}).
We assume that the mass of the companion star is not significantly affected by the explosion, and that the orbit, once modified by the supernova, remains unchanged until today  (\ref{sec:postsn orbit did not change}). 
To model the kicks, we sample them from an isotropic Maxwellian distribution (a standard choice in the literature), unless otherwise stated.   
We account for observational selection effects by limiting the periastron distances to systems wide enough for Be star disks to form \cite{panoglou_be_2016,panoglou_be_2018} and tight enough for significant accretion rates, allowing X-ray detection \cite{liu_population_2024}.   

To compare the simulated post-supernova orbits to the observations, we calculate the likelihood of obtaining the data given a set of model parameters, which describe the distributions of the pre-supernova periods, masses and kicks.  We estimate the posteriors of the model parameters using Markov Chain Monte Carlo sampling (\ref{sec:methods_point_estimates}). Throughout the paper, we report median values with 90\% credible intervals. For illustration, we show model predictions based on the posterior medians of the free parameters, which we refer to as the best-fit model. We report Bayes factors, p-values of posterior predictive checks and Kolmogorov-Smirnov (KS) tests, where appropriate (\ref{sec:model selection}).  

A detailed description of our method and a discussion of our assumptions is provided in sec.~\ref{sec:methods}. In sec.~\ref{sec:supplementary_results} we show that our results are not sensitive to the choices of parameters and shapes of the unknown distributions. 

\subsection*{Results} 
\begin{figure}
    \centering
    \includegraphics[width=0.47\textwidth]{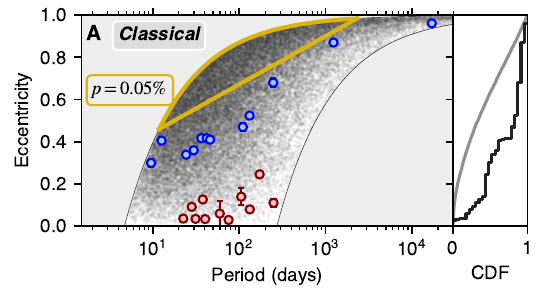}%
    \includegraphics[width=0.47\textwidth]{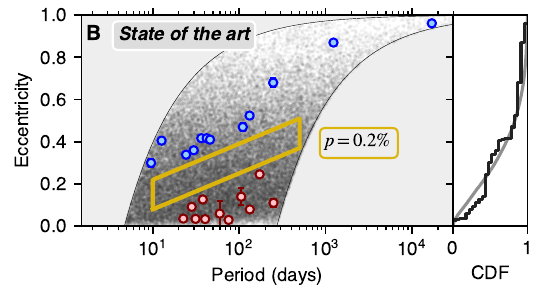}
    \includegraphics[width=\textwidth]{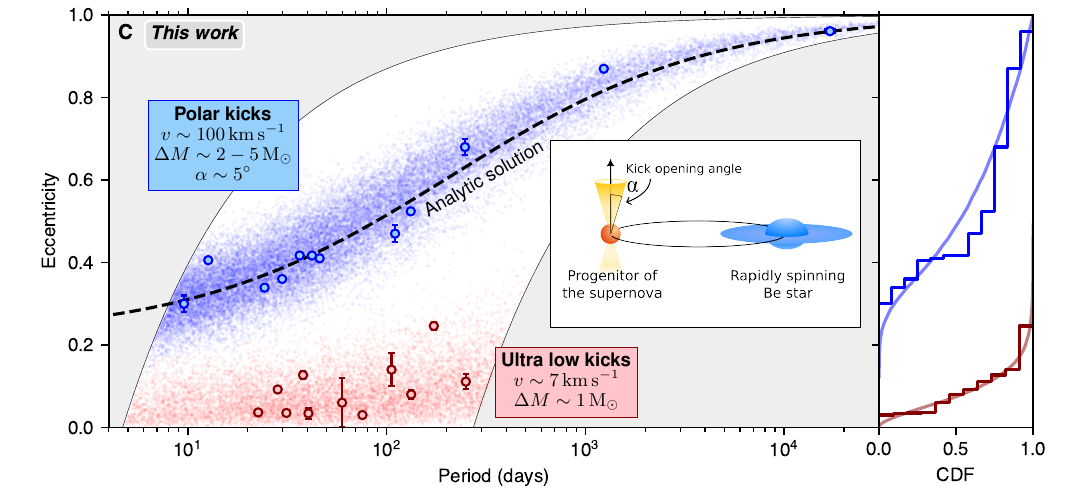}
    \caption{\textbf{Models of Be X-ray binaries.} The panels compare different supernova kick models and their resulting period-eccentricity distributions for Be X-ray binaries. The large circles are the observed systems, the low-e group in red, and the high-e sequence in blue.
   \textbf{Panel A}: the classical isotropic Maxwellian kick distribution based on isolated pulsars \cite{hobbs_statistical_2005}. \textbf{Panel B}: as in panel A, but including a second low-velocity component to account for low-mass progenitors and electron-capture supernovae, as often assumed in recent works. The yellow contours in panel A and B mark regions that are highly populated by the models, but actually devoid of observations. Next to them, we report the probability that those regions would be empty, when randomly sampling 23 systems from the respective models. \textbf{Panel C}: our preferred model, which applies two distinct kick prescriptions: ultralow isotropic kicks for the low-e group (red) and polar kicks for the high-e sequence (blue). The dashed line represents the analytic fit for the high-e sequence (eq.~\ref{eq:ecc_from_polar_kicks}). The inset illustrates the polar kick scenario.  The dot clouds are random samples drawn from the models. Grey shaded areas cover the region where systems are not expected to be observable, and are excluded from the models (\ref{sec:methods_model}). The right panels compare cumulative eccentricity distributions (CDFs) for the observed (dark line) and model-predicted (light line) samples.}
    \label{fig:main_plot}
\end{figure}

Our results are visualized in Fig.~\ref{fig:main_plot}, where we compare our findings with two standard approaches commonly used in the literature.  The {\it ``classical'' approach} assumes kick velocities drawn from a single isotropic Maxwellian. The most widely used version is the ``Hobbs distribution'' with a dispersion $\sigma_{\rm Hobbs} = 265\,\kms$ derived from isolated pulsars \cite{hobbs_statistical_2005}. The best-fit model, shown in panel A, assumes this distribution while fitting for the progenitor period and mass distributions. According to this model, over 90\% of binaries break apart. The few systems that remain bound are predicted to have eccentricities higher than the ones observed. This kick model fails to explain the abundance of low-eccentricity systems (red), as noted by \cite{pfahl_new_2002}, and over-predicts systems with high eccentricities (blue): the probability for the region marked with yellow boundaries to be empty for a sample of this size is extremely small (0.05\%) according to this model. A very low p-value from a 2D KS test ($ < 10^{-4}$) further highlights the discrepancy (\ref{sec:hobbs fit}).

\begin{figure}
    \centering
    \includegraphics[]{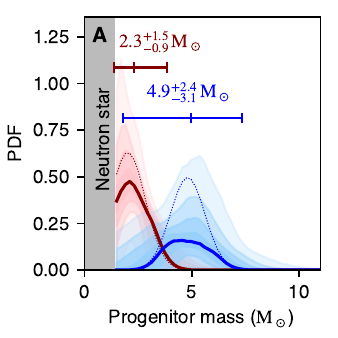}%
    \includegraphics[]{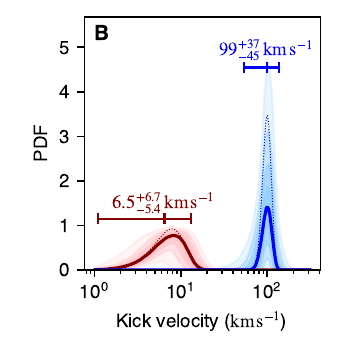}
    \includegraphics[]{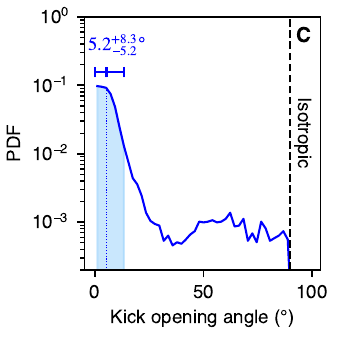}%
    \includegraphics[]{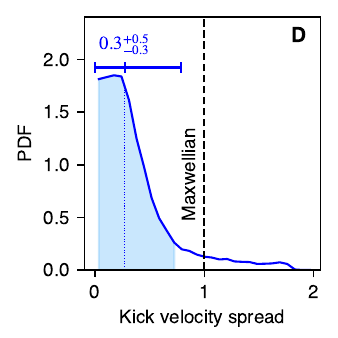}
    \caption{\textbf{Probability distributions of kick and progenitor parameters.} The distributions are shown for the low-e group (red) and the high-e sequence (blue). The top panels display the progenitor mass (\textbf{A}) and kick velocity (\textbf{B}), marginalized over the posterior. Solid lines represent median probabilities, while shaded bands indicate quantiles from 0.1 to 0.9 in steps of 0.1. Dotted lines show the best fit using median posterior values. The bottom panels show the posterior distributions of the kick opening angle $\alpha$ (\textbf{C}) and the kick velocity spread (\textbf{D}), defined as the ratio of the standard deviation to the mean relative to a Maxwellian distribution (vertical dashed line). In the bottom panels, the vertical dotted lines mark the median, and the shaded regions indicate the 90\% credible interval. The same information (median and 90\% credible interval) is also shown with horizontal bars above the distributions.}
    \label{fig:posteriors_summary}
\end{figure}

The {\it ``state-of-the-art'' approach} is to  assume a two-component kick distribution where, in addition to Hobbs, kicks can be drawn from a second Maxwellian with a dispersion typically taken to be $\sigma_2 = 15-45\,\kms$ \cite{podsiadlowski_effects_2004, giacobbo_impact_2019, vigna-gomez_formation_2018,  igoshev_combined_2021}. This second component is commonly called the electron-capture supernova contribution, even though it is not necessarily related to electron-capture-driven explosions \cite{jones_electron-capture_2016}.  In panel B, we show our results for this model, fitting for the velocity dispersion and the progenitor population. We find $\sigma_2 = 7.6^{+9.8}_{-7.6}\,\kms$, which is only marginally consistent with the lower range adopted in current studies. Despite comprising two components, this model predicts an effectively unimodal distribution in the period-eccentricity plane because only a very small fraction of the kicks drawn from the Hobbs distribution keep the binary systems bound. The distribution peaks at intermediate eccentricities, highlighted by the box shown in yellow, which is devoid of data. The probability that the marked region would be found empty according to this model is very small (0.2\%, \ref{sec:state-of-the-art fit}). 

The discrepancies between the data and the model predictions resulting from both the ``classical'' and the ``state-of-the-art'' approach, and their failure to recover the dichotomy of the two groups, motivated us to explore the data further, analyzing and discussing both groups separately.

 \vspace{12px}
\noindent \textit{\textbf{1. Low-e group---Evidence for extremely low kicks from very low-mass progenitors:}}
The low-eccentricity systems (hereafter low-e group, shown in red) can be explained well with a model in which kicks are drawn from a single isotropic Maxwellian distribution (\ref{sec:low-e group fit}). Our best-fit model is shown with red shading in Fig.~\ref{fig:main_plot}C. We find $\sigma = 4.5^{+2.8}_{-3.2}\,\kms$, much lower than typically assumed in population synthesis simulations.  We exclude $\sigma > 10\,\kms$ with 99.8\,\% probability for this group, implying that this subpopulation originates from ultralow kicks that are currently not accounted for in most population synthesis studies. 
The resulting kick velocities (Fig.~\ref{fig:posteriors_summary}B) are also lower than those currently reported in state-of-the-art supernova simulations, which range from a few tens up to a thousand $\kms$ \cite{janka_interplay_2024,burrows_theory_2024,wang_supernova_2024,sykes_long-time_2024}.  

We infer a progenitor mass distribution with a typical mass of $M_{\rm prog}= 2.3^{+1.5}_{-0.9}\,\Msun$ (Fig.~\ref{fig:posteriors_summary}A). These values are much lower than expected from single-star evolution \cite{sukhbold_core-collapse_2016}, confirming that the progenitors indeed lost their envelope before the explosion, as we had implicitly assumed.
The inferred masses are consistent with, but at the low end of those predicted for stripped-envelope supernovae \cite{yoon_type_2010,yoon_type_2017,eldridge_death_2013} of stars with initial masses of about $10\,\Msun$ \cite{laplace_different_2021}. 
The implied ejecta masses,  $\Delta M_{\rm prog}= 0.9^{+1.5}_{-0.9}\,\Msun$,  are at the low end of the range inferred from light curve analysis for stripped-envelope supernovae \cite{taddia_early-time_2015,lyman_bolometric_2016}, more in the range of the very rapidly fading transients discovered by \cite{drout_rapidly_2014}.  

\vspace{12px}
\noindent \textit{\textbf{2. High-e sequence---Evidence for intermediate kicks correlated with the progenitor spin:}}
    The systems with high eccentricities (hereafter high-e sequence, shown in blue) fall on a very narrow relation in the period-eccentricity plane.  The narrow sequence cannot be reproduced well by an isotropic Maxwellian kick distribution.  To reproduce the observed narrow trend, kicks must come from a restricted range of directions and possibly a restricted range of velocities (\ref{sec:high-e sequence fit}). 

   A natural choice for a preferred direction is set by the stellar spin, which tends to align with the orbital angular momentum due to tides and mass exchange \cite{marcussen_banana_2024, de_mink_rotation_2013}. Rotation also breaks the spherical symmetry of the explosion and provides a natural direction for the kick \cite{mosta_magnetorotational_2014}. A simple analytical expression for polar kicks assuming a fixed kick velocity and mass loss (eq.~\ref{eq:ecc_from_polar_kicks}) reproduces the high-e sequence remarkably well (the black dashed line in Fig.~\ref{fig:main_plot}C).

    Using our full inference method, we consider kicks restricted to a cone in the polar direction, defined by a half-aperture $\alpha$, which we leave as a free parameter varying from $0$ (fully aligned) to $90^{\circ}$ (recovering our original assumption of isotropic kicks).  We draw kick magnitudes from a modified Maxwellian distribution with an additional parameter to adjust the spread.    
    We find strong evidence (\ref{sec:evidence for polar}) favoring polar kicks with a narrow opening angle ($\alpha=5.2^{+8.3}_{-5.2}\,^{\circ}$) over an isotropic distribution.  Our best fit is obtained for a kick distribution about four times narrower than a Maxwellian, but only with marginal evidence (\ref{sec:evidence for narrowness}). The best-fit kick distribution can be well approximated by a Normal distribution with the same mean and standard deviation ($100\,\kms$ and $11\,\kms$, respectively).
    For the progenitor masses, we infer  $M_{\rm prog} = 4.9^{+2.4}_{-3.1}\,\Msun$, significantly larger than the one found for the low-e group but still consistent with masses expected for stars stripped in binary systems, with initial masses of about $15\,\Msun$ \cite{laplace_different_2021}.  In Fig.~\ref{fig:posteriors_summary}, we show the inferred distributions of progenitor mass, kick velocity, and direction.

    Our proposed scenario is preferred over an isotropic Maxwellian kick distribution (\ref{sec:evidence against iso maxwellian}). It is also strongly favored over the idea that high-e systems originate from the low-velocity tail of a Hobbs-like distribution (\ref{sec:high-e with hobbs}), and provides a better fit than a model with kicks aligned with the equatorial plane (\ref{sec:polar vs equatorial}). These results establish the polar-kick scenario as the most likely explanation for the observed high-e systems.

   \vspace{12px}
\noindent \textit{\textbf{3. Strong evidence in favor of a new multimodal kick distribution}}:
The picture that emerges from our analysis of Be X-ray binaries is a multimodal distribution of kick velocities comprising two new components (I) \& (II) derived from the low-e group and high-e sequence in addition to the classical high-velocity component derived from isolated pulsar studies (III): 
\begin{align*} \label{eq:three kick components}
 v \sim &
  \setlength{\arraycolsep}{0pt}
  \renewcommand{\arraystretch}{1.2}
  \left\{\begin{array}{l @{\quad} l @{\quad}  c}
       \rm{Maxwellian}\,(\sigma=5\,\kms) & \rm{isotropic}, & (I) \\
       {\rm Normal}\,(\mu=100\,\kms, \sigma=11\,\kms)& \rm{polar}\,(\alpha = 5^{\circ}),  &(II)\\
       {\rm Maxwellian}\,(\sigma \simeq 300 \,\kms)  &\rm{isotropic}. & (III)
  \end{array}\right.
\end{align*}
The two new components (I) and (II) provide an alternative to a single ``electron capture'' Maxwellian, which can be considered state-of-the-art in the field. Our new model, by design, reproduces the observed dichotomy, which is where the current state-of-the-art fails. We find decisive evidence (Jeffreys' scale, Bayes Factor of $10^7$) favoring our new model over the ``state-of-the-art'' approach (\ref{sec:state-of-the-art vs preferred model}). 

In sec.~\ref{sec:prior work}, we compare with previous constraints inferred from a range of systems, adopting different methods, and observables. While these earlier results may appear inconsistent with one another, they probe different regimes of kick velocities. Combining all these constraints, a coherent picture emerges that supports our new three-component model.

\subsection*{Additional support from independent observables}
The multi-modal kick distribution we infer has implications for a wide range of astrophysical systems. We test the consequences of our findings and find consistency across several independent observables, providing compelling support for our conclusions. We discuss the most illustrative cases, with further examples discussed in sec.~\ref{sec:independent observables}.

\begin{figure*}
    \centering
    \includegraphics[width=0.33\textwidth]{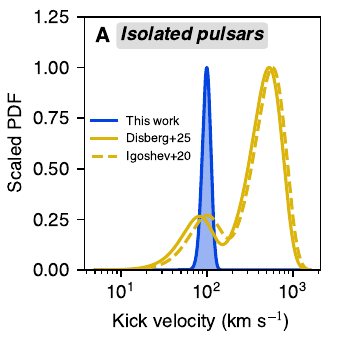}%
    \includegraphics[width=0.33\textwidth]{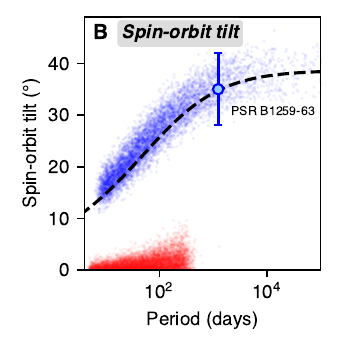}%
    \includegraphics[width=0.33\textwidth]{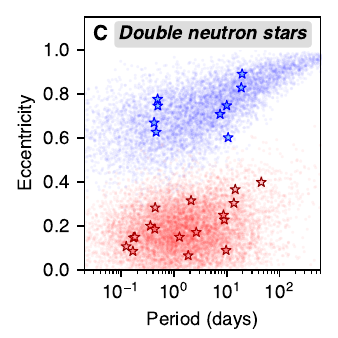}%
    \caption{\textbf{Independent observables.} We test our multi-component kick model by exploring the implications for several independent observables. We demonstrate consistency between the predictions from our model and observations across a variety of astrophysical objects affected by kicks, providing additional support for our findings.
    (\textbf{A}) The higher kicks component in our model disrupts wider binaries and thus predicts isolated pulsars with velocities of about 100\,\kms. This coincides with the lower velocity components observed in very young pulsars \cite{igoshev_observed_2020, disberg_kinematically_2025}, cf.~\ref{sec:isolated pulsars}. 
    (\textbf{B}) The polar kicks from our higher-kick component predicts spin-orbit tilts that  match the tilt measurement of PSR B1259-63 \cite{shannon_kinematics_2014}, cf.~\ref{sec:spin-orbit misalignment}.
    (\textbf{C}) The orbital properties of double neutron stars is naturally reproduced by our model when we fit for the pre-supernova mass and period distributions, cf.~\ref{sec:dns}.}
    \label{fig:additional_predictions}
\end{figure*}

\vspace{12px}
\noindent {\textit{\textbf{Velocities of isolated pulsars:}}}
    The kicks inferred from the high-e sequence of about $100\,\kms$ are strong enough to unbind binaries with sufficiently wide orbits. These systems should produce a population of isolated pulsars that escape at a velocity comparable to the received kicks. In Fig.~\ref{fig:additional_predictions}A, we show recent measurements of the velocity distribution of young isolated pulsars \cite{igoshev_observed_2020}. The distribution is dominated by a high-velocity component similar to \cite{hobbs_statistical_2005}, but also shows a lower velocity component peaking at $100\,\kms$. It has been questioned whether this feature is due to low-number statistics \cite{disberg_kinematically_2025}, but the remarkable match to the kick velocity we derive independently for the high eccentricity sequence of Be X-ray binaries supports the notion that this feature is real (\ref{sec:isolated pulsars}). 

\vspace{12px}
    \noindent {\textit{\textbf{Spin-orbit misalignment:}}}
    An out-of-plane supernova kick will tilt the orbit with respect to the original orbital plane. The Be star's spin remains unaffected and can still be used as a tracer of the pre-explosion orbit.  For the Be X-ray binary PSR B1259-63 (a member of the high-e sequence), the spin-orbit misalignment has been measured to  $\theta=35^{\circ} \pm 7^{\circ}$  \cite{shannon_kinematics_2014}. It is the only Be X-ray system with such a measurement to date, but it is in excellent agreement with our predictions for the polar kicks we inferred, as shown in blue in Fig.~\ref{fig:additional_predictions}B. For systems in the low-e group we expect no significant misalignment (\ref{sec:spin-orbit misalignment}).

\vspace{12px}
 \noindent {\textit{\textbf{Orbits of double neutron stars:}}}
    Our two new modes of kicks can also account for the properties of double neutron star (DNS) systems \cite{tauris_formation_2017}, whose population also appears to be divided in two groups \cite{beniamini_formation_2016}. To model the second supernova in a DNS systems, we set the companion mass to the canonical neutron star value of $1.4\,\Msun$ and leave the progenitor mass and orbital period distributions free. Crucially, we adopt the same kick velocities and directions inferred from Be X-ray binaries, using their posterior medians. As shown in Fig.~\ref{fig:additional_predictions}C, these kicks reproduce the two observed branches in the eccentricity–period plane of DNS systems. The fit favors lower ejecta masses and shorter pre-supernova periods compared to Be X-ray binaries, consistent with expectations for DNS progenitors (\ref{sec:dns}).

\subsection*{Conclusions and implications}
Our results reveal a richer diversity in the outcomes of core-collapse supernovae than previously recognized.  We identify two new distinct types of explosions: one producing neutron stars with ultralow kicks (under $10\,\kms$) and low ejecta masses, and another with polar-aligned kicks, narrowly distributed around $100\,\kms$.

The ejecta masses inferred from our analysis provide direct, independent evidence that Be X-ray binaries originate from stripped-envelope supernovae \cite{van_den_heuvel_centaurus_1972}.
The bimodality in ejecta masses may reflect whether the progenitor underwent an additional phase of mass stripping, which occurs preferentially in less massive stripped stars \cite{laplace_expansion_2020,ercolino_mass-transferring_2024}, thus plausibly creating a gap in the pre-supernova mass distribution.
The low ejecta masses inferred for the ultralow kick systems are expected to result in fast and faint explosions that are likely to escape detection in standard transient surveys, unless they are optimized for high-cadence searches \cite{drout_rapidly_2014}.

These new kick modes have important consequence for a wide variety of questions in astrophysics. Kicks determine which binary systems and even higher-order multiple systems remain bound after a supernova and set the systemic velocities of those that survive. They contribute to neutron-star retention in globular clusters \cite{pfahl_comprehensive_2002}, influence the location of supernova explosions \cite{wagg_delayed_2025}, and affect the contribution of massive binaries to chemical enrichment in galaxies \cite{beniamini_natal_2016}.
Crucially, the classical high-velocity kicks disrupt binaries so effectively that nearly all bound systems containing a neutron star are expected to originate from one of the two low-kick modes identified here. As a consequence, the two low-kick modes govern the formation of X-ray binaries and double neutrons stars, and set the delay time of gravitational wave sources and associated electromagnetic transients \cite{giacobbo_impact_2019}.

At a deeper level, these discoveries provide insight into the complexity of core-collapse supernovae. The ultralow kick velocities we infer are significantly smaller than those reported in supernova simulations for this type of progenitors. Producing such weak kicks requires the suppression of not only hydrodynamical asymmetries, but also anisotropic neutrino emission \cite{janka_interplay_2024}.
The polar-aligned, narrow-velocity kicks we identify pose a distinct challenge, since they are not reproduced by current 3D supernova simulations \cite{powell_three_2023,janka_interplay_2024}. Although aligned kicks could, in principle, be imparted by asymmetric spin-down radiation \cite{lai_orbital_1996,hirai_neutron_2024}, 
this would require neutron-star birth spin periods of about $1\,\rm{ms}$, much shorter than those inferred from both models and observations \cite{du_initial_2024,ma_angular_2019}.

How the collapse of massive stellar cores leads to an explosion has remained one of the central challenges in theoretical and computational astrophysics for decades. The existence of an ultralow kick mode of neutron star formation, together with evidence for preferentially polar-aligned kicks, pose significant and unexpected questions for our current understanding of the death of massive stars.

\section*{Acknowledgments}

The authors thank Dietrich Baade, Julia Bodensteiner, Andrea Ercolino, Alessia Franchini, Jim Fuller, Marat Gilfanov, Matteo Guardiani, Andrei Igoshev, Thomas Janka, Harim Jin, David Kaltenbrunner, Jakub Klencki, Michael Kramer, Hans Ritter, Jakob Stegmann and Irene Tamborra for useful discussions.

\paragraph*{Funding:}
C.J.~acknowledges funding from the Royal Society through the Newton International Fellowship funding scheme (project No. NIF$\backslash$R1$\backslash$242552).
D.K.~acknowledges support by the German Research Foundation (DFG) through the Collaborative Research Center “Neutrinos and Dark Matter in Astro- and Particle Physics (NDM),” Grant No. SFB-1258-283604770. 
\paragraph*{Author contributions:} R.V. contributed to conceptualization, data  curation, formal analysis, methodology, software, supervision, visualization, writing of the original draft, review and editing.
S.E.D.M. contributed to conceptualization, funding acquisition, methodology, project administration, supervision, visualization, writing of the original draft, review and editing.
S.J. contributed to conceptualization, methodology, writing, review and editing.
T.C., C.J., D.K., N.L., A.C.R., A.V.-G., C.W. contributed to the conceptualization, review and editing.
\paragraph*{Competing interests:} The authors have no competing interests to declare.
\paragraph*{Data and materials availability:} All data are available in the manuscript or the supplementary materials. The code used to conduct the analysis will be made available as a public repository on Zenodo upon publication.

\newpage
\renewcommand\thesection{S\arabic{section}}
\setcounter{figure}{0}
\renewcommand{\thefigure}{S\arabic{figure}}
\renewcommand{\thetable}{S\arabic{table}}
\renewcommand{\theequation}{S\arabic{equation}}
\renewcommand{\thepage}{S\arabic{page}}

\renewcommand{\contentsname}{Contents Supplementary Materials}
\tableofcontents

\newpage
\section{Data} \label{sec:data}

The data we use in this study are the orbital parameters of a sample of observed Be X-ray binaries. In this section, we first discuss why Be X-ray binaries are suitable for studying kicks in supernovae (Sec.~\ref{sec:Be X-ray binaries}), what their typical properties are (Sec.~\ref{sec:observational properties}), and how they are detected and characterized. In sec.~\ref{sec:selection criteria} and \ref{sec:two groups in the sample}, we then present an overview of the sample we use in this study, and state our selection criteria, followed by  
a discussion and assessment of the possible biases present in the sample (Sec.~\ref{sec:selection biases}).

\subsection{Be X-ray binaries} \label{sec:Be X-ray binaries}

\begin{figure}
    \centering
    \includegraphics[width=\textwidth]{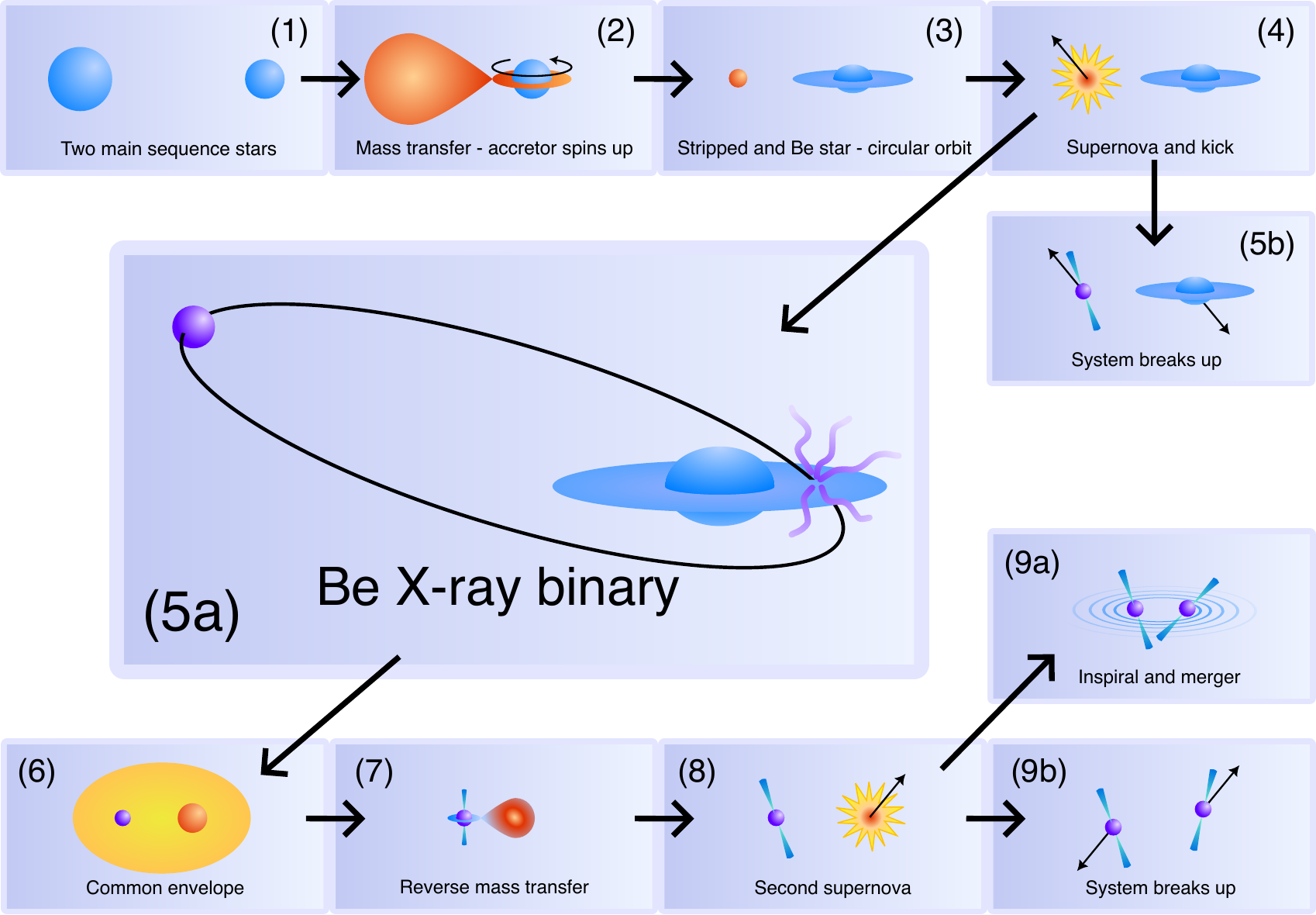}
    \caption{\textbf{Past and future evolution of a Be X-ray binary}. We depict the evolutionary history of a Be X-ray binary and one possible future evolution into a double neutron star \cite{tauris_formation_2017}.}
    \label{fig:evolution_cartoon}
\end{figure}

Be X-ray binaries are a class of high-mass X-ray binaries, consisting of a compact object---typically a neutron star---and a classical Be star (see central panel of fig.~\ref{fig:evolution_cartoon} for a cartoon depiction). The neutron star is the compact remnant formed from the collapsed core left behind by a massive star after a supernova explosion. Classical Be stars \cite{rivinius_classical_2013, rivinius_classical_2024} are rapidly rotating main sequence early-type stars with an equatorial viscous decretion disk. When the neutron star accretes material from the disk, X-rays are produced, often showing pulsations, that can be observed and used to characterize the orbit.

Be X-ray binaries are ideal systems for studying supernova kicks because of their properties and formation history.
The most widely accepted formation scenario involves isolated binary evolution \cite{van_den_heuvel_centaurus_1972,rappaport_x-ray_1982,verbunt_origin_1993,pols_formation_1991, vinciguerra_be_2020, misra_x-ray_2023, rocha_be_2024, liu_multiwavelength_2024}, as illustrated in  fig.~\ref{fig:evolution_cartoon}.
The progenitor of the neutron star is initially the most massive star in the system (panel 1 in fig.~\ref{fig:evolution_cartoon}), which evolves on a shorter timescale than the less massive companion star. It expands, fills its Roche lobe, and initiates mass transfer (panel 2). The process strips the progenitor of most of its hydrogen envelope, circularizes the orbit due to strong tidal forces, and spins up the companion \cite{packet_spin-up_1981,de_mink_rotation_2013}. The additional angular momentum facilitates the mass loss events that lead to the formation of the Be star's decretion disk. After mass transfer, the system consists of a Be star and a stripped star in an almost circular orbit (panel 3).

Depending on the orbital separation and the further evolution of the stripped star, the system may undergo a second mass transfer phase, but ultimately the stripped star explodes as a supernova (panel 4) and forms a neutron star. Asymmetries in the supernova explosion impart a natal kick to the newly formed neutron star, which, combined with the sudden mass loss, can significantly perturb the orbit of the system. If the explosion does not unbind the system, the neutron star may accrete material from the disk of the Be star, resulting in emission of X-rays. The system will then appear as a Be X-ray binary (panel 5a).

The future evolution can take different paths. Figure~\ref{fig:evolution_cartoon} depicts a possible scenario \cite{tauris_formation_2017}, where the Be star eventually fills its Roche lobe, but instead of undergoing stable mass transfer again, it triggers a common-envelope or unstable mass transfer phase due to the large mass ratio between the two stars (panel 6). If the envelope is successfully expelled, the system transitions into a short-period binary with a neutron star and a stripped star companion. The stripped star may undergo reverse mass transfer (panel 7), further reducing its mass \cite{dewi_evolution_2002,dewi_late_2003}, before it explodes as an ultra-stripped supernova \cite{tauris_ultra-stripped_2015}, creating a second neutron star (panel 8).

If the system survives this second supernova kick, the two neutron stars’ orbits gradually decay due to gravitational wave emission, leading to an eventual inspiral and merger (panel 9a). Such events, observable in both electromagnetic radiation and gravitational waves, provide a unique opportunity to study the physics of compact objects \cite{abbott_multi-messenger_2017}.

\subsection{Observation and characterization} \label{sec:observational properties}

X-ray observations have been crucial for the discovery and characterization of Be X-ray binaries.  Some systems show persistent low luminosity X-ray emission ($L_X \sim 10^{34}\,\rm{erg}\,\rm{s}^{-1}$), while others alternate between outbursts and quiescent phases.
The outbursts can be divided into two classes. Normal, or Type I outbursts ($L_X \sim 10^{36}-10^{37}\,\rm{erg}\,\rm{s}^{-1}$) occur close to the periastron passage, often happening in series and lasting a small fraction of the orbital period.
Giant or Type II outbursts ($L_X > 10^{37}\,\rm{erg}\,\rm{s}^{-1}$) generally start shortly after a periastron passage, can last for multiple orbital periods and are typically an isolated occurrence.
These peculiar behaviors stem from the complex interactions between the neutron star and the Be disk \cite{okazaki_natural_2001, franchini_type_2019, martin_frequency_2019}.
Moreover, Be stars are often highly variable, and can grow their disk or disperse it completely in a matter of weeks to months \cite{haubois_dynamical_2012}; some Be X-ray binaries can remain quiescent for decades between outbursts.

Once an outburst is detected and X-ray pulses are identified, pulse timing analysis can be used to accurately measure the orbit of a Be X-ray binary, as the binary motion causes Doppler shifts in the pulse frequency.
However, this technique requires multiple observations with good coverage of the orbital phase.
The transient nature of these sources further complicates the scheduling of the observations, as they may not be visible for most of the time. As a result, orbital periods and eccentricities have been determined for only about 30 Be X-ray binaries using this method. Despite these challenges, the technique provides very precise measurements, typically determining the orbital period within one percent and the eccentricity within an absolute uncertainty of less than 0.05 \cite{doroshenko_orbit_2018}.

\subsection{Sample selection} \label{sec:selection criteria}
This work is based on a sample of Be X-ray binaries taken from the high-mass X-ray binary (HMXB) catalog by \cite{fortin_catalogue_2023}.
We included in our sample all the systems that satisfy the following criteria:
\begin{itemize}
    \item \textbf{Presence of a neutron star.} We select systems where there is evidence for the presence of a neutron star, for example, from X-ray pulsations. We excluded those systems for which the nature of the companion is not known, and could be a white dwarf or hot sub-dwarf, which means that no supernova explosion has taken place in the system.

    \item \textbf{Presence of a Be star.}  We select systems containing a classical Be star, as they are typically well within their Roche-lobe and the orbit is not expected to have evolved significantly since the supernova (see sec.~\ref{sec:postsn orbit did not change}). In contrast, X-ray binaries with a giant donor are near Roche lobe filling, and their orbit is affected by tides.

    \item \textbf{Availability of accurate orbital parameters.} We carefully vetted each system in our sample to ensure that the measurements of period and eccentricity are of high quality and reliable: we verified that the orbital parameters are from a sufficient number of epochs (at least $15$, usually more than $30$) and with a well-defined solution.
    
    \item \textbf{They are in the Galaxy.} To keep the sample homogeneous, we focused on the systems in our galaxy, and did not include Be X-ray binaries in the Small Magellanic Cloud (SMC) in our main analysis. Systems in the SMC are considerably more distant, and harder to characterize. Additionally, the lower metallicity can influence the properties of the progenitors, for instance, by leaving a more massive envelope on the stripped stars \cite{gotberg_ionizing_2017,yoon_type_2017}. See also sec.~\ref{sec:smc}.
\end{itemize}
The sample of selected systems is reported in Tab.~\ref{tab:galacticBeX}.
Here, we list the specific corrections and updates we made to the base catalog from \cite{fortin_catalogue_2023} version 2024-08.

\begin{figure}
    \centering
    \includegraphics{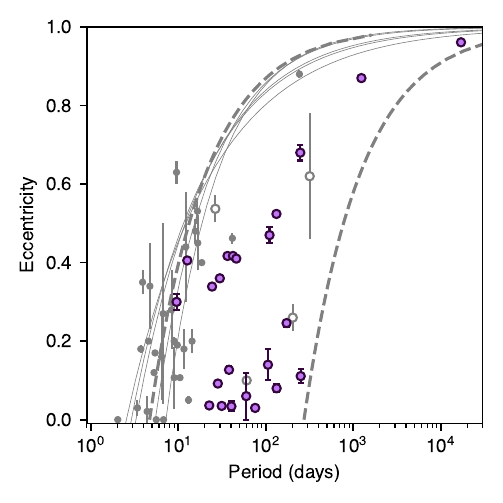}
    \caption{\textbf{Eccentricities and periods of the Be X-ray binaries in our sample.} The purple dots indicate the Be X-ray binaries in our sample. For comparison, we show the known position in the diagram of all the other HMXBs in the Galaxy\cite{fortin_catalogue_2023}. Empty circles indicate Be X-ray binaries that do not have a neutron star companion, or for which the nature of the compact companion is unknown. Gray dots indicate HMXBs that do not host a Be star, or for which the nature of the visible star is unknown.  Gray solid lines are tidal envelopes from \cite{bashi_features_2023}, gray dashed lines are lines of constant periastron distance, corresponding, from left to right, to $r_{\rm peri} = 30\,\Rsun$ and $450\,\Rsun$, for a system composed of a $15\,\Msun$ Be star and a $1.4\,\Msun$ neutron star.}
    \label{fig:elogPsample}
\end{figure}

\begin{itemize}
    \item \textbf{2S 1417-264}: updated the value of eccentricity from $e=0.4169$ \cite{finger_reappearance_1996} to the new measurement $e=0.446$ \cite{raichur_apsidal_2010}.
    \item \textbf{2S 1553-542}: adopted the measurement of period and eccentricity by  \cite{tsygankov_nustar_2016} not reported in the base catalog.
    \item \textbf{EXO 2030+375}: we update the values of period and eccentricity from $P=(46.0214 \pm 0.0005)\,\rm{d}$ and $e=0.419 \pm 0.002$ \cite{wilson_decade_2002} to the new measurements $P=(46.02217 \pm 0.00035)\,\rm{d}$ and $e=0.4102 \pm 0.0008$ \cite{fu_timing_2023}.
    \item \textbf{4U 2206+543}: the uncertainty on the orbital period was incorrectly reported as $0.04$ days in \cite{fortin_catalogue_2023}, while the measured value is $0.004$ days \cite{stoyanov_orbital_2014}. This system's optical component presents peculiarities that are unusual for Be stars, but it does show evidence of fast rotation and mass loss through a rotationally supported structure \cite{blay_multiwavelength_2006}.
    \item \textbf{SAX J0635.2+0533}: the orbital parameters of this system were deduced based on only 7 data points spanning less than one orbital period \cite{kaaret_x-ray_2000}. Therefore, we deemed the quality of the orbital solution insufficient and decided to exclude it from the sample. However, we note that with the reported $P=11.2\,\rm{d}$ and $e=0.29$ it would have followed the trend of the other systems in high-e sequence, had we decided to include it.
    \item \textbf{4U 1145-619}: the eccentricity of $e=0.8$ is often reported in the literature without any uncertainty, but it is only a rough estimate based on the accretion luminosity variations \cite{watson_2s_1981}. We did not find a more recent estimate of the eccentricity of the system and we deemed the quality of the original estimate to be insufficient. We decided to exclude it from our sample.
    \item \textbf{LS 5039}: The catalog by \cite{fortin_catalogue_2023} classifies it as a Be star with unknown companion. While the compact object is likely to be a neutron star \cite{yoneda_sign_2020,makishima_further_2023}, the spectrum of the optical component does not show evidence of a Be disk \cite{mcswain_n_2004} and the extremely short orbital period (3.9 days) would hardly leave space for a disk to form. Indeed, the orbit of the system has likely been affected significantly by tides since the supernova, and we have therefore excluded it from the sample.
    \item \textbf{HD 259440}: A gamma ray source composed of a Be star and a companion that is likely a neutron star, but no X-ray pulses have ever been detected. Authors that have studied it report very different orbital solutions, with eccentricity ranging from $0.4$ to $0.83$ \cite{casares_binary_2012,moritani_orbital_2018,adams_observation_2021,matchett_new_2025}. Given the uncertainty on the orbital configuration, we have decided to exclude it from the sample. If we were to include it, it would likely sit in the long period tail of the upper branch.
    \item \textbf{RX J2030.5+4751}: The catalog by \cite{fortin_catalogue_2023} erroneously reports the eccentricity and period of EXO 2030+375 \cite{sidoli_integral_2018}. To our knowledge, the period and eccentricity of this system has not been measured \cite{harvey_galactic_2022}.
\end{itemize}

\begin{table}
    \centering
    \caption{The sample of Be X-ray binaries considered in this work.}
    \begin{tabular}{|l|cc|l|}
    \hline
        Name & Period (day) & Eccentricity & Refs. \\ \hline
        \multicolumn{4}{|c|}{Low-e group} \\
        \hline
        Swift J0243.6+6124 & $28.3\pm0.2$ & $0.092\pm0.007$ & \cite{doroshenko_orbit_2018,reig_optical_2020} \\ 
        X Per & $250.3\pm0.6$ & $0.111\pm0.018$ & \cite{delgado-marti_orbit_2001,zamanov_spectral_2019} \\ 
        SGR 0755-2933 & $59.69\pm0.15$ & $0.06\pm0.06$ & \cite{doroshenko_sgr_2021,richardson_high-mass_2023} \\ 
        GS 0834-430 & $105.8\pm0.4$ & $0.14\pm0.04$ & \cite{wilson_sequence_1997,israel_discovery_2000} \\ 
        XTE J1543-568 & $75.56\pm0.25$ & $<0.03$ & \cite{int_zand_discovery_2001} \\ 
        2S 1553-542 & $31.303\pm0.027$ & $0.0351\pm0.0022$ & \cite{tsygankov_nustar_2016, lutovinov_2s_2016} \\ 
        SWIFT J1626.6-5156 & $132.89\pm0.03$ & $0.08\pm0.01$ & \cite{baykal_orbital_2010,reig_multi-frequency_2011} \\ 
        XTE J1859+083 & $37.97\pm0.09$ & $0.127\pm0.009$ & \cite{bissinger_observations_2016,salganik_nature_2022} \\ 
        4U 1901+03 & $22.5827(2)$ & $0.0363\pm0.0003$ & \cite{galloway_discovery_2005,strader_optical_2019} \\ 
        XTE J1946+274 & $172.7\pm0.6$ & $0.246\pm0.009$ & \cite{verrecchia_identification_2002,ozbey_arabaci_detection_2015} \\ 
        KS 1947+300 & $40.415\pm0.010$ & $0.034\pm0.013$ & \cite{negueruela_bex-ray_2003,galloway_frequency_2004} \\ 
        \hline
        \multicolumn{4}{|c|}{high-e sequence} \\
        \hline
        4U 0115+634 & $24.3174(4)$ & $0.339\pm0.005$ & \cite{negueruela_bex-ray_2001,raichur_apsidal_2010}\\ 
        V 0332+53 & $36.5\pm0.29$ & $0.417\pm0.007$ & \cite{negueruela_bex-ray_1999,raichur_apsidal_2010} \\ 
        1A 0535+262 & $110.3\pm0.3$ & $0.47\pm0.02$ & \cite{finger_quasi-periodic_1996,wang_constraints_1998,okazaki_natural_2001} \\ 
        GRO J1008-57 & $247.8\pm0.4$ & $0.68\pm0.02$ & \cite{coe_discovery_1994,coe_now_2007} \\ 
        GX 304-1 & $132.189\pm0.022$ & $0.524\pm0.007$ & \cite{parkes_shell_1980,sugizaki_luminosity_2015}\\ 
        PSR B1259-63 & $1236.724526(6)$ & $0.86987970(6)$ & \cite{negueruela_astrophysical_2011,shannon_kinematics_2014} \\ 
        2S 1417-624 & $42.12\pm0.03$ & $0.4169\pm0.0033$ & \cite{grindlay_optical_1984,finger_reappearance_1996,raichur_apsidal_2010}\\ 
        GRO J1750-27 & $29.806\pm0.001$ & $0.360\pm0.002$ & \cite{scott_discovery_1997,shaw_accretion_2009} \\ 
        PSR J2032+4127 & $17000\pm1000$ & $0.961\pm0.002$ & \cite{lyne_binary_2015,ho_multiwavelength_2017} \\ 
        EXO 2030+375 & $46.02217(35)$ & $0.4102\pm0.0008$ & \cite{motch_accretion_1991,fu_timing_2023} \\ 
        SAX J2103.5+4545 & $12.66536(88)$ & $0.4055\pm0.0032$ & \cite{baykal_timing_2007,reig_discovery_2004} \\ 
        4U 2206+543 & $9.558\pm0.004$ & $0.30\pm0.02$ & \cite{blay_multiwavelength_2006,stoyanov_orbital_2014}\\ \hline
    \end{tabular}
    \label{tab:galacticBeX}
\end{table}

\subsection{Definition of the two eccentricity groups} \label{sec:two groups in the sample}
We divide this sample into two groups based on eccentricity: a low-eccentricity group (low-e, $e \le 0.25$) and a high-eccentricity group (high-e, $e > 0.25$). This separation is visually striking in the period-eccentricity plane and was already recognized when the sample was half its current size, two decades ago \cite{pfahl_new_2002}, although no further studies have investigated the matter in detail. We also recover this division using the Spectral Clustering algorithm from the \texttt{scikit-learn} library \cite{yu_multiclass_2003, pedregosa_scikit-learn_2011, damle_simple_2019}. However, we note that the irregular shape of the high-eccentricity group makes it difficult to recover with less sophisticated clustering algorithms, such as K-means or Gaussian mixtures. 

\subsection{Selection biases} \label{sec:selection biases}
Most systems were discovered serendipitously during bright outbursts. The availability of well-measured parameters relies on follow-up observations, which have not been carried out systematically and homogeneously across the sample.
Consequently, the sample is affected by selection biases, and here we discuss the most relevant ones.

\begin{itemize}
    \item The X-ray luminosity is larger for larger mass loss from the Be star, and is smaller for larger distance from the observer and periastron separation \cite{brown_modelling_2019}. This favors massive Be stars, which tend to feed the disk at a higher rate \cite{granada_populations_2013,vieira_life_2017}, close by systems, and systems with a periastron distance smaller than a few hundred solar radii. 
    
    \item We only observe systems that are still bound, favoring short periods, and low kick velocities.
    
    \item Systems that have a periastron distance below a few times the stellar radius do not have space to form a disk, or the disk is tidally truncated \cite{panoglou_be_2016,panoglou_be_2018}. Additionally, the Be star would be quickly spun down by tides.
    
    \item To measure the orbital properties of a system, it needs to remain X-ray bright for a significant fraction of its orbit, or to have frequent outbursts. This selects for systems that have a high duty cycle, and are X-ray bright most of the time.
\end{itemize}

We show the effect of some of the above selection biases in fig.~\ref{fig:elogPsample}, where we plot the period and eccentricity of our sample of Be X-ray binaries and compare it with the wind-fed supergiant X-ray binaries. The wind-fed systems are located close to the tidal envelope, the line below which the tidal circularization time is much longer than the lifetime of the systems \cite{bashi_features_2023}, and therefore their orbit may have been significantly altered by tides since the supernova. In contrast, Be X-ray binaries are typically found below this line.
At long orbital periods instead, the density of the Be disk drops and the accretion luminosity becomes fainter. Indeed, no system is observed with a periastron distance larger than about $450\,\Rsun$.

\begin{figure}[ph!]
    \centering
    \includegraphics{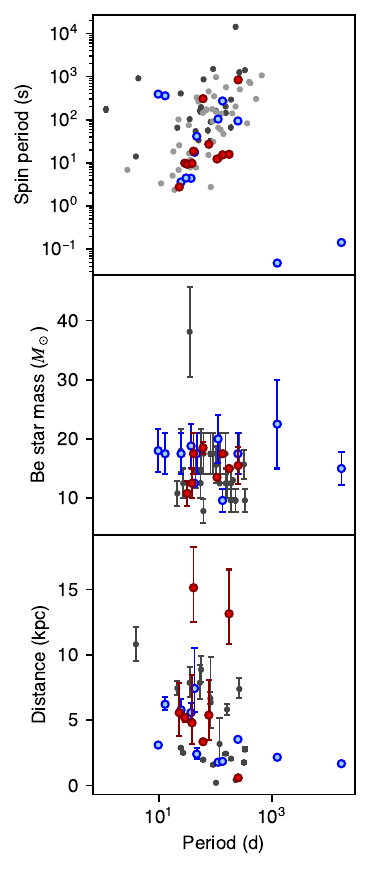}
    \includegraphics{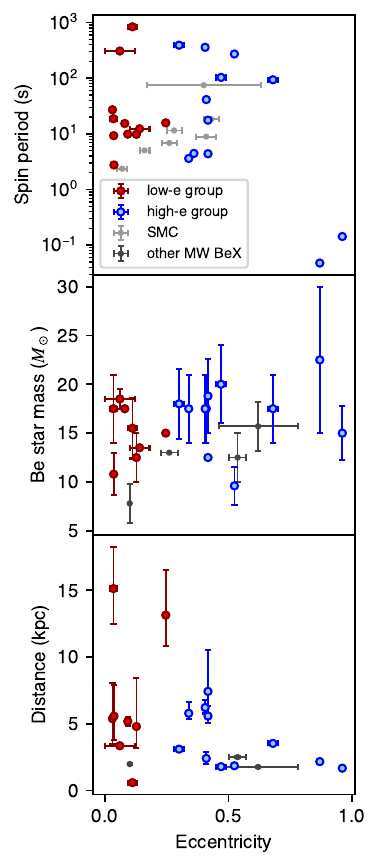}
    \caption{\textbf{Overview of the properties of the Be X-ray binaries included in our sample.} The low-e group is shown in red and high-e sequence in blue. For comparison, we also show the Galactic Be X-ray binaries that are not included in our sample (typically because the eccentricity is not known) and the Be X-ray binaries in the SMC \cite{coe_catalogue_2015}.  We show the spin period, Be star mass and distance as a function of orbital period (left column) and eccentricity (right column). For the Be star masses that are derived from spectral types, we display a 20\% uncertainty. We conclude that there are no unexpected trends that could suggest that the separation of our sample in the low-e and high-e groups could be explained by a selection effect.}
    \label{fig:trends_in_observations}
\end{figure}

\paragraph{Investigating selection biases as the origin of the two groups}
We investigate the properties of our sample in  fig.~\ref{fig:trends_in_observations}, to ensure that there are no evident trends that could explain the separation in two groups with a selection bias. We show the two groups in our sample against the other Galactic Be X-ray binaries \cite{fortin_catalogue_2023} and the Be X-ray binaries in the SMC \cite{coe_catalogue_2015}. We look for trends in the distance from the systems, Be star masses, and pulsar spin periods. We find no significant difference between the two groups in the sample and between the groups and the other Be X-ray binary populations. The known correlation between neutron star spin period and orbital periods \cite{corbet_beneutron_1984} extends over both groups and does not cause a bias that could explain the split in the two groups.

\begin{figure}[]
    \centering
    \includegraphics{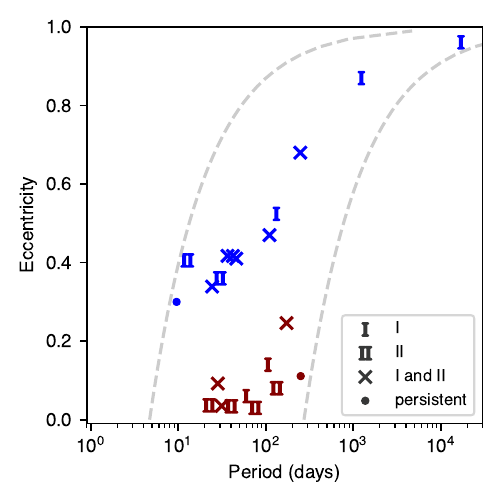}
    \caption{\textbf{Outburst behavior of the Be X-ray binaries in our sample.} The outburst types do not seem to show a preference for either of the two groups we consider.}
    \label{fig:outburst_type}
\end{figure}

In fig.~\ref{fig:outburst_type} we show the type of outburst displayed by each system in our sample. We classify them based on visual inspection of the systems X-ray light curves, noting that the distinction is not always clear-cut, and this introduces some degree of subjectivity. Systems experiencing Type I or Type II outbursts are evenly distributed among the two groups, excluding the possibility that the separation in two groups arises from a selection bias induced by a different outburst behavior.

\newpage

\section{Methods} \label{sec:methods}
\begin{figure}
    \centering
    \includegraphics[width=\textwidth]{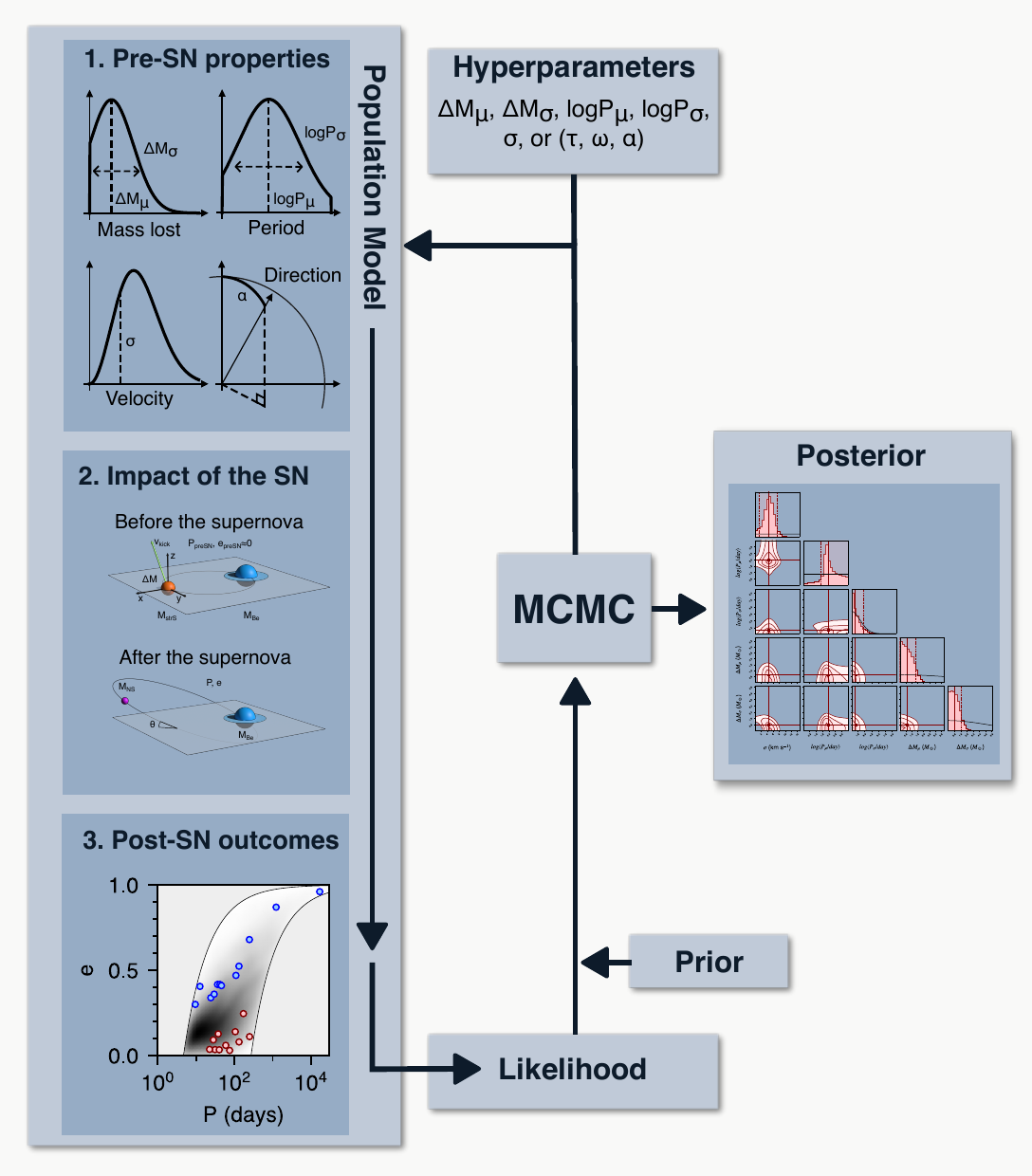}
    \caption{\textbf{Illustration of the methods.} On the left, the population model: starting from distributions of pre-supernova properties, we calculate how supernova kicks and mass loss affect the orbital parameters. We then compare the resulting post-supernova period-eccentricity distribution to observations. This population model is integrated into an MCMC framework to infer the posterior distributions of the hyperparameters that describe the pre-supernova conditions and kick properties.}
    \label{fig:methods_schematics}
\end{figure}

To constrain the properties of supernova kicks, we model a population of binary stars and compute the effect of different supernova kick distributions on their orbital configuration.
We then compare our model with the observed systems and perform a Bayesian inference with a Markov Chain Monte Carlo (MCMC) method to obtain the distribution of kick velocities and mass loss. In this section we describe the population model (\ref{sec:methods_model}), the inference method (\ref{sec:methods_inference}) and discuss the main assumptions (\ref{sec:method_model_assumptions}). A graphical overview is shown in fig.~\ref{fig:methods_schematics}.

\subsection{Modelling a Be X-ray binary population} \label{sec:methods_model}
We start with a pre-supernova population of binary systems consisting of a stripped star and a main sequence Be star, modelling the supernova explosion as an instantaneous event that removes the mass from the stripped star, turns it into a neutron star and imparts a velocity kick to it. We then calculate the resulting post-supernova orbit and apply observational selection effects to identify which systems would be detected as Be X-ray binaries.

\paragraph{Computing post-supernova orbital properties}
Multiple authors have derived the effect of supernova kicks on the orbit of a binary \cite{hills_effects_1983,brandt_effects_1995,kalogera_orbital_1996,tauris_runaway_1998,pfahl_comprehensive_2002}. We report here the equations that relate the orbital configuration before and after the supernova.

\begin{figure}
    \centering
    \includegraphics[width=0.4\textwidth]{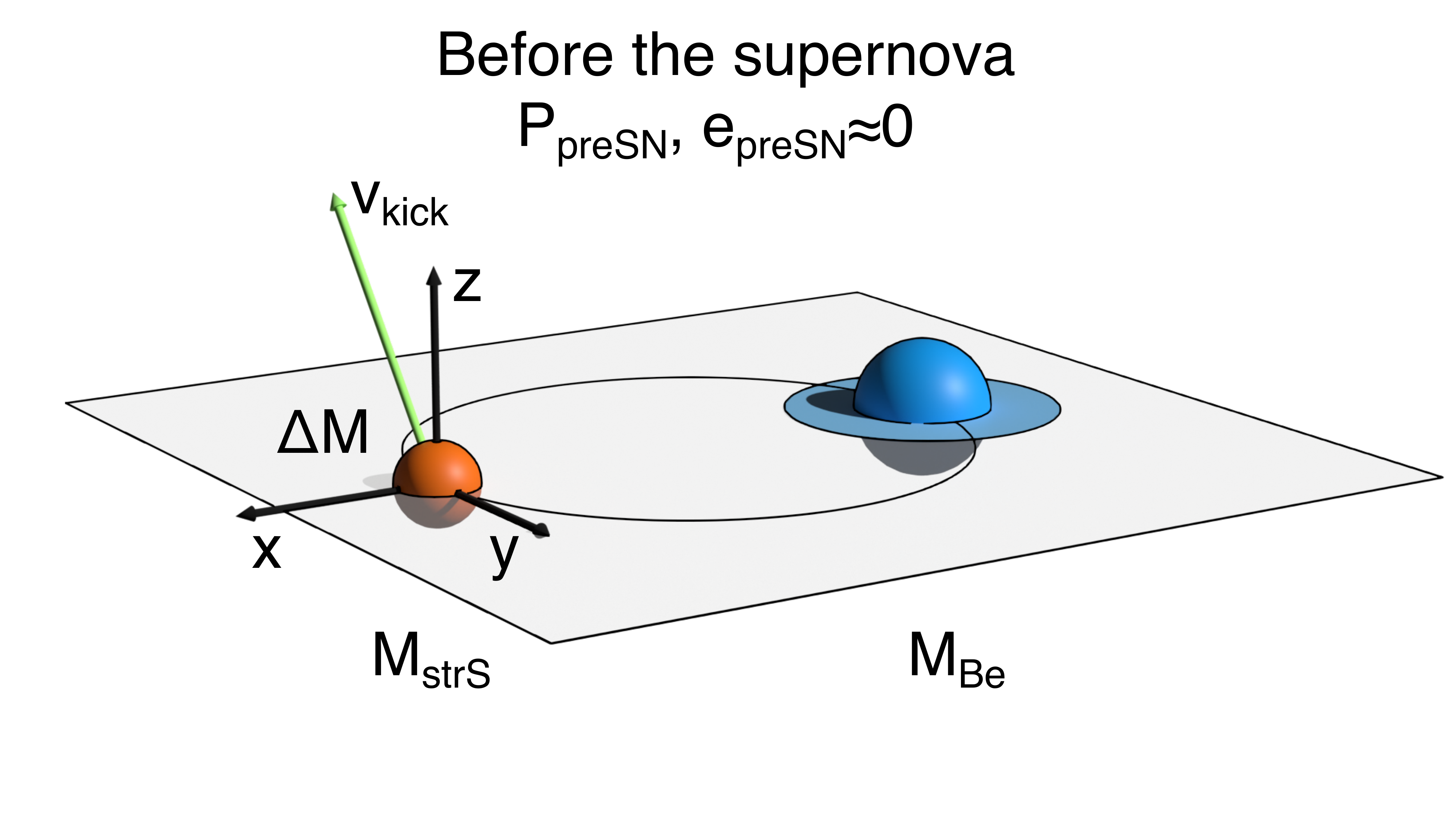}
    \includegraphics[width=0.4\textwidth]{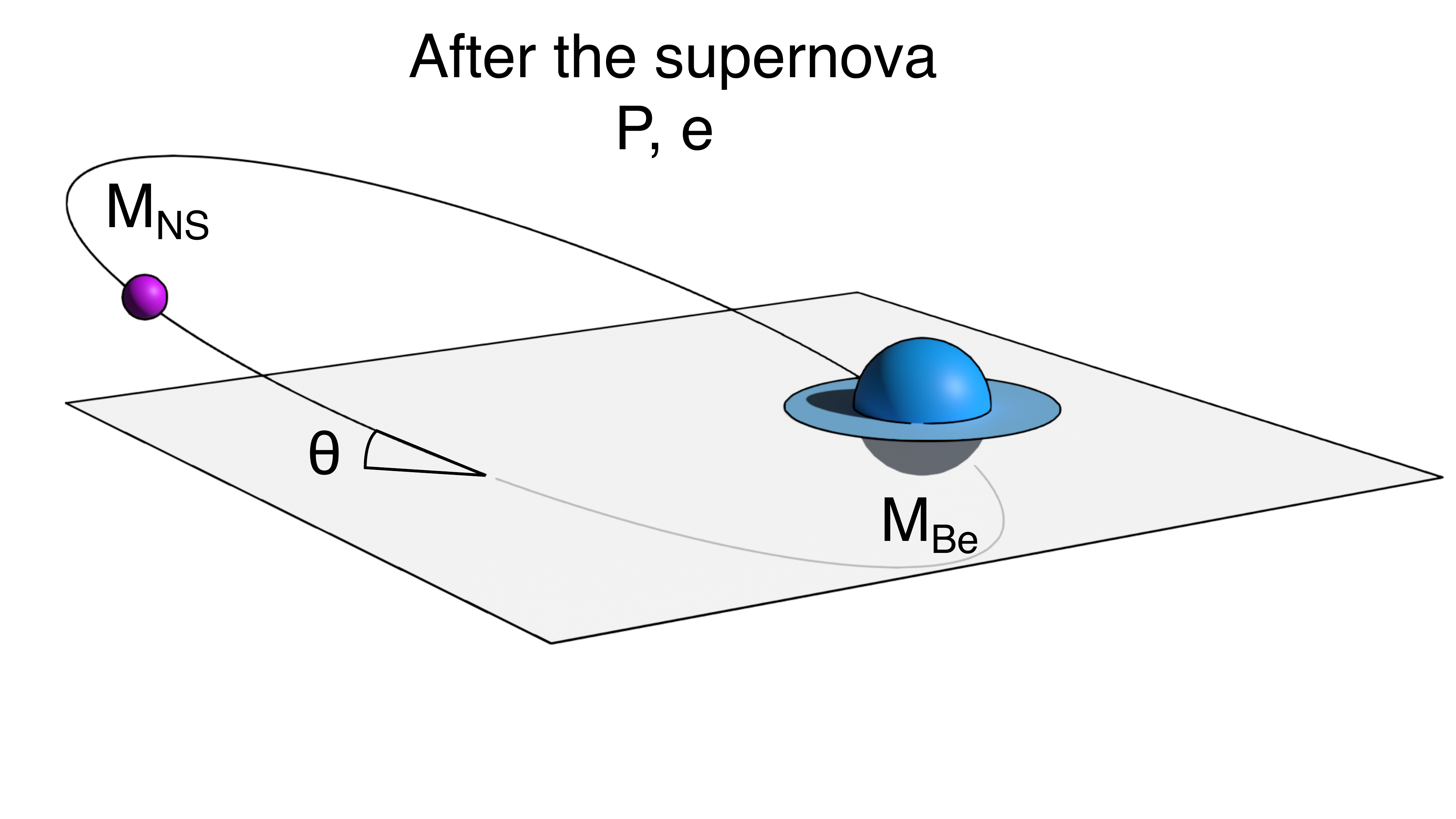}
    \caption{\textbf{Schematic of the configuration before and after the supernova.} Before the supernova (left panel), the binary is composed of a stripped star (orange) and a Be star (blue) in a circular orbit. At the moment of the explosion the stripped star loses a mass $\Delta M$ and receives a kick velocity $\bm{v}_{\rm kick}$. We describe the kick in a frame of reference centered on the exploding star, with $x$-axis along the line connecting the two stars, $y$-axis directed as the orbital velocity of the exploding star, and $z$-axis parallel to the orbital angular momentum. After the explosion (right panel), the stripped star has collapsed into a neutron star (violet), the orbit has become eccentric and tilted at an angle $\theta$ with respect to the pre-supernova orbital plane, while the Be star is unaffected.}
    \label{fig:before_after_schematics}
\end{figure}

A binary system before the supernova is characterized by the mass of the stripped star that will explode ($M_{\rm str}$), the mass of the Be star $M_{\rm Be}$ and the orbital period $P_{\rm preSN}$. We denote the total mass before the explosion as $M = (M_{\rm str} +  M_{\rm Be})$. At the moment of the explosion, the stripped star becomes a neutron star of mass $M_{\rm NS}$, losing a mass $\Delta M = M_{\rm str} - M_{\rm NS}$ in the process. We assume that the mass of the Be star is not changed by the explosion.

We show here the case of a pre-supernova circular orbit, illustrated in fig.~\ref{fig:before_after_schematics}. For eccentric orbits, see for example \cite{pfahl_comprehensive_2002} and sec.~\ref{sec:effect of initial eccentricity}. We adopt the reference frame shown in fig.~\ref{fig:before_after_schematics}, centered in the exploding star, with the $x$-axis along the line connecting the two stars, the $y$-axis pointing in the direction of the orbital motion and the $z$-axis parallel to the orbital angular momentum.
Before the supernova, the stars have a relative orbital velocity $v_r = (2 \pi G M/P_{\rm preSN})^{1/3}$, where $G$ is the gravitational constant.
The supernova imparts an instantaneous kick of velocity ${\bm v}_{\rm kick} = (v_{x},v_{y},v_{z})$ to the exploding star. We denote $V_{j} = v_{j}/v_r$. The orbital period $P$ and eccentricity $e$ after the supernova are then given by \cite{kalogera_orbital_1996}
\begin{align}
    P/P_{\rm preSN} &= \beta^{1/6}\left[2\beta - V_{x}^2 - (V_{y}+1)^2 - V_{z}^2\right]^{-3/2},\\
    1-e^2 &= \frac{V_{z}^2 + (V_{y}+1)^2}{\beta^2}\left[2\beta - V_{x}^2 - (V_{y}+1)^2 - V_{z}^2\right],
\end{align} 
where $\beta = (M-\Delta M)/M$.

For kicks constrained to the direction perpendicular to the orbital plane, or polar kicks, a simple analytical expression relates the post-supernova eccentricity and periods (de Mink et al., in prep): 
\begin{equation}
    e = \frac{p^{2/3} - \beta + 1}{p^{2/3} + \beta},
    \label{eq:ecc_from_polar_kicks}
\end{equation}
where $p = P/P_{\rm kick}$ is the post-supernova period normalized to $P_{\rm kick} = 2\pi G M/v_{\rm kick}^3$, the maximum period for which a system can remain bound under a polar kick of velocity $v_{\rm kick}$ in the absence of mass loss.

Similarly, the expression for the tilt angle $\theta$ of the orbital plane caused by the supernova is 
\begin{equation}
        \cos^2{\theta} = \frac{1+(p/\beta)^{2/3}}{1+2\,p^{2/3}\beta^{1/3}}.
        \label{eq:spin-orbit angle from polar kicks}
\end{equation}

\paragraph{Initial conditions}
We sample the pre-supernova binary systems assuming that $\log P$ follows a Normal distribution with mean $\log P_\mu$ and standard deviation $\log P_\sigma$. We truncate this distribution below 1 day (where the stars would merge) and above 3000 days (where they would not interact). For the mass $\Delta M$ lost by the exploding star we also assume a Normal distribution with mean $\Delta M_\mu$, and standard deviation $\Delta M_\sigma$. We truncate the distribution below 0. These four parameters describe the distributions from which the individual systems are sampled, and they are left free in our Bayesian analysis.
We fix the pre-supernova eccentricity to $e_{\rm preSN} = 0$, the neutron star mass to $M_{\rm NS} = 1.4\,\Msun$ and the Be star mass to $M_{\rm Be} = 15\,\Msun$, where $\Msun$ denotes the mass of the Sun. 

\paragraph{Kick distribution}
For the magnitude of the supernova kick distribution, we consider two options. For the first option, we draw the magnitude of the kick from a Maxwellian distribution \cite{hobbs_statistical_2005} described by a single parameter $\sigma$.  Alternatively, we consider a modified Maxwellian distribution that allows us to explore the effect of varying the width with two parameters:
 $\tau$, which determines the location of the peak of the distribution and $\omega$, which controls the spread. The distribution is given by
\begin{equation}
    f(x, \tau, \omega) = \frac{1}{2^{\frac{k}{2}-1} \Gamma(\frac{k}{2})} \frac{1}{s} \left(\frac{x}{s}\right)^{k-1} \exp\left(-\frac{1}{2}\left(\frac{x}{s}\right)^2\right),
    \label{eq:modified maxwellian}
\end{equation}
where $k = 1/\omega^2$, and $s=\omega \tau$.
When $\omega \ll 1$, the distribution has a narrow peak around $\tau$, with standard deviation $\omega\tau/\sqrt{2}$. Increasing $\omega$ broadens the distribution to the point that, when $\omega = 1/\sqrt{3}$ it recovers a Maxwellian with dispersion $\sigma=\tau$. Finally, when $\omega = 1$, the distribution becomes a Half-Normal with parameter $\sigma=\tau$.

For the direction of the kicks, we consider three different options (a) isotropic (b) in a cone aligned with the polar axis with half-aperture $\alpha$ and (c) in a wedge centered on the equatorial plane with a half-aperture angle $\alpha_{\rm equator}$.

\paragraph{Selection effects}

We account for observational selection effects by excluding systems where the orbital configuration would likely prevent their detection as Be X-ray binaries. We exclude systems with a periastron passage closer than $30\,\Rsun$ because in closer systems tidal effects would become significant and there would not be space for a Be star disk to form  \cite{panoglou_be_2016,panoglou_be_2018}. This choice closely follows the tidal envelope observed for massive binary systems \cite{almeida_tarantula_2017,mahy_tarantula_2020, shenar_tarantula_2022}.
We also exclude systems with periastron passages larger than $450\,\Rsun$, to account for the fact that these would accrete at too low rates and are therefore unlikely to be observed in X-ray surveys \cite{liu_population_2024}. These thresholds serve as practical limits to capture the observed population, acknowledging that actual selection effects likely transition gradually rather than abruptly. 
In reality, the biases are complex and the data set is not homogeneous: the systems have been discovered and characterized by different groups, using various instruments, across several decades, which makes it effectively impossible to accurately model the selection effects.   Our treatment is simple but we consider it to be effective, as it includes the essential aspects of the physical picture. Experimenting with variations on these choices shows that the results are not sensitive to them.

\subsection{Inference method \label{sec:methods_inference}}
We consider a population of astrophysical systems (in our case, Be X-ray binaries), each characterized by a set of parameters\footnote{In this section, we use the following notation conventions.
We indicate all vectors in boldface, conditional probabilities of $x$ given $y$ as $p(x|y)$, likelihoods as $\mathcal{L}(y;x) = p(x|y)$, priors as $\pi$ and posteriors as $P$.}  $\bm{\theta}$ (in our case, the present-day orbital period and eccentricity). 
We further consider a set of hyper-parameters $\bm{\Lambda}$, which do not describe the individual systems themselves, but the distributions that were used to generate them (in this case, for example, parameters that describe the shape of the distribution of supernova kicks or parameters that describe the shape of the initial mass distribution of the supernova progenitors.) 
We denote the probability distribution of the parameters of individual systems $\bm{\theta}$  given a specific choice of the hyper parameters $\bm{\Lambda}$ as $\rho(\bm{\theta}|\bm{\Lambda})$.  

Let $N$ be the number of systems that are observed.  The likelihood of having obtained these $N$ independent measurements is given by

\begin{equation}
    \mathcal{L}(\bm{\Lambda}; \{\bm{\theta}\}, N) \propto
    \prod_i^N \rho'(\bm{\theta}_i|\bm{\Lambda}),
    \label{eq:likelihood}
\end{equation}
where 
\begin{equation}
    \rho'(\bm{\theta}|\bm{\Lambda}) =  \frac{\rho(\bm{\theta}|\bm{\Lambda}) D(\bm{\theta})}{\int \rho(\bm{\theta}'|\bm{\Lambda}) D(\bm{\theta}') \rm{d}\bm{\theta}'}
\end{equation}
is the probability density of the observable population: the probability density of the intrinsic population $\rho$  after accounting for the selection effects $D(\bm{\theta})$. Here, $D(\bm{\theta})$ is the probability that a system with parameters $\bm{\theta}$ is detected by the survey, and it encodes, for example,  the selection criteria and biases. 

When writing the likelihood in eq.~\ref{eq:likelihood}, we are assuming that the parameters $\bm{\theta}$ are measured exactly, with no uncertainty. This assumption is justified by the small measurement uncertainties in the periods and eccentricities of almost all our Be X-ray binary sample. Additionally, we marginalize over the unknown expected number of detected systems, which depends on the underlying population and survey sensitivity, applying a log-flat prior on this quantity \cite{mandel_parameter_2010, mandel_extracting_2019,abbott_population_2021}.

Our main objective is the posterior on the hyper-parameters (in this case, for example, the parameter $\sigma$ that describes the kick velocity distribution). This is given by Bayes' theorem
\begin{equation}
    P(\bm{\Lambda}) = \frac{\mathcal{L}(\bm{\Lambda}; \{\bm{\theta}\}) \pi(\bm{\Lambda})}{\int \mathcal{L}(\bm{\Lambda}; \{\bm{\theta}\}) \pi(\bm{\Lambda}) {\rm{d}} \bm{\Lambda}},
\end{equation}
Here, $\pi(\bm{\Lambda})$ is the prior we assume for the hyper-parameters.

In our case, $\rho'(\bm{\theta}_i|\bm{\Lambda})$ has no analytic form, so we compute an approximate likelihood instead. We estimate $\rho'(\bm{\theta}_i|\bm{\Lambda})$ by randomly drawing 10,000 systems and generating a mock observation in the period-eccentricity plane. We then compute a 2D kernel density estimate (KDE) of the period-eccentricity distribution, using the Scott rule \cite{scott_multivariate_1992} to chose the bandwidth. Using a KDE ensures that the distribution is smooth and is a convenient way of overcoming the problem of not having an analytical expression for the likelihood. However, it should be kept in mind that it only represents a numerical approximation of the actual likelihood.  

We compute the posterior $P(\bm{\Lambda})$ by MCMC sampling with the Python package \texttt{emcee} \cite{goodman_ensemble_2010, foreman-mackey_emcee_2013}.  We run 30 MCMC walkers for at least 50,000 and at most 800,000 steps. We chose these settings after testing the convergence of the chain by ensuring the length of the chain is larger than 50 times the integrated auto-correlation length $\tau_c$ \cite{goodman_ensemble_2010}. For each converged chain, we discard the first $3\tau_c$ steps as burn-in to eliminate memory of the starting point. We also thin the chain by selecting one sample every $\tau_c/4$ steps, to reduce the auto-correlation between samples.
We choose broad uninformative priors (shown in Fig~\ref{fig:lowB_maxwellian_corner} and \ref{fig:uppB_vwm_polar_corner}) and we verified that the posteriors are predominantly informed by the data and not by the priors.

\paragraph{Reporting posterior point-estimates and credible intervals} \label{sec:methods_point_estimates}
Bayesian inference produces a full multidimensional posterior distribution that encapsulates all the information derived from the analysis. Despite its completeness, this distribution is often challenging to visualize, interpret, and impractical to communicate in its entirety. Therefore, it is customary to condense the information into a few numerical values, such as point estimates and credible intervals. In this study, every time we report a point estimate, we are indicating the median of the posterior distribution, and provide the 90\% credible interval as uncertainty.

For the model parameters $\bm\Lambda$ (e.g., those in the bottom row of Fig.~\ref{fig:posteriors_summary}), we extract the marginal posterior distributions from the MCMC samples. We compute the credible intervals using the algorithm proposed by \cite{chen_monte_1999}, which guarantees to find the smallest interval containing the desired probability. When the distribution is unimodal, this interval corresponds to the highest density interval (HDI).

For derived quantities $q$, such as the ones in the top row of Fig.~\ref{fig:posteriors_summary}, we compute the marginal probability as
\begin{equation}
    P(q) = \int P(q | \bm\Lambda) {\rm{d}} \bm{\Lambda} \approx \frac{1}{N_s}\sum_i^{N_s} P(q|\bm\Lambda_i),
\end{equation}
where $N_s=3000$ and $\bm\Lambda_i$ is sampled from the MCMC posterior. Since $P(q|\bm\Lambda_i)$ is often not analytical, we approximate it by drawing samples from it.  

\subsection{Model selection} \label{sec:model selection}
We employ different statistical techniques to quantify how well a model describes the data, or to compare alternative models.

\paragraph{Bayes factors}
Given two models $M_1$ and $M_2$, we quantify their ability to describe the data $d$ via the Bayes factor, which is the odds ratio

\begin{equation}
    B_{12} = \frac{P(d|M_1)}{P(d|M_2)}.
\end{equation}
The higher the Bayes factor, the greater is the preference for model $M_1$ over $M_2$. The Bayes factor naturally accounts for model complexity, favoring models that achieves the best fit with the least freedom.
If the models are nested, meaning that $M_1$ reduces to $M_2$ for a particular choice of one of the parameters, $\lambda = \lambda^*$, and if the priors are separable, then the Bayes factor can be expressed as a Savage-Dickey ratio \cite{dickey_weighted_1971}

\begin{equation}
    B_{12} = \frac{P(\lambda|d)}{\pi_1(\lambda)}\Bigg\rvert_{\lambda = \lambda^*},
\end{equation}

\noindent where $P(\lambda|d)$ is the marginal posterior of the parameter $\lambda$, $\pi_1(\lambda)$ is its prior, and they are both computed at the value  $\lambda^*$.
We exploit this property to compute the Bayes factor of different models, by constructing a nested version of the models we want to compare.
All our models have analytical and separable priors, which are straightforward to compute. We construct the marginal posterior $P(\lambda|d)$ as a histogram over the posterior samples obtained with our MCMC. We interpolate between the values of the histogram bins using the \texttt{PCHIP} method \cite{fritsch_method_1984} as implemented in \texttt{scipy}, and then integrate the interpolating function numerically to obtain the normalization. The resulting estimate of the Bayes factor depends to some extent on the binning of the histogram; a finer binning improves the quality of the interpolation, but also increases the Poisson noise.
To mitigate this uncertainty, we compute the Bayes factor with different bin sizes, and report the mean of the results.

\paragraph{Posterior predictive checks}
Posterior predictive checks assess a model's validity by comparing simulated data from its posterior distribution to observations \cite{gelman_bayesian_2014}. They are particularly useful for diagnosing issues such as poor fit, or missing physical effects. We use both qualitative and quantitative approaches: qualitatively, we visually check if the simulated distributions are consistent with observations; quantitatively, we identify regions of the period-eccentricity plane that lack observed counterparts but are populated by the model. We then compute the probability that this happens by chance. A low probability suggests the model may not fully capture the data's morphology, though it does not necessarily warrant rejection (\ref{sec:hobbs fit},\ref{sec:state-of-the-art fit}).

\paragraph{Kolmogorov–Smirnov test}
The two-sample Kolmogorov–Smirnov (KS) test is a widely used non-parametric method for assessing whether two sets of observations are consistent with being drawn from the same underlying distribution. It returns a p-value, which represents how likely it would be to obtain two sets of observations that differ as much as the observed ones, assuming that they were actually drawn from the same distribution. If the p-value is below a chosen significance threshold (typically 0.05), the sets of observations are considered statistically different. We use a 2D generalization of the two-sample KS test \cite{peacock_two-dimensional_1983,fasano_multidimensional_1987,ndtest} to compare the observed period-eccentricity distribution with model predictions. While the KS test does not rank compatible models (higher p-values do not indicate a better fit), it is ideal for identifying models that are unlikely be compatible with the observed data (\ref{sec:hobbs fit}).

\subsection{Discussion of the key model assumptions} \label{sec:method_model_assumptions}
Two key assumptions of our model are that the orbit of of the binary was nearly circular before the supernova, and that the orbit has remained largely unchanged since then, meaning that the present-day orbit directly reflects the post-explosion configuration. Below, we assess the validity of these assumptions.

\subsubsection{Eccentricity before the supernova} \label{sec:presn orbit was circular}
Be X-ray binaries had a binary interaction in the past, where the neutron star progenitor filled its Roche lobe and was stripped of its hydrogen envelope.
Strong tidal interactions during the process are expected to circularize the orbit of the binary, and indeed this is reflected in observations of Be X-ray binary progenitors.

Before the supernova explosion, Be X-ray binary progenitors consisted of a B or Be star with a stripped star (strS) companion in a mass range between an O-type subdwarf (sdO) and a Wolf-Rayet (WR) star (about $2$ to $8\,\Msun$). Such stripped stars are predicted to be a common outcome of binary interactions, but they are challenging to observe \cite{gotberg_spectral_2018, schootemeijer_clues_2018, yungelson_elusive_2024}, and only a handful have been recently discovered \cite{drout_observed_2023, gotberg_stellar_2023, ramachandran_partially_2023, villasenor_b-type_2023, ramachandran_x-shooting_2024, rivinius_newborn_2025}. Much better studied are the lower and higher masses analogues of the Be X-ray binary progenitors.

\begin{figure}
    \centering
    \includegraphics{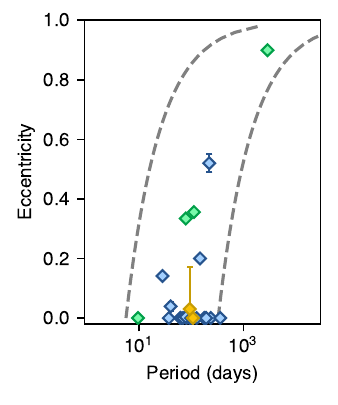}%
    \includegraphics{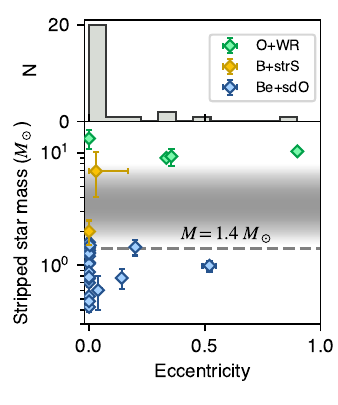}
    \caption{\textbf{Orbit and masses of Be X-ray binary progenitors.} Prior to the supernova explosion, the progenitors of Be X-ray binaries consisted of an OB star paired with a stripped star companion. We present observations of several systems, including Be stars plus type-O subdwarfs (Be+sdO) shown in blue \cite{peters_detection_2008, peters_far-ultraviolet_2013, mourard_spectral_2015, chojnowski_remarkable_2018, shenar_hidden_2020, gies_h_2020, harmanec_v1294_2022, wang_orbital_2023, klement_chara_2024,muller-horn_hip_2025}, B stars plus stripped stars (Be+strS) in yellow \cite{villasenor_b-type_2023, ramachandran_x-shooting_2024}, and O stars plus Wolf-Rayet companions (O+WR) in green \cite{north_2_2007, de_la_chevrotiere_spectroscopic_2011, richardson_first_2021, thomas_orbit_2021}. The left panel illustrates the orbital eccentricity and period of these systems. The grey dashed lines are the same as the edge of the shaded region in Fig.~\ref{fig:main_plot}. Orbital periods span from 10 to several hundred days, comparable to those of Be X-ray binaries. Except for some O+WR systems (where the effect of a third companion may become important), the eccentricities are below 0.2. In the right panel, we show a histogram of the eccentricities (top), and the mass of the stripped star versus the eccentricity (bottom). The gray dashed line indicates the Chandrasekhar limit, the maximum mass of a stable white dwarf. The dark shaded region indicates the mass range where stripped stars are expected to become neutron stars.}
    \label{fig:progenitors}
\end{figure}

In Fig~\ref{fig:progenitors} we collect observations of Be+sdO, Be+strS, and Be+WR binaries, showing their periods, eccentricities and masses. The left panel shows that these systems share a similar period range with Be X-ray binaries. The histogram in the right panel indicates that about $90\%$ of the systems have an eccentricity $e\leq0.2$, with $75\%$ consistent with $e=0$. The Be+WR systems appear to allow for higher eccentricities; however, due to their higher masses, they are also more likely to host a tertiary companion \cite{moe_mind_2017}, which could excite the eccentricity through the von Zeipel-Kozai-Lidov (ZKL) mechanism \cite{von_zeipel_sur_1910, kozai_secular_1962, lidov_evolution_1962}.

These observations support the hypothesis that most systems are in nearly circular orbits before the supernova. However, we still examine the effects of an initial eccentricity in sec.~\ref{sec:effect of initial eccentricity}, and find that it must have been below about 0.2 to reproduce the observations.

\subsubsection{Orbital evolution since the supernova}\label{sec:postsn orbit did not change}
Several mechanisms could potentially alter the period and eccentricity of a Be X-ray binary between the time of the supernova explosion and the current observation. Here, we demonstrate that these effects are negligible over the lifetime of a Be X-ray binary, and that the orbit remains effectively unchanged.

\paragraph{Tides}

Tidal interactions in binary systems can dissipate orbital energy and transfer angular momentum, leading to orbital synchronization and circularization.
A first indication that tides are unlikely to be effective on the systems in our sample is that synchronization of the stellar spin occurs typically faster than orbital circularization \cite{zahn_tidal_1977}. Since Be stars are fast rotators and far from synchronization, tides are unlikely to have significantly altered their eccentricity.

\begin{figure}
    \centering
    \includegraphics{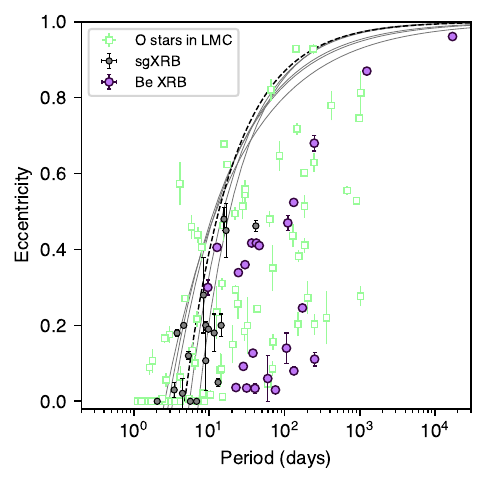}
    \caption{\textbf{The effects of tides in the period-eccentricity diagram}. The solid gray lines are tidal envelopes from \cite{bashi_features_2023}, corresponding samples of G and F stars of different effective temperatures. The light green squares are a sample of massive O binaries in the Large Magellanic Cloud (LMC) from the TMBM survey \cite{almeida_tarantula_2017,mahy_tarantula_2020, shenar_tarantula_2022}. The dark gray dots are the Galactic HMXB classified as supergiants (sgXRB) in \cite{fortin_catalogue_2023}. The purple dots are our sample of Be X-ray binaries (Be XRB). The black dashed line corresponds to a constant periastron distance $r_{\rm peri} = 30\,\Rsun$ for a typical Be X-ray binary with masses $M_{\rm Be} = 15\,\Msun$ and $M_{\rm NS} = 1.4\,\Msun$.}
    \label{fig:tidal_envelope}
\end{figure}

A second argument comes from the location of the systems in the period-eccentricity diagram (Fig.~\ref{fig:tidal_envelope}). We adopt the concept of a tidal envelope, an empirical boundary separating systems whose orbits are likely unaffected by tidal circularization over their lifetimes (below the envelope) from those where tidal effects could have significantly altered the eccentricity (above the envelope). While the tidal envelope's location can depend on the stellar population, its position is found to vary only weakly with mass and age \cite{meibom_robust_2005,zanazzi_tale_2022,bashi_features_2023}. For instance, the envelopes shown in fig.~\ref{fig:tidal_envelope}, estimated from populations of G- and F-type stars \cite{bashi_features_2023}, are also compatible with the locations of massive O-type binaries in the Tarantula Massive Binary Monitoring (TMBM) survey \cite{almeida_tarantula_2017,mahy_tarantula_2020,shenar_tarantula_2022}. 

Most Be X-ray binaries in our sample lie well below the tidal envelope, further supporting the idea that tidal effects on their eccentricities are negligible. Two systems fall near the envelope, but excluding them does not alter our conclusions. Conversely, supergiant X-ray binaries often lie near or on the tidal envelope. Their giant companions have convective envelopes, where equilibrium tides are much more effective at circularizing orbits than the dynamical tides acting on the radiative envelopes of Be stars \cite{zahn_tidal_1977}. Consequently, their eccentricities have likely been modified by tides and cannot be directly used to infer supernova kicks.
While we cannot completely rule out that complex tidal effects \cite{witte_tidal_1999} might play a role in shaping the groups we identify, the arguments outlined above suggest that tides are unlikely to significantly influence the post-supernova evolution of our sample.

\paragraph{Mass loss}
Mass loss via stellar winds or the Be disk can carry away angular momentum,but its impact on the orbital evolution of the systems in our sample is negligible. Be stars primarily lose mass through the disk at a rate between $10^{-12}$ and $10^{-9}\,\Msun yr^{-1}$ \cite{granada_populations_2013,vieira_life_2017}, resulting in a total loss of $10^{-5}$ to $10^{-2}\,\Msun$ over the lifetime of a Be X-ray binary of the order of $10\,\rm{Myr}$. This is negligible compared with the mass of the system.

\paragraph{Interaction between the neutron star and the disk}
The interaction between the neutron star and the Be star's disk can occur through two main mechanisms: direct contact with the disk material and secular gravitational torques exerted by the disk on the neutron star. Both effects are negligible in the systems we consider.
Direct contact with the disk causes the neutron star to experience dynamical friction and ram pressure. They only alter the orbit significantly once the neutron star has ploughed through a mass comparable to its own. Since the total mass that is shed by the Be star is $10^{-5}$ to $10^{-2}\,\Msun$, much smaller than the neutron star mass, these effects are negligible.
Secular gravitational torques between the disk and neutron star are similarly ineffective, since they scale with the mass of the the disk. Over an orbit, the neutron star velocity is perturbed by a factor $\Delta v/v \sim m_{\rm disk}/M$. Over the lifetime of the system $\Delta t \approx 10\,\rm{Myr}$, this amounts to $\Delta v/v \sim (\Delta t/P) (m_{\rm disk}/M)$. Assuming a disk build-up time of $1\,\rm{yr}$ to reach the steady state \cite{rimulo_life_2018}, and a constant mass feeding rate of $10^{-12}$ to $10^{-9}\,\Msun yr^{-1}$, the mass of the disk at any given time is $m_{\rm disk} \approx 10^{-12}-10^{-9}\,\Msun$. For a typical orbital period $P=0.1\, \rm{yr}$, and a system mass $M=10\,\Msun$, the velocity perturbation of the neutron star over the system lifetime is only $\Delta v/v \approx 10^{-5}-10^{-2}$.

\paragraph{Effect of a third star}
The majority of binaries containing a star more massive than $10\,\Msun$ at birth are expected to host at least one third star in a wide outer orbit \cite{moe_mind_2017}. Such a companion could influence the inner binary's eccentricity through the ZKL mechanism \cite{von_zeipel_sur_1910, kozai_secular_1962, lidov_evolution_1962}.
While the presence of tertiary companions in our sample cannot be definitively excluded, the range of feasible pre-supernova outer orbital periods is quite limited. The outer orbit would need to be wide enough to ensure the dynamical stability of the system, yet close enough to avoid disruption by the systemic velocity imparted to the inner binary during the supernova.
In our preferred model for the high-eccentricity group, systemic velocities range from $20$ to $60\,\kms$, consistent with previous estimates \cite{fortin_constraints_2022}. Under these conditions, outer periods below about $10^3$ days are generally dynamically unstable, while those exceeding about $10^4$ days are frequently disrupted by the supernova. This narrow range leaves little room for a tertiary companion to exert significant influence.

Furthermore, the presence of a tertiary companion would introduce perturbations to the orbital eccentricity. This would likely disrupt the distinct patterns observed in our sample, such as the clear separation into low- and high-eccentricity groups and the tight period-eccentricity correlation within the high-e sequence. The presence of these patterns in our data suggests that tertiary companions, if present, have minimal impact on the systems in our sample.

\paragraph{Observed orbital variation}
Direct observations support the conclusion that orbital evolution in Be X-ray binaries is negligible. For instance, in the Be X-ray binary and radio pulsar system PSR B1259-63, a change in the orbital period was measured at $\dot P = 1.4\times 10^{-8}$ \cite{shannon_kinematics_2014}. Assuming that the measured trend remains constant over $10\,\rm{Myr}$, this corresponds to a period change of approximately $4\%$. Although this is slightly higher than the above theoretical estimates, it may be attributed to the high eccentricity of this system ($e = 0.87$), which could amplify tidal effects during periastron passages. Even so, this observed change remains modest, suggesting that the orbital evolution in the majority of Be X-ray binaries, which do not have such an extreme eccentricity, would be even less significant.

\paragraph{Effect of the supernova on the Be companion}
When the supernova interacts with the Be star, it can influence the dynamics of the system in several ways. The ejecta may impart momentum to the Be star, they can strip mass via ablation and can deposit thermal energy. The first two effects are generally negligible for systems with orbital periods above 10 days, such as those in our sample \cite{liu_interaction_2015}.

The third mechanism has a less direct effect on the orbit. As the shockwave traverses the Be star, it deposits heat, leading to a significant expansion on the thermal timescale of the affected layers. This shock-induced heating can cause the star to temporarily inflate to several hundred solar radii \cite{hirai_possible_2015, hirai_comprehensive_2018, ogata_observability_2021, hirai_constraining_2023}. As a consequence, all the Be X-ray binary with a pre-supernova orbital period smaller than about 100-200 days (potentially all the systems in our sample) would have overflown their Roche lobes shortly after the supernova.

Such expansion could, in principle, alter the orbit through mass transfer or enhanced tidal interactions. However, the shockwave primarily heats the outermost $0.01 - 0.1\,M_\odot$ of the star's envelope \cite{hirai_comprehensive_2018}, a small fraction of the total mass. Additionally, this expanded state only lasts 1-100 years \cite{ogata_observability_2021}. Given the limited mass involved and the short duration of the inflated phase, according to recent models of eccentric mass transfer the effect on the orbit should be minimal \cite{akira_rocha_mass_2024}.

\newpage

\section{Supplementary Results} \label{sec:supplementary_results}
In this section, we provide additional information on our main results. We show an overview of the models in table~\ref{tab:model_summary} and provide a summary of the outcome of our analysis in table~\ref{tab:best_fit_parameters}. 

\begin{table}
    \centering
    \footnotesize
    \caption{\textbf{Description of the models.} In addition to the ones listed, the parameters $P_\mu$, $P_\sigma$, $\Delta M_\mu$, and $\Delta M_\sigma$ are common to all models. The preferred models are in bold.}
    \begin{tabular}{|p{0.3\textwidth} | p{0.15\textwidth} | p{0.5\textwidth} |}
    \hline
    Model name  & Parameters & Description \\
    \hline
    \texttt{Classical}  & --- &
        Isotropic Maxwellian with $\sigma = 265\,\kms$. Applied to the full Be X-ray binary sample.\\
    \texttt{State-of-the-art} & $\sigma_2$, $f_2$ &
        Isotropic Maxwellian with $\sigma_1 = 265\,\kms$, plus a second isotropic Maxwellian component with free $\sigma_2$ containing the fraction $f_2$ of systems. Applied to the full Be X-ray binary sample.\\
    \textbf{\texttt{Low-e preferred}} & $\sigma$ &
        Isotropic Maxwellian with free $\sigma$. Applied to the low-e group.\\
    \textbf{\texttt{High-e preferred}} & $\tau$, $\omega$, $\alpha$ &
        Modified Maxwellian with free $\tau$ and $\omega$, direction restricted to a polar cone of opening angle $\alpha$. Applied to the high-e sequence.\\
    \hline
    \end{tabular}
    \label{tab:model_summary}
\end{table}

\renewcommand{\arraystretch}{1.5}
\begin{table}
    \centering
    \caption{Overview of the inferred parameters for the models shown in Fig.~\ref{fig:main_plot}. The two preferred models are marked in bold. For each parameter we provide the posterior median and the 90\% credible interval. $\sigma$ is the dispersion of a Maxwellian distribution. When two Maxwellians are considered $\sigma_1$ and $\sigma_2$ are their dispersion, and $f_2$ is the fraction of systems receiving a kick from the second Maxwellian.  $v_{\rm kick}$ is the kick velocity, $\Delta M$ is the mass loss, $v_{\rm kick,mean}$ is the mean of the kick velocity distribution, $v_{\rm kick,spread}$ is the ratio between the standard deviation and the mean of the velocity distribution. $\tau$ and $\omega$ are the two parameters of the modified Maxwellian distribution, as defined in section~\ref{sec:methods_model} and $\alpha$ is the maximum angle between the kick direction and the polar axis.}
    \begin{tabular}{|ll|}
        \hline
        Model                          & \texttt{Classical}\\
        Sample                         & full sample\\
        \multicolumn{2}{|l|}{Isotropic, Hobbs Maxw.}\\
        \hline
        $\sigma$                       & $265\kms$ (fixed)\\
        $\Delta M$                     & $3.9^{+9.3}_{-3.9}\Msun$\\
        $\log(P_{\rm preSN}/\rm{day})$ & $2.0^{+1.6}_{-0.9}$\\
        
        \hline
        \hline

        Model                          & \texttt{State-of-the-art}\\
        Sample                         & full sample\\
        \multicolumn{2}{|l|}{Isotropic, Hobbs plus a second Maxw.}\\
        \hline
        $\sigma_1$                     & $265\kms$ (fixed)\\
        $\sigma_2$                     & $7.6^{+9.8}_{-7.6}\kms$\\
        $f_2$                          & $0.32^{+0.40}_{-0.32}$\\
        $v_{\rm kick}$                 & $21^{+358}_{-21}\kms$\\
        $\Delta M$                     & $4.1^{+4.9}_{-4.1}\Msun$\\
        $\log(P_{\rm preSN}/\rm{day})$ & $1.6^{+1.4}_{-1.5}$\\
        \hline
    \end{tabular}\hspace{5pt}%
    \begin{tabular}{|ll|}
        \hline
        Model                          & \textbf{\texttt{Low-e preferred}}\\
        Sample                         & Low-e group\\
        \multicolumn{2}{|l|}{Isotropic, Maxwellian}\\
        \hline
        $\sigma$                       & $4.5^{+2.8}_{-3.2}\kms$\\
        $v_{\rm kick,mean}$            & $7.2^{+4.5}_{-5.1}\kms$\\
        $v_{\rm kick,spread}$          & $0.42$ (fixed)\\
        $v_{\rm kick}$                 & $6.5^{+6.7}_{-5.4}\kms$\\
        $\Delta M$                     & $0.9^{+1.5}_{-0.9}\Msun$\\
        $\log(P_{\rm preSN}/\rm{day})$ & $1.9^{+1.6}_{-1.0}$\\

        \hline 
        \hline
        Model                          & \textbf{\texttt{High-e preferred}}\\
        Sample                         & high-e sequence\\
        \multicolumn{2}{|l|}{Polar, Modified Maxwellian}\\
        \hline
        $\tau$                         & $100^{+29}_{-31}\kms$\\
        $\omega$                       & $0.16^{+0.29}_{-0.16}$\\
        $v_{\rm kick,mean}$            & $99^{+28}_{-33}\kms$\\
        $v_{\rm kick,spread}$          & $0.11^{+0.22}_{-0.11}$\\
        $v_{\rm kick}$                 & $99^{+37}_{-45}\kms$\\
        $\Delta M$                     & $3.5^{+2.4}_{-3.1}\Msun$\\
        $\alpha$                       & $5.2^{+8.3}_{-5.2}\,^{\circ}$\\
        $\log(P_{\rm preSN}/\rm{day})$ & $1.6^{+1.4}_{-1.6}$\\
        \hline
    \end{tabular}
    \label{tab:best_fit_parameters}
\end{table}
\renewcommand{\arraystretch}{1.0}

\subsection{Are Be X-ray binaries consistent with a Hobbs-like kick?} \label{sec:hobbs fit}
The \texttt{Classical} model, derived from observations of isolated pulsar velocities \cite{hobbs_statistical_2005}, consists in isotropic kicks following a Maxwellian with $\sigma = 265\,\kms$. We assess the ability of this model to reproduce the observed properties of Be X-ray binaries by fixing the velocity distribution to this Maxwellian, while allowing the progenitor mass and pre-supernova period distributions to vary. The best fit is shown in Fig.~\ref{fig:main_plot}A and the results reported in table~\ref{tab:best_fit_parameters}. Over 90\% of the systems become unbound due to the high kick velocities in our best fit model, while the remaining systems populate the high eccentricity region, which is devoid of observations.
We perform a posterior predictive check by calculating that the probability that a randomly drawn sample of 23 systems (matching the observed sample size) would produce no points within the yellow-contoured region in Fig~\ref{fig:main_plot}A is only $5 \times 10^{-5}$. This suggests an inconsistency between the model and the observed data.
Additionally, a 2D KS test comparing the \texttt{Classical} model to observations yields a p-value less than $10^{-4}$, confirming that the \texttt{Classical} kick distribution is statistically inconsistent with the data. We conclude that this model fails to reproduce the observed properties of Be X-ray binaries.

\subsection{Is a second component necessary to explain the observations? Is it sufficient?} \label{sec:state-of-the-art fit}
A common approach to address the shortcomings of the \texttt{Classical} model is to introduce a second, low velocity Maxwellian component to the kick distribution, typically with $\sigma_2 = 15-45\,\kms$ \cite{podsiadlowski_effects_2004, giacobbo_impact_2019, vigna-gomez_formation_2018, igoshev_combined_2021}.
This two-component model has been proposed to simultaneously explain the high-velocity of isolated pulsars and the formation of bound binaries, which require smaller kicks.
We test this \texttt{State-of-the-art} model by extending the \texttt{Classical} model to include a second isotropic Maxwellian with free dispersion $\sigma_2$, applied to a fraction $f_2$ of systems, also left as a free parameter.
The best-fit results are reported in table~\ref{tab:best_fit_parameters}, and illustrated in Fig.~\ref{fig:main_plot}B. The addition of a low-velocity component significantly improves the fit (confirmed by a 2D KS test with $p>0.10$), and proves to be essential: the low-velocity component accounts for 99\% of the bound systems predicted by the model. However, this single newly added component attempts to fit both observed groups of Be X-ray binaries simultaneously, which causes the model to overpopulate the intermediate region between them, where no systems are observed. A posterior predictive check shows that the probability of this region (the yellow box in Fig.~\ref{fig:main_plot}B) remaining empty by chance is only 0.2\%, indicating that the \texttt{State-of-the-art} model, while an improvement over the \texttt{Classical} one, fails to capture the bimodal morphology of the data.

\subsection{What kick distribution fits the low-e group?} \label{sec:low-e group fit}
\begin{figure}
    \centering
    \includegraphics[width=\textwidth]{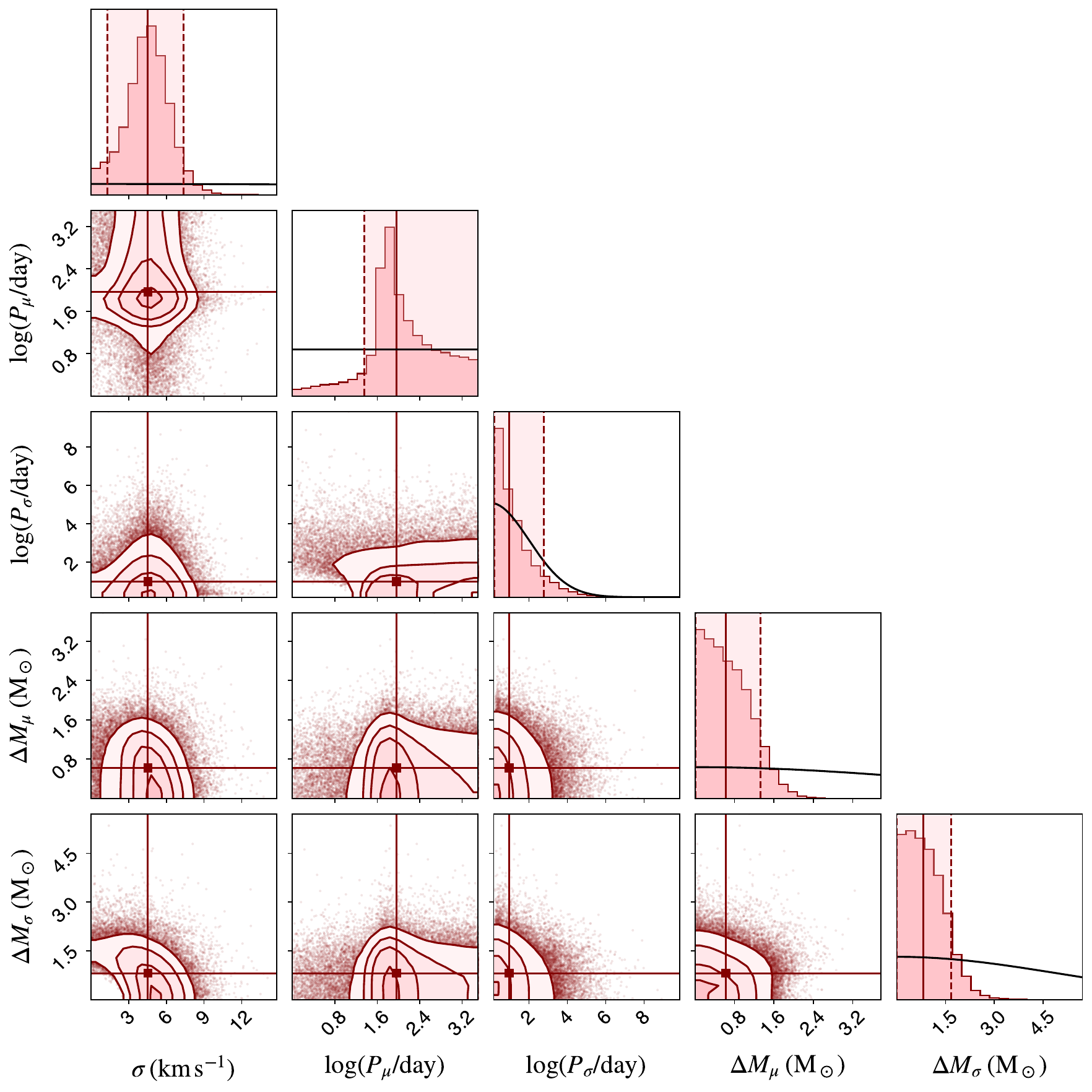}
    \caption{\textbf{Cornerplot of the posterior of our preferred model for low-e group.}
    The model assumes that kicks are drawn from a Maxwellian distribution described by the parameter $\sigma$, and that the direction is isotropically distributed.
    The 1D histograms on the diagonal are the marginalized posterior for each parameter, the solid vertical lines indicate the median, and the dashed lines the 90\% credible interval. The black solid curve is the prior distribution.
    The panels below the diagonal are 2D joint posterior distributions for any 2-parameter combination. The contours mark the 0.5, 1, 1.5 and 2-sigma levels, containing 11.8\%, 39.3\%, 67.5\% and 86.4\% of the probability mass, respectively.
    From left to right, the plot shows posteriors for the parameter $\sigma$ of the Maxwellian distribution, the parameters describing the period distribution $\log P_\mu$ and $\log P_\sigma$, and those describing the mass loss distribution $\Delta M_\mu$ and $\Delta M_\sigma$. 
    }
    \label{fig:lowB_maxwellian_corner}
\end{figure}

We model the low-e population by assuming that the systems received isotropic kicks drawn from a Maxwellian distribution, with the velocity dispersion $\sigma$ treated as free parameter. As in the \texttt{Classical} and \texttt{State-of-the-art} models, we assume Gaussian distributions for the pre-supernova period $\log P_{\rm preSN}$ and mass loss $\Delta M$. We report our results in table~\ref{tab:best_fit_parameters} under the model name \texttt{Low-e preferred}, and the posterior distributions of the model parameters are shown in fig.~\ref{fig:lowB_maxwellian_corner}.
As illustrated in red in Fig.~\ref{fig:main_plot}C, this simple model successfully reproduces the observed population without requiring additional assumptions.

\paragraph{Is there evidence for anisotropy or deviations from a Maxwellian in the low-e group?}
We also test a more complex model---similar to the one used for the high-e sequence---which involves a modified Maxwellian kick distribution restricted to a polar cone. This model fits the data equally well, yielding kick velocities and mass-loss estimates consistent with those of the simpler isotropic model. We find no evidence in favor or against anisotropy or a velocity spread different from the Maxwellian one. Given the comparable performance and fewer assumptions, the isotropic Maxwellian is preferred for describing the low-e group.

\subsection{What kick distribution fits the high-e sequence?} \label{sec:high-e sequence fit}

\begin{figure}
    \centering
    \includegraphics{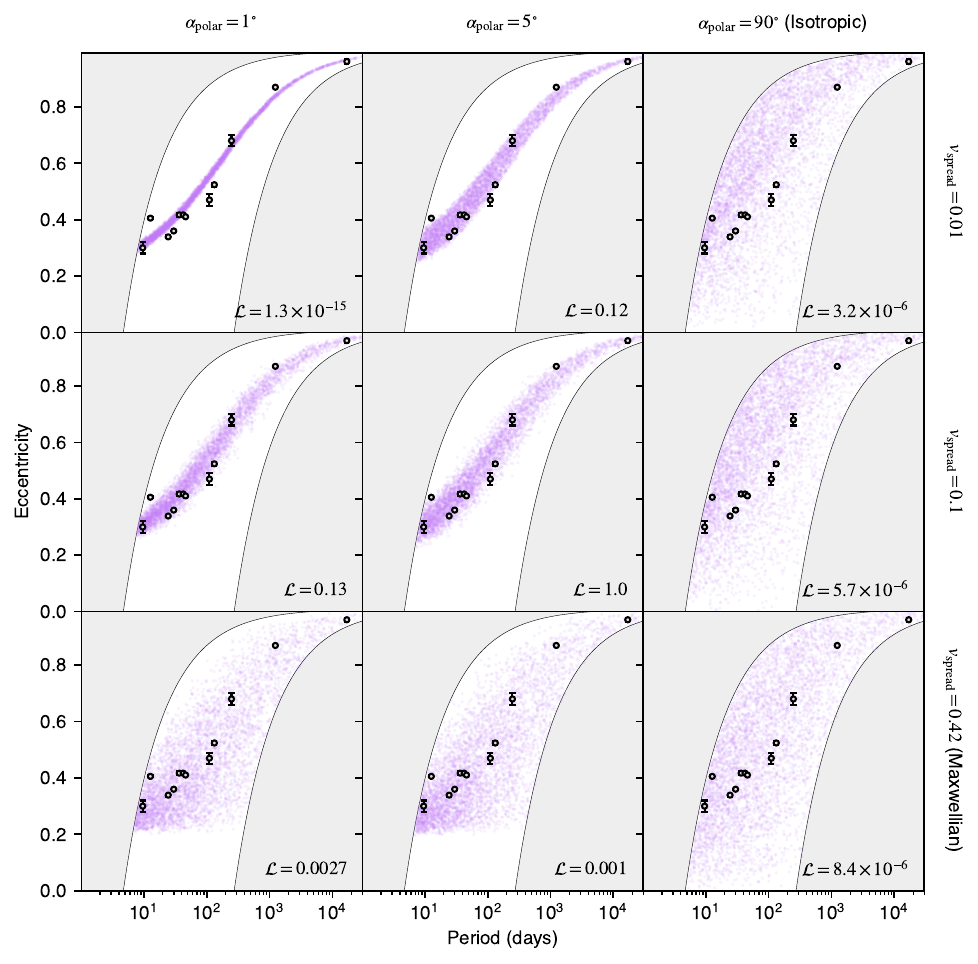}
    \caption{\textbf{Effect of kick velocity and direction on the period-eccentricity distribution.} We show in purple the result of a modified Maxwellian kick distribution with peak at $\tau=100\,\kms$. From left to right we vary the opening angle $\alpha$ of the allowed kick directions around the polar axis. From top to bottom we vary the ratio $v_{\rm spread}$ between the standard deviation and the mean of the velocity distribution. The black points indicate the observed systems in the high-e sequence. For each panel we report the relative likelihood $\mathcal{L}$ that the observed points originate from the shown distribution, normalized to the value for the central panel. The observations are best reproduced by the models in the central panel, where the direction is restricted to a $5^\circ$ cone along the polar axis, and the velocity distribution has a spread of $10\%$.}
    \label{fig:spread_in_angle_and_velocity}
\end{figure}

The high-e group traces a narrow sequence in the period eccentricity plane, suggesting that both the direction and magnitude of the kicks must be constrained to reproduce its morphology. This is illustrated in fig.~\ref{fig:spread_in_angle_and_velocity}, where we vary the kick opening angle around the polar axis $\alpha$ and the velocity spread, defined as the ratio of the standard deviation to the mean kick velocity, $v_{\rm kick, spread}$. We compute the likelihood $\mathcal{L}$ of the observations in each of the panels, defined as the product of the probability density at the observed points. The relative likelihood we present here is normalized such that $\mathcal{L}=1$ for the central panel. The systems are best reproduced when kicks have a 10\% spread in velocity, and when the direction of the kicks is restricted to a narrow cone (central panel). In contrast, an isotropic Maxwellian distribution (bottom right panel) does not adequately reproduce the observed data, with a likelihood $\mathcal{L} < 10^{-5}$.

\begin{figure}
    \centering
    \includegraphics[width=\textwidth]{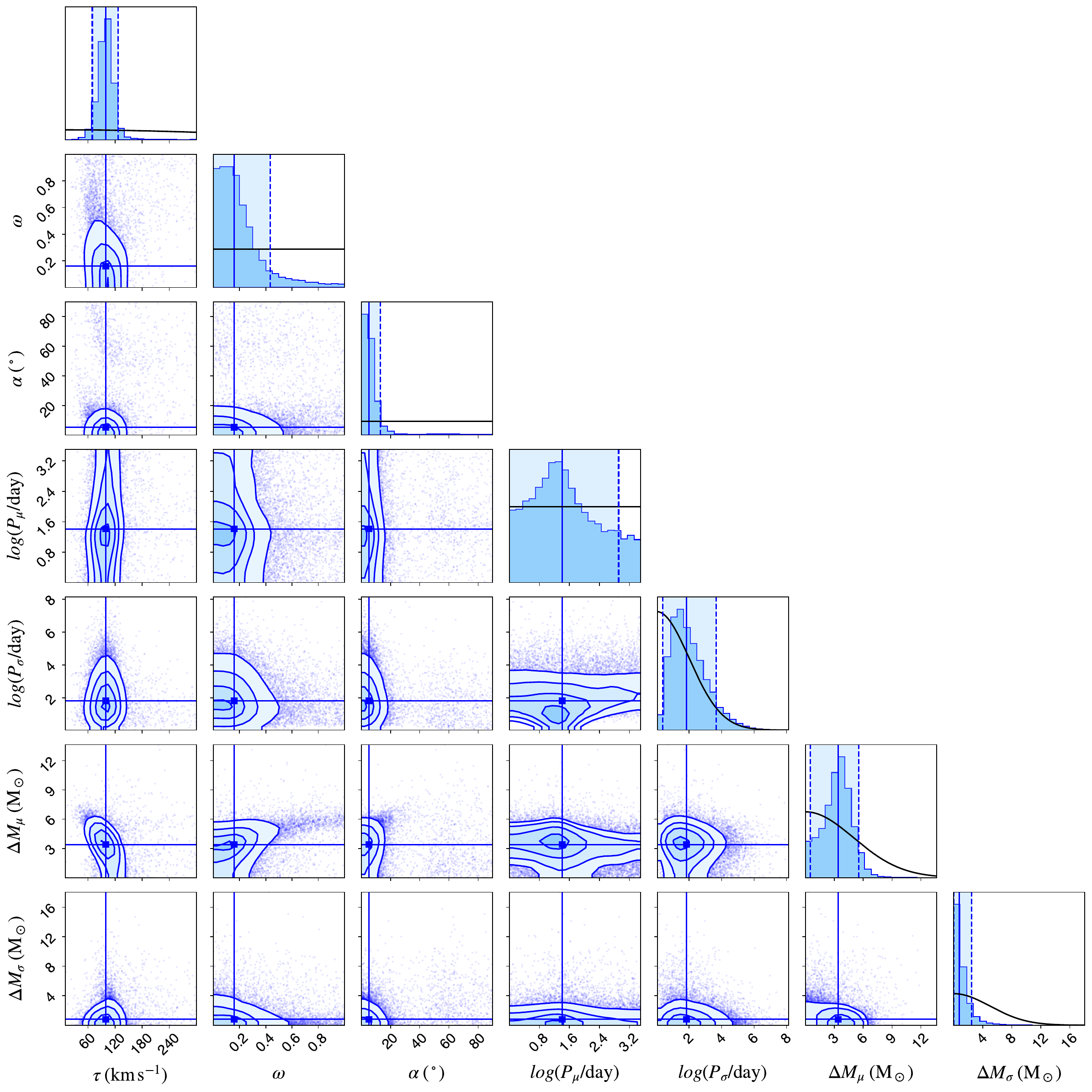}
    \caption{\textbf{Cornerplot of the posterior of our preferred model for the high-e sequence.} The model assumes that kicks are drawn from a modified Maxwellian distribution described by the parameters $\tau$ and $\omega$. The kick direction is confined within a cone of half-aperture $\alpha$ around the polar axis. See fig.~\ref{fig:lowB_maxwellian_corner} for a more detailed description of the figure.}
    \label{fig:uppB_vwm_polar_corner}
\end{figure}

Motivated by these findings, we construct the \texttt{High-e preferred} model, in which kicks follow a modified Maxwellian distribution (eq.~\ref{eq:modified maxwellian})---allowing independent control over the peak and width---and are confined within a polar cone with free opening angle $\alpha$.
Figure~\ref{fig:main_plot}A displays the best-fit model in blue, with the full posterior distributions shown in fig.~\ref{fig:uppB_vwm_polar_corner}. Summary statistics are reported in table~\ref{tab:best_fit_parameters}.

\subsubsection{Evidence for polar kicks} \label{sec:evidence for polar}
The \texttt{High-e preferred} model constrains kick directions to within $15^\circ$ of the polar axis with more than 90\% probability (fig~\ref{fig:posteriors_summary}C). Compared to an otherwise identical model with isotropic kick directions, it is favored by a Bayes factor of 16 (computed using the Savage-Dickey ratio, sec.~\ref{sec:model selection}), providing strong support for anisotropic kicks.

\subsubsection{Evidence for a narrow velocity distribution} \label{sec:evidence for narrowness}
The marginal posterior for the velocity spread in the \texttt{High-e preferred} model shows that, with over 90\% probability, the kick distribution is narrower than a Maxwellian with the same mean (fig~\ref{fig:posteriors_summary}D). This model is favored over one with a fixed Maxwellian distribution by a Bayes factor of 4.5, providing moderate evidence for a narrower spread in kick velocities.

\subsubsection{Comparison between the preferred model for the high-e sequence and an isotropic Maxwellian} \label{sec:evidence against iso maxwellian}
We have so far compared individually the hypotheses of isotropy versus kicks restricted to a polar cone, and of a Maxwellian distribution versus one with a variable spread. However, we are also interested in gauging the evidence for the \texttt{High-e preferred} model, which features both of these ingredients, against the simpler isotropic Maxwellian scenario. In fig.~\ref{fig:alternative high-e models}A, we show how the isotropic Maxwellian model struggles to reproduce the narrow band where the observed systems are located, predicting a much broader distribution.

For a quantitative comparison, we compute the Bayes factor. To use the Savage-Dickey ratios, we formulate a nested model: we introduce an intermediate model that assumes a Maxwellian kick velocity distribution but restricts the direction of the kicks within a polar cone, with the opening angle $\alpha$ as a free parameter. This model contains, nested in it, the isotropic Maxwellian case, when $\alpha = 90^\circ$. At the same time, the intermediate model is itself nested into the \texttt{High-e preferred} model, when the velocity spread parameter $\omega = 1/\sqrt{3}$, which recovers the Maxwellian distribution. This nesting structure allows us to compute Bayes factors for each step using the Savage-Dickey ratio, and to obtain the total Bayes factor for the full comparison by multiplying the two. We find a combined Bayes factor of 30 in favor of the \texttt{High-e preferred} model, indicating strong evidence against the isotropic Maxwellian scenario.

\subsubsection{Can the high-e sequence arise from Hobbs-like kicks?} \label{sec:high-e with hobbs}
We have already shown that the full Be X-ray binary sample cannot be explained by a Hobbs-like kick distribution in sec.~\ref{sec:hobbs fit}. However, there is the possibility that this tension is driven primarily by the low-e group. Perhaps these systems have a different origin, while the high-e sequence could still represent the low-velocity tail of the Hobbs distribution.
To test this possibility, we fit the \texttt{Classical} model (Isotropic Maxwellian kicks with $\sigma = 265\,\kms$) to the high-e sequence alone. We show the best fit in fig.~\ref{fig:alternative high-e models}B. The model presents the same shortcomings it had when we applied it to both groups: most systems are unbound, and those that remain tend to cluster at high eccentricities, well above the observed range. 

To quantitatively compare the \texttt{Classical} and \texttt{High-e preferred} models, we estimate the Bayes factor. To this aim, we construct a mixture model in which a fraction $f_{\rm Hobbs}$ of systems follows the Hobbs distribution, while the rest follow our preferred model. This nested setup allows us to compute a Bayes factor via the Savage-Dickey ratio. We find a Bayes factor of 77 in favor of the \texttt{High-e preferred} model, strongly disfavoring a Hobbs-like origin for the high-e sequence.

\subsubsection{Comparison between polar and equatorial kicks} \label{sec:polar vs equatorial}
Our \texttt{High-e preferred} model assumes kicks are confined in the polar direction. This choice is motivated by symmetry considerations: in a wide binary where tidal distortions are negligible, the only natural axis that breaks spherical symmetry is the stellar rotation axis, which is expected to align with the orbital angular momentum---i.e., the polar direction \cite{de_mink_rotation_2013, marcussen_banana_2022, marcussen_banana_2024}.
However, if the rotation axis is indeed the source of symmetry breaking, an alternative scenario is possible: kicks might preferentially lie in the equatorial plane, orthogonal to the spin axis. Assuming spin-orbit alignment, this corresponds to kicks confined within the orbital plane.

\begin{figure}
    \centering
    \includegraphics[width=0.33\textwidth]{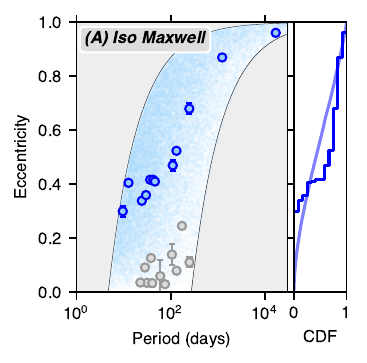}%
    \includegraphics[width=0.33\textwidth]{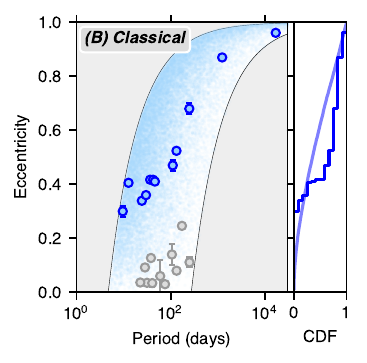}%
    \includegraphics[width=0.33\textwidth]{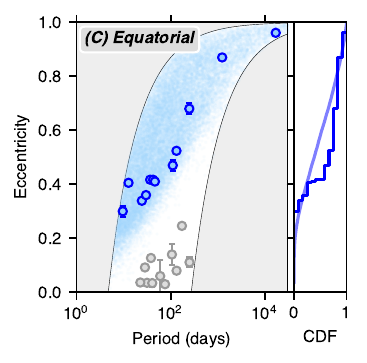}%
    \caption{\textbf{Alternative models for the high-e sequence.} Panel A: kicks follow an isotropic Maxwellian distribution, where $\sigma$ is inferred from the data; Panel B: kicks follow an isotropic Maxwellian distribution with $\sigma=265\,\kms$; Panel C: kicks follow a modified Maxwellian distribution, like for the \texttt{High-e preferred} model, but instead of being directed along the polar axis, they are constrained within a wedge centered on the equator.}
    \label{fig:alternative high-e models}
\end{figure}

To explore this scenario, we construct a variant of the \texttt{High-e preferred} model where the kick direction is restricted to a wedge centered on the equatorial plane, with a free half-opening angle $\alpha_{\rm equator}$. The best fit is shown in fig.~\ref{fig:alternative high-e models}C. The model does not exhibit a preference for kicks constrained to the equatorial plane, and it fails to reproduce the narrowness of the high-e sequence as effectively as the polar-kick scenario.

Both the polar and equatorial models reduce to the isotropic case when their respective opening angles reach $90^\circ$, allowing us to compute Bayes factors relative to isotropy using the Savage-Dickey ratio. Taking the ratio of these two Bayes factors gives the Bayes factor comparing the polar and equatorial models directly. We find a Bayes factor of 7.7 in favor of the polar-kick model, indicating that polar kicks are preferred over equatorial ones.

\subsection{Comparison between our preferred model and the state-of-the-art} \label{sec:state-of-the-art vs preferred model}
To assess how our preferred models perform relative to the current state-of-the-art, we construct a combined model that interpolates between the two approaches and enables the use of the Savage–Dickey method (\ref{sec:model selection}).
The combined model is formulated as a mixture of likelihoods. For each system in the sample, the total likelihood is a weighted sum of the likelihoods from the state-of-the-art and preferred models:

\begin{equation}
    \mathcal{L} = \prod_i \left[f_{\rm SotA}\mathcal{L}_{\rm{SotA},i} + (1-f_{\rm SotA})\mathcal{L}_{\rm{preferred}, i}\right],
\end{equation}
where $f_{\rm SotA} \in [0,1]$ is a free parameter that controls the relative contribution of the two models, and $\mathcal{L}_{\rm{SotA},i}$ and $\mathcal{L}_{\rm{preferred}, i}$  are the likelihoods evaluated at the period and eccentricity of the $i$-th system under the two respective models.
The likelihood $\mathcal{L}_{\rm{SotA},i}$ is computed using a single isotropic Maxwellian distribution (omitting the Hobbs component, which we have shown to be negligible for this population).
The likelihood $\mathcal{L}_{\rm{preferred}, i}$ is taken from the \texttt{Low-e preferred} model if the $i$-th system belongs to the low-e group, and from the \texttt{High-e preferred} model otherwise.
In addition to the period and eccentricity of the systems in the sample, this approach requires the information of which system belongs to which group. However, since the separation between the two groups is unambiguous, this classification is straightforward.
The model includes all parameters from the three submodels, plus the mixing fraction $f_{\rm SotA}$.

We compute the marginal posterior of $f_{\rm SotA}$ in the two limiting cases: $f_{\rm SotA}=0$, which corresponds to using the state-of-the-art model alone, and $f_{\rm SotA}=1$, which corresponds to the combination of the two preferred models, each applied to their respective group. The Bayes factor comparing these two cases is approximately $10^7$, providing overwhelmingly strong evidence in favor of the preferred models over the state-of-the-art.

This result confirms that modeling the low-e and high-e populations separately, with our preferred models, provides a significantly better description of the observed properties of Be X-ray binaries, even when accounting for the additional complexity of our preferred model.

\subsection{What is the impact of the chosen neutron star and Be star masses?}

\begin{figure}
    \centering
    \includegraphics{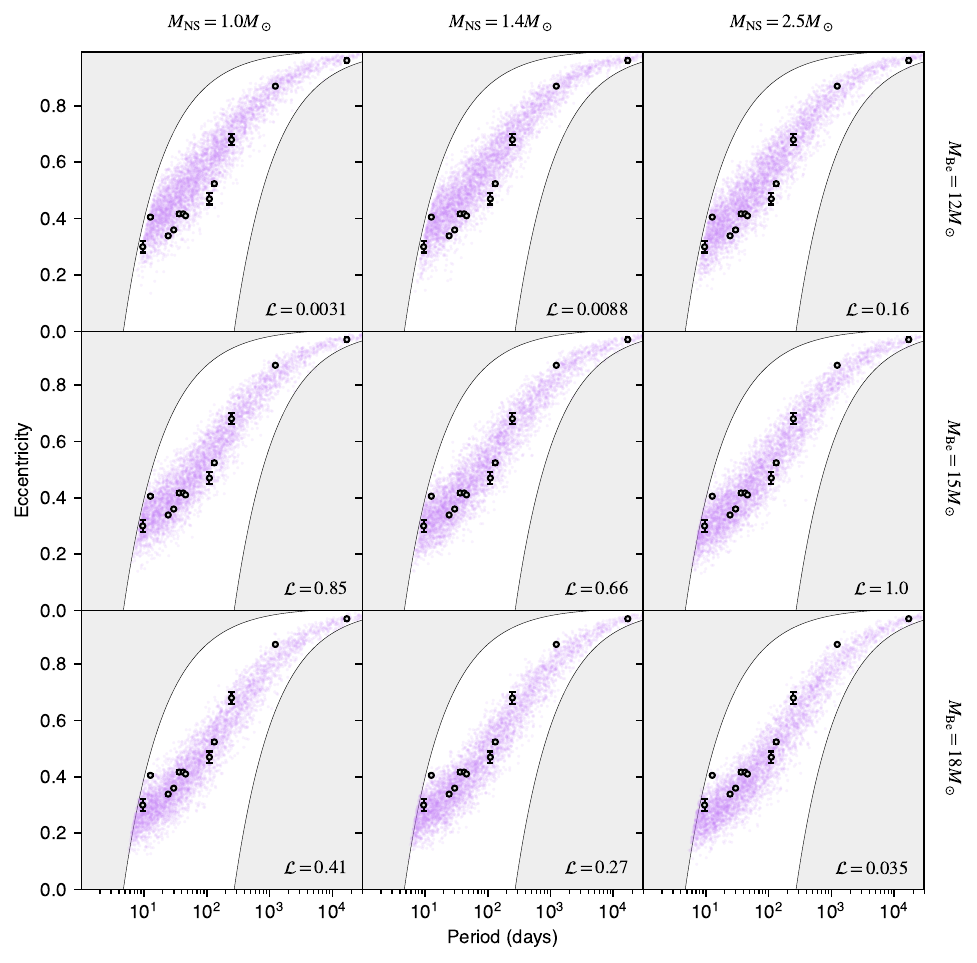}
    \caption{\textbf{Effect of varying the Be star mass and neutron star mass}. As in fig.~\ref{fig:spread_in_angle_and_velocity}, but we show in purple the result of our preferred model \texttt{High-e preferred}, varying the assumed Be star mass $M_{\rm Be}$ (from top to bottom) and neutron star mass $M_{\rm NS}$ (from left to right). The black points indicate the observed systems in the high-e sequence. The predicted distribution does not appear to change appreciably when varying the neutron star mass, while there is a weak dependence on the mass of the Be star.}
    \label{fig:exploring_MBe_MNS}
\end{figure}

The neutron star masses in our sample of Be X-ray binaries have not been measured, and the Be star masses are typically estimated from spectral types, a procedure affected by significant uncertainties and biases. Given this lack of precise mass measurements, we adopt representative values of $M_{\rm NS} = 1.4\,\Msun$ and $M_{\rm Be} = 15\,\Msun$, consistent with typical expectations. In fig.~\ref{fig:exploring_MBe_MNS}, we explore how these mass assumptions influence the predicted orbital periods and eccentricities. Using the best-fit parameters of the \texttt{High-e preferred} model, we vary the neutron star and Be star masses across the panels. The results show that reasonable variations in these masses, within the expected observational ranges \cite{ozel_masses_2016}, have only a modest impact on the model predictions.

\subsection{What is the impact of a pre-supernova eccentricity?} \label{sec:effect of initial eccentricity}
\begin{figure}
    \centering
    \includegraphics{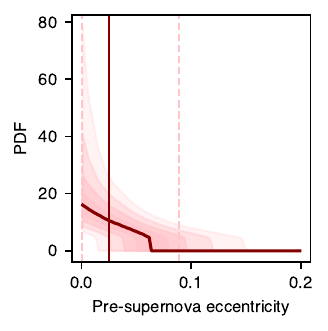}
    \includegraphics{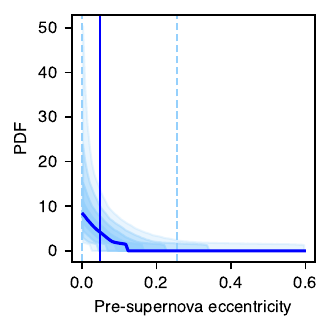}
    \caption{\textbf{Inferred distribution of pre-supernova eccentricity}.
    Left panel: low-e group; right panel: high-e sequence. These distributions are obtained by extending the preferred models for each group to allow for a pre-supernova eccentricity drawn from a uniform distribution between 0 and a maximum value $e_{\rm max}$, with a flat prior on $e_{\rm max}$ between 0 and 1. Solid curves show the median posterior probability, while shaded bands represent deciles from the 10th to the 90th percentile. Vertical solid lines mark the median eccentricity, and dashed lines indicate the 90\% credible interval.}
    \label{fig:pre-sn eccentricity probability}
\end{figure}

Throughout this work, we have assumed that the pre-supernova orbit was circular. This assumption is supported by both theoretical expectations and observations of analogous systems that have not yet undergone a supernova (\ref{sec:presn orbit was circular}).
To test the robustness of our results, we now consider the possibility of a non-zero pre-supernova eccentricity. Specifically, we assume that eccentricities were uniformly distributed between 0 and a maximum value $e_{\rm max}$, left as a free parameter with a flat prior between 0 and 1. We compute the post-supernova orbit assuming that the explosion occurs at a random orbital phase.
We find that allowing for pre-supernova eccentricity has a negligible effect on our main conclusions: the inferred kick velocities and mass loss remain consistent with those obtained from the \texttt{Low-e preferred} and \texttt{High-e preferred} models, which assume circular orbits. Figure~\ref{fig:pre-sn eccentricity probability} shows the resulting posterior distribution of pre-supernova eccentricities for the low-e group and the high-e sequence. In both cases, over 90\% of the probability lies below $e = 0.25$, in line with measurements of Be+sdO and Be+stripped star systems, whose observed eccentricities are typically below 0.2 (see fig.~\ref{fig:progenitors}).

\subsection{What is the impact of the pre-supernova parameter distributions?}
\label{sec:effect_of_pre-SN_distributions}

To test the impact of our assumptions about pre-supernova distributions, we explored an alternative set of models in which the progenitor mass and orbital period are drawn from power-law distributions, rather than the (log)normal forms used in our preferred models. These power laws, defined by a minimum, maximum, and slope, offer greater flexibility at the cost of additional parameters. We find that the resulting kick velocity and ejecta mass distributions are fully consistent with those derived from the original parametrization, indicating that our main conclusions are robust to these modeling choices.

\section{Comparison with independent observables} \label{sec:independent observables}

\subsection{Velocity of isolated pulsars} \label{sec:isolated pulsars}

In Fig.~\ref{fig:additional_predictions}A, we compare the kick velocity distribution inferred for the high-e sequence with estimates from young isolated pulsars (ages $< 3\,\Myr$), whose velocity is least affected by the interaction with the Galactic potential. All distributions are normalized to 1 at their maximum. The two curves for the isolated pulsars are based on the same pulsar sample but rely on different methods: \cite{igoshev_observed_2020} derive present-day velocities from proper motion and parallax measurements, while \cite{disberg_kinematically_2025} infer the birth velocity from the eccentricity of their Galactic orbits.
We note that, when older pulsars are included in the sample, the bimodality is no longer recovered. The reason for this difference remains unclear \cite{disberg_kinematically_2025}, although it is possible that larger uncertainties on the kicks of the older systems could mask the feature. When restricted to young pulsars, both methods yield consistent results, showing a bimodal distribution whose low-velocity component matches the one we infer for the high-e sequence, suggesting a possible common origin (See also sec.~\ref{sec:prior work:kick velocity})

\subsection{Spin-orbit misalignment} \label{sec:spin-orbit misalignment}

Before the supernova, the spin axis of the Be star is expected to be aligned with the system's orbital angular momentum axis \cite{marcussen_banana_2022, marcussen_banana_2024}. The supernova kick then tilts the orbital plane, but the Be star's spin axis remains unchanged. 
Consequently, the angle between the Be star's spin axis and the new orbital angular momentum axis directly reflects the tilt induced by the supernova. This spin-orbit misalignment provides a measurable constraint on the magnitude and direction of the kick imparted by the explosion. 
Figure~\ref{fig:additional_predictions}B shows predictions from the \texttt{High-e preferred} model in blue, which assumes supernova kicks are confined to a narrow cone around the polar axis. The black dashed line represents the spin-orbit tilt predicted for strictly polar kicks, from Eq.\ref{eq:spin-orbit angle from polar kicks}, with its two free parameters obtained by fitting eq.~\ref{eq:ecc_from_polar_kicks} to the period and eccentricity distribution of the high-e sequence. For comparison, predictions from the \texttt{Low-e preferred} model are shown in red.

The only available measurement of spin-orbit tilt comes from PSR B1259–63 \cite{shannon_kinematics_2014}, a system that is particularly valuable for being both an X-ray binary and a radio pulsar. This made it possible to measure the orbital precession caused by spin-orbit coupling, from which the spin-orbit misalignment angle was inferred. The measurement is in excellent agreement with the high-e model prediction. In addition, the system's center-of-mass velocity relative to its likely birthplace, Cen OB1, measured at $34\pm13\,\kms$ \cite{miller-jones_geometric_2018}, also matches our model predictions (\ref{sec:systemic velocity}). Although this conclusion is based on a single object, it provides further evidence for the preferential polar alignment of supernova kicks. Additional measurements of spin-orbit tilt in other systems would significantly enhance confidence in the polar-kick model, should they support the same trend. 

\begin{figure}
    \centering
    \includegraphics{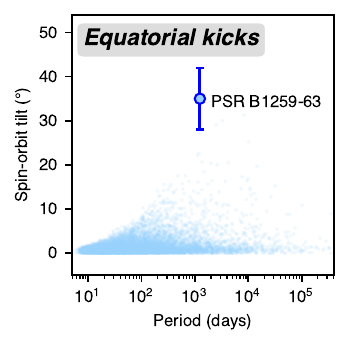}
    \caption{\textbf{Spin-orbit misalignment angle induced by equatorial kicks.} We show predictions for the spin-orbit tilt of a model with kicks restrained within a wedge around the equatorial plane. Except for the opening angle of the wedge, which we fix at $5^\circ$, we assume the median posterior values of the parameters of the equatorial kick model discussed in sec.~\ref{sec:polar vs equatorial}.}
    \label{fig:spin-orbit tilt equator}
\end{figure}

The observed tilt angle also strongly disfavors kick scenarios confined to the equatorial plane. This is illustrated in fig.~\ref{fig:spin-orbit tilt equator}, where predictions for kicks restricted to a $5^\circ$ wedge around the equator yield tilt angles inconsistent with the observed value.

\subsection{Double neutron stars} \label{sec:dns}
Double neutron star (DNS) systems can offer an additional constraint on supernova kicks, and they have been extensively studied in this context \cite{fryer_double_1997,hughes_constraints_1999,dewi_spin_2005,beniamini_formation_2016,chruslinska_constraints_2017,tauris_formation_2017,vigna-gomez_formation_2018,andrews_double_2019,disberg_kinematic_2024}. While their properties and formation channels differ from Be X-ray binaries, both classes are expected to have been in circular orbits that become eccentric following a supernova explosion. We therefore investigate if the same kick velocities we infer from Be X-ray binaries can also reproduce the observed orbital properties of DNSs.

To this aim, we collect a sample of DNSs from the \texttt{ATNF} pulsar catalog (v2.5.1) \cite{manchester_australia_2005}. We select all systems classified as binaries with a neutron star companion and with measured orbital period and eccentricity, yielding a sample of 25 DNS.
Four are located in globular clusters and could have formed dynamically, while three have uncertain companion classification. Excluding these systems does not affect our overall conclusions.

We take into account gravitational-wave induced orbital decay, by following \cite{tauris_formation_2017,andrews_double_2019}: we integrate the orbits backward in time using the equations of \cite{peters_gravitational_1964}, for a duration equal to the pulsar’s characteristic age $\tau_c=P_s/(2\dot P_s)$, where $P_s$ is the pulsar spin period. 
Similar to Be X-ray binaries, the DNS population also appears to separate into two distinct groups: a low-e group ($e < 0.5$) and a high-e group ($e > 0.5$), separated by a noticeable gap (Fig.~\ref{fig:additional_predictions}C).

We model these two groups using our \texttt{Low-e preferred} and \texttt{High-e preferred} models, but keeping the kick velocity distributions fixed to the median posterior values inferred from Be X-ray binaries. We set the mass of the companion to $1.4\,\Msun$ (neutrons stars masses in DNSs range between $1.1\,\Msun$ and $1.6\,\Msun$ \cite{ozel_masses_2016}). We allow the pre-supernova period and progenitor mass distributions to vary, since these are expected to differ from those of Be X-ray binaries. DNSs tend to have shorter periods, and their progenitors may be more deeply stripped prior to collapse \cite{tauris_ultra-stripped_2015}.

Figure~\ref{fig:additional_predictions}C shows that the best-fit models, which assume the same velocity distributions inferred from Be X-ray binaries, are also able to reproduce the DNS populations.
We find ejecta masses of $\Delta M = 0.5^{+0.4}_{-0.5}\,\Msun$ for the low-e group and $\Delta M = 1.9^{+0.8}_{-0.8}\,\Msun$ for the high-e group, lower than the ones we found for Be X-ray binaries and indicative of a deeper stripping of the supernova progenitors. The latter value agrees with estimates from \cite{andrews_double_2019}, however, the kick velocities we have assumed, around $100\,\kms$, are much higher than the $25\,\kms$ inferred in their work. The reason is that \cite{andrews_double_2019} assumed an isotropic kick distribution, which produces a broad spread in the period-eccentricity plane, while the observations occupy a relatively small region. By restricting the kicks to the polar direction, we find that the observed period-eccentricity distribution can be well reproduced, even with the higher kick velocities we infer from Be X-ray binaries. The kick velocity and direction inferred for each system individually by \cite{tauris_formation_2017} is also compatible with the scenario we propose, with the two exceptions of PSR B1534+12 and PSR B1913+16, which require kicks above $200\,\kms$, and may thus have received a ``Hobbs'' kick, or may have been dynamically formed and then ejected by a cluster \cite{andrews_double_2019}.

Despite their different properties and evolutionary histories, the fact that the same kick distributions can account for the orbital properties of both Be X-ray binaries and DNSs suggests that our results may capture a more universal aspect of neutron star formation.

\subsection{Role of metallicity: analog systems in the Small Magellanic Cloud} \label{sec:smc}

\begin{figure}
    \centering
    \includegraphics{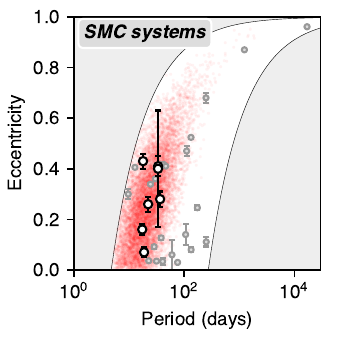}
    \caption{\textbf{Be X-ray binaries in the SMC.} Only seven Be X-ray binary systems in the SMC have reliable solutions for their eccentricities and orbital periods, shown here as large dots. We explore the possibility that low metallicity stripped stars may retain part of their hydrogen envelope: the red point cloud is the best fit population, obtained by assuming the posterior median kick velocity distribution that we derived for low-e group, and leaving the progenitor masses and pre-supernova period distributions free.}
    \label{fig:smc_fit}
\end{figure}

At low metallicity, weaker winds \cite{vink_mass-loss_2001, gotberg_stellar_2023} and less efficient stripping during mass transfer result in stars with more massive envelopes than their higher metallicity counterparts \cite{gotberg_ionizing_2017,yoon_type_2017}. As more material is ejected during the supernova explosion, the eccentricity of the system is boosted. Higher eccentricities are indeed observed in Be X-ray binaries in the SMC, which has a metallicity five times lower than the solar value. 
Only seven Be X-ray binaries in the SMC have a measurement of eccentricity \cite{coe_catalogue_2015}, and they are shown as large circles in fig.~\ref{fig:smc_fit}. Although this sample is small and the uncertainties are, in some cases, large, these systems offer valuable insights, as they probe progenitors at lower metallicity.

We show in fig.~\ref{fig:smc_fit}, that we can reproduce the observations by modeling an SMC population with the same kick parameters derived from the \texttt{Low-e preferred} model, but with a larger mass loss $\Delta M= 4.4^{+4.5}_{-4.4}\,\Msun$, as expected for lower metallicity systems. However, we caution that the current sample of SMC systems with well-determined parameters is very limited, and this analysis should be revisited once a larger, more complete sample becomes available.

\subsection{Systemic velocities} \label{sec:systemic velocity}
\begin{figure}
    \centering
    \includegraphics{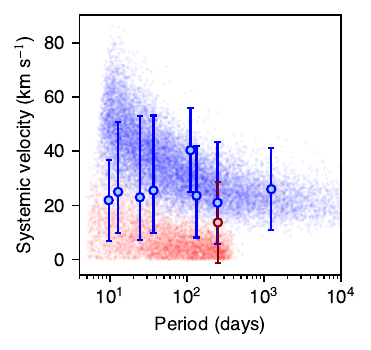}
    \caption{\textbf{Systemic velocities.} We compare the systemic velocity (the velocity of the center of mass) predicted by our preferred models (small dot clouds) with the ones obtained by \cite{fortin_constraints_2022} (large circles). In red is the low-e group, with the high-e sequence in blue.}
    \label{fig:systemic_velocity}
\end{figure}

At the time of a supernova explosion, the combination of sudden mass loss and natal kick can impart a velocity to the center of mass of the binary, which can be used as an additional probe of supernova kicks.
In fig.~\ref{fig:systemic_velocity}, we compare the observed systemic velocities of Be X-ray binaries in our sample with our preferred model's best-fit prediction.
The measurements are taken from \cite{fortin_constraints_2022}, who estimated the 3D systemic velocity by combining the Gaia proper motion and the radial velocity, when available. They then subtracted the mean Galactic motion at the location of the binary to obtain the systemic velocity induced by the supernova.

The model appears compatible with the observations. However, only a subset of systems have measured systemic velocities, and only one (X Per) belongs to the low-e group. In addition, the measurements carry large uncertainties, mainly due to the local stellar velocity dispersion, which may blur the separation between the low-e and high-e groups. This could partly explain why \cite{fortin_constraints_2022} did not identify the low-velocity component we associate with the low-e group (See also sec.~\ref{sec:prior work:kick velocity}).

\subsection{Neutron stars that inherit their spin axis from the progenitor} \label{sec:spin-velocity alignment}

Our model predicts that, for at least one class of progenitors, neutron star kicks are preferentially directed along the rotation axis of the progenitor star. If the neutron star also inherits its spin axis from the progenitor’s core rotation---a long debated scenario \cite{spruit_birth_1998,heger_presupernova_2005,ott_spin_2006,coleman_kicks_2022,janka_supernova_2022}---this would naturally lead to an alignment between the spin and velocity of isolated neutron stars.

Indeed, such alignment has been observed in individual systems \cite{dodson_vela_2003, kaplan_precise_2008, yao_evidence_2021} and supported by statistical studies of pulsar populations \cite{johnston_evidence_2005, johnston_evidence_2007, wang_neutron_2006, rankin_further_2007, noutsos_pulsar_2012, noutsos_pulsar_2013, rankin_toward_2015, biryukov_evidence_2024}. These findings are consistent with our prediction and suggest a connection between the direction of the natal kick and the progenitor’s spin axis (See also sec.~\ref{sec:prior work:preferential direction}).

\subsection{Velocities of unbound Be star companions} \label{sec:runaway Be stars}
In systems that become unbound, the Be star companion is ejected with a velocity close to its linear orbital velocity before the supernova explosion. For polar kicks, the systems that are disrupted are those where the relative orbital velocity is $v_{\rm orb} \le v_{\rm kick} \sim 100\kms $. The orbital velocity of the Be star is simply given by $v_{\rm orb,Be} = v_{\rm orb}/(1+M_{\rm Be}/M_{\rm str})$. For typical values of $M_{\rm Be}=15\,\Msun$ and $M_{\rm str} = 5\,\Msun$, we obtain $v_{\rm orb,Be} \le 25\kms$, consistent with observed velocities of runaway Be stars \cite{berger_search_2001, boubert_kinematics_2018}.

\subsection{Radio pulsars with massive companions}
Our sample consists of systems detected in X-rays. Two of the systems also show radio pulses (PSR B1259-63 and PSR J2032+4127).  Beyond these, further radio pulsars with massive OB-type companions have been identified. Those additional systems were not included in our sample, since we are not aware that they have X-ray detections, but here we briefly compare their properties with our sample. 

The properties of two of the systems seem to naturally support the existence of the two groups we identify. 
PSR J1740-3052 ($e\approx0.58$, period 231\,days) has a late O-type or early B-type companion, and falls neatly on the high-e sequence \cite{stairs_psr_2001, madsen_timing_2012}.
PSR J2108+4516 ($e \approx 0.09$, period 269\,days) fits neatly in the low-eccentricity group \cite{andersen_chime_2023}. 
PSR J1954+2529 ($e \approx 0.11$, period 82.7\,days) would also fall in the low-eccentricity group but seems likely to have a low-mass companion \cite{parent_study_2022}. 
%
These systems would also be interesting candidates for testing the any tilt between the binary orbit and the orientation of the B-type star (\ref{sec:spin-orbit misalignment}). 

We further note PSR J0045-7319 ($e \approx 0.81$, period 51\,days), which is thought to have a B star companion \cite{kaspi_massive_1994, kaspi_evidence_1996}. This system is sufficiently eccentric that it falls in the region where tides strongly affect the orbit, and does not fit within the two groups we identify.  This system may result from a larger, Hobbs-like, kick (mode III) where the direction of the kick was in the opposite direction of the pre-supernova orbital motion, the outcome of which would also fit with the suggestion that the orbit is counterrotating with respect to the spin of the B-type star \cite{lai_orbital_1996,kumar_orbital_1998,su_dynamical_2022}.

\section{Comparison with previous studies} \label{sec:prior work}
The study of supernova kick properties has driven significant observational and theoretical efforts for over the past decades. Data on kick magnitude, direction, and velocity distribution have been collected from various classes of systems, each providing a unique perspective on the phenomenon. We summarize the key findings from previous studies and compare them with our current results. 

\subsection{Multiple neutron star populations as evidence of different supernova mechanisms} \label{sec:prior work:kick velocity}

Numerous studies have attempted to infer the distribution of natal kicks imparted to neutron stars, particularly to determine whether there is evidence for multiple components in the distribution. We categorize these studies based on the primary observable they use: the velocity of isolated pulsars, the orbital properties of binary systems, and the systemic velocity of binaries. We illustrate how each of these observables is sensitive to different velocity ranges. When combined, these results come together to reveal the presence of the three distinct components in the velocity distribution that we propose in this work.

\subsubsection{Constraints from velocities of isolated pulsars}
Early suggestions that neutron stars receive natal kicks during the supernova explosion were prompted by the observation of a peculiar source (the neutron star) with a large proper motion in the Crab Nebula \cite{trimble_motions_1968} and by the spatial distribution of pulsars in our galaxy \cite{gunn_nature_1970}. 

Over time, this has led to multiple efforts to measure the proper motions of various pulsars. A major contribution was given by \cite{hobbs_statistical_2005}. Based on the 2D velocities of 73 young pulsars, they find that the magnitude of their 3D velocity can be described by a Maxwellian distribution with $\sigma=265\,\kms$.

The shape of the distribution, in particular the question of whether there is evidence for multiple components, has been a matter of debate. Early claims of bimodality \cite{cordes_neutron_1998, arzoumanian_velocity_2002, brisken_proper-motion_2003}, were later argued to be the effect of small sample size and systematic uncertainties \cite{hobbs_statistical_2005, faucher-giguere_birth_2006}.  
However, in the following decade, continued improvements in the distance determinations and measurements of the proper motions of isolated pulsars allowed to separate two peaks in their velocity distribution, corresponding to Maxwellian distributions with $\sigma=56\,\kms$ and $\sigma=336\,\kms$ \cite{verbunt_observed_2017, igoshev_observed_2020}.
The low velocity component of this distribution is compatible with the velocity we find for our high-e sequence, hinting that the low velocity isolated pulsars and the neutron stars in our high-e sequence may belong to the same underlying population.

The velocity of old pulsars may have been significantly influenced by the interaction with the Galactic potential. For this reason, most studies focus on young pulsars, less than a few million years old, for which the velocity is more closely related to the kick velocity, but this reduces substantially the sample size. \cite{disberg_kinematically_2025} overcome this limitation estimating the eccentricity of the Galactic orbit, which remains correlated with the original kick velocity for a longer time. If they apply this method to young pulsars only, they recover the bimodal distribution of \cite{verbunt_observed_2017, igoshev_observed_2020}. However, when they include older pulsars, they do not find a bimodal distribution anymore. 

The velocity of older pulsars can be significantly altered by interactions with the Galactic potential, which is why most studies restrict their analyses to young pulsars (less than a few million years old) whose present-day velocities more directly reflect their natal kicks. However, this approach considerably limits the sample size. \cite{disberg_kinematically_2025} propose a way around this by using the eccentricity of the pulsars' Galactic orbits, a quantity that remains correlated with the original kick velocity over longer timescales. When applying this method to a young pulsar sample, they recover the bimodal velocity distribution reported by \cite{verbunt_observed_2017, stockinger_three-dimensional_2020}. Yet, when the analysis is extended to include older pulsars, the bimodality is no longer recovered, reigniting the ongoing debate over the true shape of the neutron star kick velocity distribution.

Isolated pulsar velocities are helpful in constraining the higher end of the supernova kick velocity distribution. However, these studies inherently lack sensitivity to the lower-velocity end, and they exclude neutron stars whose kick was not strong enough to unbind the binary. To better investigate the low-velocity tail of the kick distribution, it is more effective to study bound systems such as X-ray binaries and double neutron stars. In these systems, the primary observables that can be used to constrain the kicks are the orbital parameters (period, eccentricity, and spin-orbit misalignment) and the system's center-of-mass velocity, or systemic velocity.

\subsubsection{Constraints from the orbits of binary systems}
The orbits of a sample of Be X-ray binaries were analyzed by \cite{pfahl_new_2002}, who identified a sub-population of systems with long orbital period ($P>30\,{\rm d}$) and small eccentricity ($e<0.3$). They argue that these systems must have received a kick of less than $50\,\kms$, smaller than the typical velocity of isolated pulsars. They speculate that stars whose envelope is removed during the main sequence retain a fast rotation in the core and explode with a small kick, while stars that are single or that transfer mass after the main sequence lose most of their angular momentum in the giant phase, have a slowly rotating core and explode with a large kick. However, the physical mechanism linking the core's rotation to the magnitude of the kick was not clearly established in their argument.
A different interpretation comes from \cite{podsiadlowski_effects_2004, van_den_heuvel_x-ray_2004, van_den_heuvel_double_2007}, who argue that the low natal kicks are a signature of electron-capture supernovae. These supernovae are expected to feature a faster explosion than core-collapse supernovae, during which asymmetries in the ejecta do not have time to develop, resulting in smaller kicks. While electron-capture supernovae may indeed provide small kick velocities, it is now believed that kicks of tens of $\kms$ are a characteristic of all supernovae with a low mass progenitor ($8$--$10\,\Msun$) independently of whether the collapse is triggered by electron capture \cite{janka_interplay_2024, wang_supernova_2024}.
Further evidence for the two supernovae hypothesis is provided by \cite{knigge_two_2011}, who showed that the spin distribution of the neutron stars in Be X-ray binaries is bimodal and correlated with the eccentricity.
However, the present spin of the neutron star does not reflect the post-supernova value, but is the result of an equilibrium between accretion and spin-down \cite{waters_relation_1989}. This has led to the interpretation that the observed spin distribution arises from differences in accretion modes rather than from intrinsic differences in the supernovae themselves \cite{cheng_spin_2014, xu_bimodal_2019}.

Like Be X-ray binaries, double neutron star systems containing a pulsar provide valuable insights into supernova physics due to the extreme precision with which their orbital elements can be measured.
Similarly to Be X-ray binaries, DNSs can be divided in two groups according to their eccentricity \cite{van_den_heuvel_double_2007}. Furthermore, the DNSs with low eccentricity contain neutron stars that spin faster \cite{beniamini_formation_2016}, and are less massive \cite{schwab_further_2010}. This latter point is consistent with the expectation that lower mass progenitors would produce less massive neutron stars and weaker natal kicks.

The orbital period and eccentricity were used by \cite{beniamini_formation_2016} to infer the single best value of kick magnitude and mass loss that reproduce the two groups. Their findings suggest that the neutron stars in low eccentricity systems formed from low-mass progenitors (ejected mass $\Delta M < 0.4\,\Msun$) with low kick velocity ($v_{\rm kick} < 12\,\kms$), whereas the more eccentric population resulted from larger mass ejections ($\Delta M < 2.2\,\Msun$) and higher kick velocities ($v_{\rm kick} \sim 150\,\kms$). These results are again consistent with our findings regarding Be X-ray binaries, suggesting that the processes producing the supernova kicks are the same between the two classes of systems.

For all the known DNSs, \cite{tauris_formation_2017} constrained individually the kick magnitude, direction, and mass loss, using the available information on the orbit, masses, and systemic velocity. They also can separate the systems in two groups, based on their kick velocity, and find a correlation with the neutron star mass.
However, they did not attempt to infer the kick distribution of the population as a whole.

Instead, \cite{andrews_double_2019} subdivided the known DNSs into three groups. They argued that the systems with short periods and high eccentricities must have received small kicks ($v_{\rm kick} < 25, \kms$) and lost around $1.8, \Msun$. Their argument is based on the realization that larger kicks would spread the systems over a broader region in the period-eccentricity plane, whereas observations seem clustered in a small region around $P = 0.3, \text{days}$ and $e = 0.6$. This conclusion holds under the assumption of isotropic kicks with a Maxwellian distribution. However, as shown in sec.~\ref{sec:dns}, larger kicks ($v_{\rm kick} \sim 100, \kms$) in the polar direction can also reproduce the observed data.

Several population synthesis studies have modeled the observed velocity of isolated pulsars \cite{bray_neutron_2016, bray_neutron_2018, giacobbo_revising_2020, igoshev_combined_2021, willcox_constraints_2021, kapil_calibration_2023}, and the orbits of DNSs \cite{andrews_evolutionary_2015, chruslinska_constraints_2017, kruckow_progenitors_2018, vigna-gomez_formation_2018, shao_role_2018}, and (Be) X-ray binaries \cite{misra_x-ray_2023, liu_population_2024, rocha_be_2024, wang_neutron_2025} to constrain kick velocities and calibrate theoretical models.
Despite the large uncertainties introduced by rapid population synthesis, these studies generally support the need for at least two different supernova prescriptions, with varying kick velocities, to match the observed properties.

\subsubsection{Constraints from systemic velocities of binary systems}

While studies focused on the eccentricity of Be X-ray binaries and DNSs support the idea of two different neutron star populations, works primarily based on systemic velocity often do not reach the same conclusion.
\cite{bodaghee_clustering_2012, zuo_displacement_2015, repetto_galactic_2017} use the distance of Galactic HMXBs from their estimated birthplaces as a proxy for systemic velocity and conclude that they have received kicks with $v_{\rm kick} > 100\,\kms$. However, \cite{bodaghee_evidence_2021} repeats a similar study for the SMC, finding significantly smaller kick velocities, $v_{\rm kick} \sim 2-34\,\kms$.

The results of \cite{prisegen_kinematic_2020} are even more puzzling: using the systemic velocities for Galactic Be X-ray binaries, and distances from the birthplace for systems in the SMC, they find that the systemic velocity is smaller for systems with larger eccentricity and longer spin period, which is the opposite of what would be expected if both the eccentricity and the systemic velocity are produced by a supernova kick.

The revolution brought by Gaia allowed for the first time to study the 3D velocity of large samples of X-ray binaries, without having to rely on proxies like the distance from the assumed birthplace. Using this method, \cite{fortin_constraints_2022} infer kick velocities and mass loss for a sample of HMXBs. They determined that the kick velocity distribution of their sample has a mean of about $116\,\kms$, and do not find evidence for a second low-kick population. The main reason for the difference with our results is that only one of the systems in their sample, X Per, belongs to our low-eccentricity group. Indeed, they find that X Per received a low-velocity kick. However, a single system is insufficient for them to reach a statistical conclusion about the existence of a low-kick group. A second reason for the discrepancy is that they introduce a $\sigma=15\,\kms$ uncertainty in the systemic velocity estimate to account for the typical velocity dispersion of massive OB binaries. Additionally, they add a 20\% uncertainty on the measurements of periods and eccentricities to account for tidal circularization after the supernova. While this approach is justified for wind-fed supergiant X-ray binaries, the effect of tides is negligible in Be X-ray binaries. These added uncertainties may have been sufficient to obscure the presence of a low-kick population in their analysis.

A similar study on low-mass X-ray binaries (LMXB) was conducted by \cite{odoherty_observationally_2023}, also finding no evidence for a lower-velocity population. However, modeling LMXBs is challenging because, like isolated pulsars, these systems can be several gigayears old. This requires backward integration of their orbits within the Galactic potential, which can introduce significant uncertainties in estimating their original kick velocities.

A comparable study by \cite{zhao_evidence_2023} analyzes the systemic velocities of HMXBs, LMXBs, and binary pulsars, revealing an anti-correlation between the system's peculiar velocity and both its total mass and orbital period. This finding aligns with the expectation that the kick velocity is independent of the companion star's mass and the orbital period: a given kick will result in a higher systemic velocity when applied to a system with a low-mass companion than to one with a massive star. Similarly, the same kick will produce a greater systemic velocity in a system with a shorter orbital period.

Lastly, \cite{disberg_kinematic_2024} explores the systemic velocities of DNS systems. By exploiting a correlation between kick velocity and the eccentricity of their Galactic orbits, they infer the systemic velocity of each system at birth. However, due to the small sample size and large uncertainties in velocity determination, identifying multiple velocity modes remains difficult. Additionally, the systemic velocity of DNS systems is influenced by the kicks from two supernovae, which can combine in complex ways, further complicating the analysis.

\subsubsection{Drawing a unified picture}
At first glance, constraints on natal kick velocities from isolated pulsars, binary orbital properties, and systemic motions may appear inconsistent, ranging from a few tens to several hundreds of kilometers per second.
However, we argue that these differences are only due to the sensitivity of these studies to different ranges of velocities. Each method preferentially probes distinct regions of the kick velocity distribution, effectively highlighting different peaks within the same underlying structure. Once these selection effects are accounted for, a global picture emerges consisting of the three velocity components listed in Eq.~\ref{eq:three kick components}.

\begin{figure}
    \centering
    \includegraphics{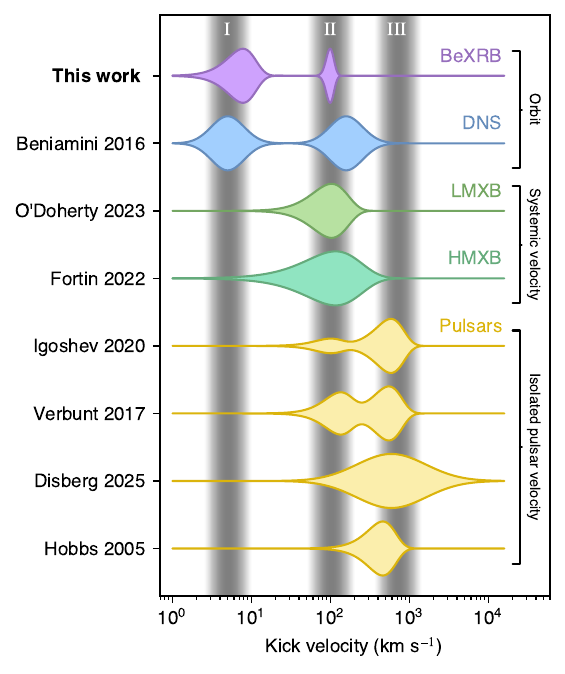}
    \caption{\textbf{Kick velocity distributions from previous studies and this work.} The studies shown focus on various classes of systems, represented by different colors, ranging from isolated pulsars \cite{hobbs_statistical_2005, disberg_kinematically_2025, verbunt_observed_2017, igoshev_observed_2020} to low-mass X-ray binaries (LMXB) \cite{odoherty_observationally_2023}, high-mass X-ray binaries (HMXB) \cite{fortin_constraints_2022}, double neutron stars (DNS) \cite{beniamini_formation_2016}, and Be X-ray binaries (BeXRB). The observables used to infer the kick velocities are annotated on the left, spanning from isolated pulsar velocities to the systemic velocities of binaries and their orbital configurations (period and eccentricity). Each distribution is normalized to its peak value. The three vertical bands mark the approximate positions of three modes that emerge from the comparison across studies (Eq.~\ref{eq:three kick components}). The figure illustrates that these three modes appear consistently across a broad range of astrophysical systems, despite their differing properties and evolutionary histories. However, the method of inference is sensitive to different kick velocity ranges: pulsar velocities predominantly capture the two higher velocity modes, systemic velocities mainly detect the intermediate mode, and orbital configurations are sensitive to the two lower velocity modes.}
    \label{fig:previous work}
\end{figure}

In Fig~\ref{fig:previous work} we compare previous estimates of kick velocity distributions from both isolated pulsars and several types of binary systems.
Studies of isolated pulsars \cite{hobbs_statistical_2005, verbunt_observed_2017, igoshev_observed_2020, disberg_kinematically_2025} are primarily sensitive to the two higher-velocity components, due to the difficulty of measuring the velocity of slow-moving pulsars.
Conversely, analyses focused on binary systems tend to probe the two lower-velocity components, because kicks from the highest-velocity component are likely to unbind the systems.
Studies based on systemic velocities \cite{fortin_constraints_2022, odoherty_observationally_2023} are unable to identify the lowest-velocity component, because the small systemic velocities expected after a kick of below $15-20\,\kms$ are challenging to distinguish from the pre-supernova velocity dispersion of the progenitors.
On the other hand, the orbit of DNSs \cite{beniamini_formation_2016} and Be X-ray binaries, provide stronger constraints on the lowest possible kicks, and allow to identify both the low-velocity and intermediate component in the kick velocity distribution.

\subsection{Kicks with a preferential direction} \label{sec:prior work:preferential direction}

The direction of neutron star kicks, alongside their magnitude, is a crucial characteristic in understanding supernova mechanisms. A long-standing question is whether these kicks are distributed isotropically, meaning that they occur equally in all directions, or if they have a preferred direction, potentially aligned with the spin axis of the pre-supernova core.

Attempts to reproduce the velocity distribution of isolated pulsars observed by \cite{hobbs_statistical_2005} found no statistical preference between isotropic kicks, kicks aligned with the spin axis, or those orthogonal to it \cite{bray_neutron_2016, bray_neutron_2018}. This lack of distinction likely stems from the fact that isolated pulsars provide little information about the pre-supernova stellar spin and orbital plane, making them less effective for determining the direction of the kicks.

In contrast, Be X-ray binaries are more promising, since information on the kick direction is retained in the tilt angle between the Be star spin axis and the orbital angular momentum (\ref{sec:spin-orbit misalignment}).
Unfortunately, measuring this tilt angle presents significant observational challenges and has only been successfully done for one Be+NS system, PSR B1259-63 \cite{shannon_kinematics_2014}. Earlier attempts to constrain spin-kick alignment \cite{hughes_constraints_1999, martin_supernova_2009} used less ideal systems where the spin-orbit inclination had been measured. These include $\gamma$ Cas \cite{hummel_spectacular_1998}, which is now thought to host a white dwarf rather than a neutron star \cite{klement_chara_2024}, and PSR J0045-7319 \cite{kaspi_evidence_1996}, whose rapidly evolving orbit due to strong tidal interactions \cite{lai_orbital_1996,kumar_orbital_1998, su_dynamical_2022} makes it unsuitable for inferring supernova kicks with this method. For PSR B1259-63, an earlier estimate of a spin-orbit misalignment of at least $55^{\circ}$ \cite{wex_timing_1998} was later revised to around $35^{\circ}$ \cite{shannon_kinematics_2014}.

Studying the orbit of DNS systems, \cite{tauris_formation_2017} found a preference for kicks directed backwards relative to the direction of motion and out of the orbital plane. However, they attributed this result to selection bias, as such kicks are more likely to keep the binary bound.

Due to the scarcity of precise spin-orbit misalignment measurements, efforts to identify a preferred direction for supernova kicks have been largely inconclusive—until now. Our work presents the first indication that, at least for a particular class of neutron stars, kicks are aligned with the polar axis.

\section*{Software}
This work made use of the following software: \texttt{matplotlib} \cite{Hunter:2007}, \texttt{numpy} \cite{numpy}, \texttt{pandas} \cite{pandas_13819579}, \texttt{scipy} \cite{mckinney-proc-scipy-2010}, \texttt{python} \cite{python}, \texttt{chatGPT} \cite{chatGPT}, \texttt{corner.py} \cite{corner-Foreman-Mackey-2016, corner.py_4592454}, \texttt{Cython} \cite{cython:2011}, \texttt{emcee} \cite{foreman-mackey_emcee_2013, emcee_10996751}, \texttt{h5py} \cite{collette_python_hdf5_2014, h5py_7560547}, \texttt{ndtest} \cite{peacock_two-dimensional_1983,fasano_multidimensional_1987, ndtest}.

{\footnotesize
\bibliography{scibib}

\begin{thebibliography}{100}
\providecommand{\url}[1]{\texttt{#1}}
\expandafter\ifx\csname urlstyle\endcsname\relax
  \providecommand{\doi}[1]{doi:\discretionary{}{}{}#1}\else
  \providecommand{\doi}{doi:\discretionary{}{}{}\begingroup
  \urlstyle{rm}\Url}\fi

\bibitem{woosley_evolution_2002}
S.~E. Woosley, A.~Heger, T.~A. Weaver, The Evolution and Explosion of Massive
  Stars. \emph{Reviews of Modern Physics} \textbf{74}, 1015--1071 (2002),
  \doi{10.1103/RevModPhys.74.1015}.

\bibitem{burrows_core-collapse_2021}
A.~Burrows, D.~Vartanyan, Core-Collapse Supernova Explosion Theory.
  \emph{Nature} \textbf{589}, 29--39 (2021), \doi{10.1038/s41586-020-03059-w}.

\bibitem{janka_long-term_2025}
H.~T. Janka, Long-{{Term Multidimensional Models}} of {{Core-Collapse
  Supernovae}}: {{Progress}} and {{Challenges}}  (2025),
  \doi{10.48550/arXiv.2502.14836}.

\bibitem{pfahl_new_2002}
E.~Pfahl, S.~Rappaport, P.~Podsiadlowski, H.~Spruit, A {{New Class}} of
  {{High-Mass X-Ray Binaries}}: {{Implications}} for {{Core Collapse}} and
  {{Neutron Star Recoil}}. \emph{The Astrophysical Journal} \textbf{574},
  364--376 (2002), \doi{10.1086/340794}.

\bibitem{fortin_constraints_2022}
F.~Fortin, F.~Garc{\'i}a, S.~Chaty, E.~{Chassande-Mottin}, A.~Simaz~Bunzel,
  Constraints to Neutron-Star Kicks in High-Mass {{X-ray}} Binaries with {{Gaia
  EDR3}}. \emph{Astronomy and Astrophysics} \textbf{665}, A31 (2022),
  \doi{10.1051/0004-6361/202140853}.

\bibitem{fortin_catalogue_2023}
F.~Fortin, F.~Garc{\'i}a, A.~Simaz~Bunzel, S.~Chaty, A Catalogue of High-Mass
  {{X-ray}} Binaries in the {{Galaxy}}: From the {{INTEGRAL}} to the {{Gaia}}
  Era. \emph{Astronomy and Astrophysics} \textbf{671}, A149 (2023),
  \doi{10.1051/0004-6361/202245236}.

\bibitem{liu_population_2024}
B.~Liu, N.~S. Sartorio, R.~G. Izzard, A.~Fialkov, Population Synthesis of {{Be
  X-ray}} Binaries: Metallicity Dependence of Total {{X-ray}} Outputs.
  \emph{Monthly Notices of the Royal Astronomical Society} \textbf{527},
  5023--5048 (2024), \doi{10.1093/mnras/stad3475}.

\bibitem{kalogera_orbital_1996}
V.~Kalogera, Orbital {{Characteristics}} of {{Binary Systems}} after
  {{Asymmetric Supernova Explosions}}. \emph{The Astrophysical Journal}
  \textbf{471}, 352 (1996), \doi{10.1086/177974}.

\bibitem{drout_observed_2023}
M.~R. Drout, \emph{et~al.}, An Observed Population of Intermediate-Mass Helium
  Stars That Have Been Stripped in Binaries. \emph{Science} \textbf{382},
  1287--1291 (2023), \doi{10.1126/science.ade4970}.

\bibitem{panoglou_be_2016}
D.~Panoglou, \emph{et~al.}, Be Discs in Binary Systems - {{I}}. {{Coplanar}}
  Orbits. \emph{Monthly Notices of the Royal Astronomical Society}
  \textbf{461}, 2616--2629 (2016), \doi{10.1093/mnras/stw1508}.

\bibitem{panoglou_be_2018}
D.~Panoglou, \emph{et~al.}, Be Discs in Coplanar Circular Binaries:
  {{Phase-locked}} Variations of Emission Lines. \emph{Monthly Notices of the
  Royal Astronomical Society} \textbf{473}, 3039--3050 (2018),
  \doi{10.1093/mnras/stx2497}.

\bibitem{hobbs_statistical_2005}
G.~Hobbs, D.~R. Lorimer, A.~G. Lyne, M.~Kramer, A Statistical Study of 233
  Pulsar Proper Motions. \emph{Monthly Notices of the Royal Astronomical
  Society} \textbf{360}, 974--992 (2005),
  \doi{10.1111/j.1365-2966.2005.09087.x}.

\bibitem{podsiadlowski_effects_2004}
{\relax Ph}.~Podsiadlowski, \emph{et~al.}, The {{Effects}} of {{Binary
  Evolution}} on the {{Dynamics}} of {{Core Collapse}} and {{Neutron Star
  Kicks}}. \emph{The Astrophysical Journal} \textbf{612}, 1044--1051 (2004),
  \doi{10.1086/421713}.

\bibitem{giacobbo_impact_2019}
N.~Giacobbo, M.~Mapelli, The Impact of Electron-Capture Supernovae on Merging
  Double Neutron Stars. \emph{Monthly Notices of the Royal Astronomical
  Society} \textbf{482}~(2), 2234--2243 (2019), \doi{10.1093/mnras/sty2848}.

\bibitem{vigna-gomez_formation_2018}
A.~{Vigna-G{\'o}mez}, \emph{et~al.}, On the Formation History of {{Galactic}}
  Double Neutron Stars. \emph{Monthly Notices of the Royal Astronomical
  Society} \textbf{481}, 4009--4029 (2018), \doi{10.1093/mnras/sty2463}.

\bibitem{igoshev_combined_2021}
A.~P. Igoshev, M.~Chruslinska, A.~Dorozsmai, S.~Toonen, Combined Analysis of
  Neutron Star Natal Kicks Using Proper Motions and Parallax Measurements for
  Radio Pulsars and {{Be X-ray}} Binaries. \emph{Monthly Notices of the Royal
  Astronomical Society} \textbf{508}, 3345--3364 (2021),
  \doi{10.1093/mnras/stab2734}.

\bibitem{jones_electron-capture_2016}
S.~Jones, \emph{et~al.}, Do Electron-Capture Supernovae Make Neutron Stars?.
  {{First}} Multidimensional Hydrodynamic Simulations of the Oxygen
  Deflagration. \emph{Astronomy and Astrophysics} \textbf{593}, A72 (2016),
  \doi{10.1051/0004-6361/201628321}.

\bibitem{janka_interplay_2024}
H.-T. Janka, D.~Kresse, Interplay between Neutrino Kicks and Hydrodynamic Kicks
  of Neutron Stars and Black Holes. \emph{Astrophysics and Space Science}
  \textbf{369}, 80 (2024), \doi{10.1007/s10509-024-04343-1}.

\bibitem{burrows_theory_2024}
A.~Burrows, T.~Wang, D.~Vartanyan, M.~S.~B. Coleman, A {{Theory}} for {{Neutron
  Star}} and {{Black Hole Kicks}} and {{Induced Spins}}. \emph{The
  Astrophysical Journal} \textbf{963}, 63 (2024),
  \doi{10.3847/1538-4357/ad2353}.

\bibitem{wang_supernova_2024}
T.~Wang, A.~Burrows, Supernova {{Explosions}} of the {{Lowest-mass Massive Star
  Progenitors}}. \emph{The Astrophysical Journal} \textbf{969}, 74 (2024),
  \doi{10.3847/1538-4357/ad5009}.

\bibitem{sykes_long-time_2024}
B.~Sykes, B.~M{\"u}ller, Long-Time {{3D}} Supernova Simulations of Non-Rotating
  Progenitors with Magnetic Fields. \emph{arXiv e-prints}  (2024),
  \doi{10.48550/arXiv.2412.01155}.

\bibitem{sukhbold_core-collapse_2016}
T.~Sukhbold, T.~Ertl, S.~E. Woosley, J.~M. Brown, H.~T. Janka, Core-Collapse
  {{Supernovae}} from 9 to 120 {{Solar Masses Based}} on {{Neutrino-powered
  Explosions}}. \emph{The Astrophysical Journal} \textbf{821}, 38 (2016),
  \doi{10.3847/0004-637X/821/1/38}.

\bibitem{yoon_type_2010}
S.~C. Yoon, S.~E. Woosley, N.~Langer, Type {{Ib}}/c {{Supernovae}} in {{Binary
  Systems}}. {{I}}. {{Evolution}} and {{Properties}} of the {{Progenitor
  Stars}}. \emph{The Astrophysical Journal} \textbf{725}, 940--954 (2010),
  \doi{10.1088/0004-637X/725/1/940}.

\bibitem{yoon_type_2017}
S.-C. Yoon, L.~Dessart, A.~Clocchiatti, Type {{Ib}} and {{IIb Supernova
  Progenitors}} in {{Interacting Binary Systems}}. \emph{The Astrophysical
  Journal} \textbf{840}, 10 (2017), \doi{10.3847/1538-4357/aa6afe}.

\bibitem{eldridge_death_2013}
J.~J. Eldridge, M.~Fraser, S.~J. Smartt, J.~R. Maund, R.~M. Crockett, The Death
  of Massive Stars - {{II}}. {{Observational}} Constraints on the Progenitors
  of {{Type Ibc}} Supernovae. \emph{Monthly Notices of the Royal Astronomical
  Society} \textbf{436}, 774--795 (2013), \doi{10.1093/mnras/stt1612}.

\bibitem{laplace_different_2021}
E.~Laplace, \emph{et~al.}, Different to the Core: {{The}} Pre-Supernova
  Structures of Massive Single and Binary-Stripped Stars. \emph{Astronomy and
  Astrophysics} \textbf{656}, A58 (2021), \doi{10.1051/0004-6361/202140506}.

\bibitem{taddia_early-time_2015}
F.~Taddia, \emph{et~al.}, Early-Time Light Curves of {{Type Ib}}/c Supernovae
  from the {{SDSS-II Supernova Survey}}. \emph{Astronomy and Astrophysics}
  \textbf{574}, A60 (2015), \doi{10.1051/0004-6361/201423915}.

\bibitem{lyman_bolometric_2016}
J.~D. Lyman, \emph{et~al.}, Bolometric Light Curves and Explosion Parameters of
  38 Stripped-Envelope Core-Collapse Supernovae. \emph{Monthly Notices of the
  Royal Astronomical Society} \textbf{457}, 328--350 (2016),
  \doi{10.1093/mnras/stv2983}.

\bibitem{drout_rapidly_2014}
M.~R. Drout, \emph{et~al.}, Rapidly {{Evolving}} and {{Luminous Transients}}
  from {{Pan-STARRS1}}. \emph{The Astrophysical Journal} \textbf{794}, 23
  (2014), \doi{10.1088/0004-637X/794/1/23}.

\bibitem{marcussen_banana_2024}
M.~L. Marcussen, \emph{et~al.}, The {{BANANA Project}}. {{VII}}. {{High
  Eccentricity Predicts Spin}}--{{Orbit Misalignment}} in {{Binaries}}.
  \emph{The Astrophysical Journal} \textbf{975}, 149 (2024),
  \doi{10.3847/1538-4357/ad75fa}.

\bibitem{de_mink_rotation_2013}
S.~E. {de Mink}, N.~Langer, R.~G. Izzard, H.~Sana, A.~{de Koter}, The
  {{Rotation Rates}} of {{Massive Stars}}: {{The Role}} of {{Binary
  Interaction}} through {{Tides}}, {{Mass Transfer}}, and {{Mergers}}.
  \emph{The Astrophysical Journal} \textbf{764}, 166 (2013),
  \doi{10.1088/0004-637X/764/2/166}.

\bibitem{mosta_magnetorotational_2014}
P.~M{\"o}sta, \emph{et~al.}, Magnetorotational {{Core-collapse Supernovae}} in
  {{Three Dimensions}}. \emph{The Astrophysical Journal} \textbf{785}, L29
  (2014), \doi{10.1088/2041-8205/785/2/L29}.

\bibitem{igoshev_observed_2020}
A.~P. Igoshev, The Observed Velocity Distribution of Young Pulsars - {{II}}.
  {{Analysis}} of Complete {{PSR$\pi$}}. \emph{Monthly Notices of the Royal
  Astronomical Society} \textbf{494}, 3663--3674 (2020),
  \doi{10.1093/mnras/staa958}.

\bibitem{disberg_kinematically_2025}
P.~Disberg, N.~Gaspari, A.~J. Levan, A Kinematically Constrained Kick
  Distribution for Isolated Neutron Stars. \emph{arXiv e-prints}  (2025),
  \doi{10.48550/arXiv.2503.01429}.

\bibitem{shannon_kinematics_2014}
R.~M. Shannon, S.~Johnston, R.~N. Manchester, The Kinematics and Orbital
  Dynamics of the {{PSR B1259-63}}/{{LS}} 2883 System from 23 Yr of Pulsar
  Timing. \emph{Monthly Notices of the Royal Astronomical Society}
  \textbf{437}, 3255--3264 (2014), \doi{10.1093/mnras/stt2123}.

\bibitem{tauris_formation_2017}
T.~M. Tauris, \emph{et~al.}, Formation of {{Double Neutron Star Systems}}.
  \emph{The Astrophysical Journal} \textbf{846}, 170 (2017),
  \doi{10.3847/1538-4357/aa7e89}.

\bibitem{beniamini_formation_2016}
P.~Beniamini, T.~Piran, Formation of Double Neutron Star Systems as Implied by
  Observations. \emph{Monthly Notices of the Royal Astronomical Society}
  \textbf{456}, 4089--4099 (2016), \doi{10.1093/mnras/stv2903}.

\bibitem{van_den_heuvel_centaurus_1972}
E.~P.~J. {van den Heuvel}, J.~Heise, Centaurus {{X-3}}, {{Possible
  Reactivation}} of an {{Old Neutron Star}} by {{Mass Exchange}} in a {{Close
  Binary}}. \emph{Nature Physical Science} \textbf{239}, 67--69 (1972),
  \doi{10.1038/physci239067a0}.

\bibitem{laplace_expansion_2020}
E.~Laplace, Y.~G{\"o}tberg, S.~E. {de Mink}, S.~Justham, R.~Farmer, The
  Expansion of Stripped-Envelope Stars: {{Consequences}} for Supernovae and
  Gravitational-Wave Progenitors. \emph{Astronomy and Astrophysics}
  \textbf{637}, A6 (2020), \doi{10.1051/0004-6361/201937300}.

\bibitem{ercolino_mass-transferring_2024}
A.~Ercolino, H.~Jin, N.~Langer, L.~Dessart, Mass-Transferring Binary Stars as
  Progenitors of Interacting Hydrogen-Free Supernovae. \emph{arXiv e-prints}
  (2024), \doi{10.48550/arXiv.2412.09893}.

\bibitem{pfahl_comprehensive_2002}
E.~Pfahl, S.~Rappaport, P.~Podsiadlowski, A {{Comprehensive Study}} of
  {{Neutron Star Retention}} in {{Globular Clusters}}. \emph{The Astrophysical
  Journal} \textbf{573}~(1), 283--305 (2002), \doi{10.1086/340494}.

\bibitem{wagg_delayed_2025}
T.~Wagg, \emph{et~al.}, Delayed and {{Displaced}}: {{The Impact}} of {{Binary
  Interactions}} on {{Core-collapse SN Feedback}}  (2025).

\bibitem{beniamini_natal_2016}
P.~Beniamini, K.~Hotokezaka, T.~Piran, Natal {{Kicks}} and {{Time Delays}} in
  {{Merging Neutron Star Binaries}}: {{Implications}} for r-Process
  {{Nucleosynthesis}} in {{Ultra-faint Dwarfs}} and in the {{Milky Way}}.
  \emph{The Astrophysical Journal} \textbf{829}, L13 (2016),
  \doi{10.3847/2041-8205/829/1/L13}.

\bibitem{powell_three_2023}
J.~Powell, B.~M{\"u}ller, D.~R. {Aguilera-Dena}, N.~Langer, Three Dimensional
  Magnetorotational Core-Collapse Supernova Explosions of a 39 Solar Mass
  Progenitor Star. \emph{Monthly Notices of the Royal Astronomical Society}
  \textbf{522}, 6070--6086 (2023), \doi{10.1093/mnras/stad1292}.

\bibitem{lai_orbital_1996}
D.~Lai, Orbital {{Decay}} of the {{PSR J0045-7319 Binary System}}: {{Age}} of
  {{Radio Pulsar}} and {{Initial Spin}} of {{Neutron Star}}. \emph{The
  Astrophysical Journal} \textbf{466}, L35 (1996), \doi{10.1086/310166}.

\bibitem{hirai_neutron_2024}
R.~Hirai, P.~Podsiadlowski, A.~Heger, H.~Nagakura, Neutron {{Star Kicks}} plus
  {{Rockets}} as a {{Mechanism}} for {{Forming Wide Low-eccentricity Neutron
  Star Binaries}}. \emph{The Astrophysical Journal} \textbf{972}, L18 (2024),
  \doi{10.3847/2041-8213/ad6e77}.

\bibitem{du_initial_2024}
S.-S. Du, \emph{et~al.}, On the {{Initial Spin Period Distribution}} of
  {{Neutron Stars}}. \emph{The Astrophysical Journal} \textbf{968}, 105 (2024),
  \doi{10.3847/1538-4357/ad4450}.

\bibitem{ma_angular_2019}
L.~Ma, J.~Fuller, Angular Momentum Transport in Massive Stars and Natal Neutron
  Star Rotation Rates. \emph{Monthly Notices of the Royal Astronomical Society}
  \textbf{488}, 4338--4355 (2019), \doi{10.1093/mnras/stz2009}.

\bibitem{rivinius_classical_2013}
T.~Rivinius, A.~C. Carciofi, C.~Martayan, Classical {{Be}} Stars. {{Rapidly}}
  Rotating {{B}} Stars with Viscous {{Keplerian}} Decretion Disks.
  \emph{Astronomy and Astrophysics Review} \textbf{21}, 69 (2013),
  \doi{10.1007/s00159-013-0069-0}.

\bibitem{rivinius_classical_2024}
T.~Rivinius, R.~Klement, Classical {{Be}} Stars. \emph{arXiv e-prints}  (2024),
  \doi{10.48550/arXiv.2411.06882}.

\bibitem{rappaport_x-ray_1982}
S.~Rappaport, E.~P.~J. {van den Heuvel}, X-Ray Observations of {{Be}} Stars, in
  \emph{Proceedings of the {IAU Symposium}} (D. Reidel, Dordrecht), vol.~98
  (1982), pp. 327--344.

\bibitem{verbunt_origin_1993}
F.~Verbunt, Origin and Evolution of {{X-ray}} Binaries and Binary Radio
  Pulsars. \emph{Annual Review of Astronomy and Astrophysics} \textbf{31},
  93--127 (1993), \doi{10.1146/annurev.aa.31.090193.000521}.

\bibitem{pols_formation_1991}
O.~R. Pols, J.~Cote, L.~B. F.~M. Waters, J.~Heise, The Formation of {{Be}}
  Stars through Close Binary Evolution. \emph{Astronomy and Astrophysics}
  \textbf{241}, 419 (1991).

\bibitem{vinciguerra_be_2020}
S.~Vinciguerra, \emph{et~al.}, Be {{X-ray}} Binaries in the {{SMC}} as
  Indicators of Mass-Transfer Efficiency. \emph{Monthly Notices of the Royal
  Astronomical Society} \textbf{498}, 4705--4720 (2020),
  \doi{10.1093/mnras/staa2177}.

\bibitem{misra_x-ray_2023}
D.~Misra, \emph{et~al.}, X-Ray Luminosity Function of High-Mass {{X-ray}}
  Binaries: {{Studying}} the Signatures of Different Physical Processes Using
  Detailed Binary Evolution Calculations. \emph{Astronomy and Astrophysics}
  \textbf{672}, A99 (2023), \doi{10.1051/0004-6361/202244929}.

\bibitem{rocha_be_2024}
K.~A. Rocha, \emph{et~al.}, To {{Be}} or {{Not To Be}}: {{The Role}} of
  {{Rotation}} in {{Modeling Galactic Be X-Ray Binaries}}. \emph{The
  Astrophysical Journal} \textbf{971}, 133 (2024),
  \doi{10.3847/1538-4357/ad5955}.

\bibitem{liu_multiwavelength_2024}
W.~Liu, \emph{et~al.}, Multiwavelength Observation of {{1A}} 0535+262={{HD}}
  245770 from 2010 to 2021. \emph{Astronomy and Astrophysics} \textbf{681}, A10
  (2024), \doi{10.1051/0004-6361/202347622}.

\bibitem{packet_spin-up_1981}
W.~Packet, On the Spin-up of the Mass Accreting Component in a Close Binary
  System. \emph{Astronomy and Astrophysics} \textbf{102}, 17--19 (1981).

\bibitem{dewi_evolution_2002}
J.~D.~M. Dewi, O.~R. Pols, G.~J. Savonije, E.~P.~J. {van den Heuvel}, The
  Evolution of Naked Helium Stars with a Neutron Star Companion in Close Binary
  Systems. \emph{Monthly Notices of the Royal Astronomical Society}
  \textbf{331}, 1027--1040 (2002), \doi{10.1046/j.1365-8711.2002.05257.x}.

\bibitem{dewi_late_2003}
J.~D.~M. Dewi, O.~R. Pols, The Late Stages of Evolution of Helium Star-Neutron
  Star Binaries and the Formation of Double Neutron Star Systems. \emph{Monthly
  Notices of the Royal Astronomical Society} \textbf{344}, 629--643 (2003),
  \doi{10.1046/j.1365-8711.2003.06844.x}.

\bibitem{tauris_ultra-stripped_2015}
T.~M. Tauris, N.~Langer, P.~Podsiadlowski, Ultra-Stripped Supernovae:
  Progenitors and Fate. \emph{Monthly Notices of the Royal Astronomical
  Society} \textbf{451}, 2123--2144 (2015), \doi{10.1093/mnras/stv990}.

\bibitem{abbott_multi-messenger_2017}
B.~P. Abbott, \emph{et~al.}, Multi-Messenger {{Observations}} of a {{Binary
  Neutron Star Merger}}. \emph{The Astrophysical Journal} \textbf{848}, L12
  (2017), \doi{10.3847/2041-8213/aa91c9}.

\bibitem{okazaki_natural_2001}
A.~T. Okazaki, I.~Negueruela, A Natural Explanation for Periodic {{X-ray}}
  Outbursts in {{Be}}/{{X-ray}} Binaries. \emph{Astronomy and Astrophysics}
  \textbf{377}, 161--174 (2001), \doi{10.1051/0004-6361:20011083}.

\bibitem{franchini_type_2019}
A.~Franchini, R.~G. Martin, Type {{I Outbursts}} in {{Low-eccentricity
  Be}}/{{X-Ray Binaries}}. \emph{The Astrophysical Journal} \textbf{881}, L32
  (2019), \doi{10.3847/2041-8213/ab3920}.

\bibitem{martin_frequency_2019}
R.~G. Martin, A.~Franchini, The Frequency of {{Kozai-Lidov}} Disc Oscillation
  Driven Giant Outbursts in {{Be}}/{{X-ray}} Binaries. \emph{Monthly Notices of
  the Royal Astronomical Society} \textbf{489}, 1797--1804 (2019),
  \doi{10.1093/mnras/stz2250}.

\bibitem{haubois_dynamical_2012}
X.~Haubois, A.~C. Carciofi, {\relax Th}.~Rivinius, A.~T. Okazaki, J.~E.
  Bjorkman, Dynamical {{Evolution}} of {{Viscous Disks}} around {{Be Stars}}.
  {{I}}. {{Photometry}}. \emph{The Astrophysical Journal} \textbf{756}, 156
  (2012), \doi{10.1088/0004-637X/756/2/156}.

\bibitem{doroshenko_orbit_2018}
V.~Doroshenko, S.~Tsygankov, A.~Santangelo, Orbit and Intrinsic Spin-up of the
  Newly Discovered Transient {{X-ray}} Pulsar {{Swift J0243}}.6+6124.
  \emph{Astronomy and Astrophysics} \textbf{613}, A19 (2018),
  \doi{10.1051/0004-6361/201732208}.

\bibitem{gotberg_ionizing_2017}
Y.~G{\"o}tberg, S.~E. {de Mink}, J.~H. Groh, Ionizing Spectra of Stars That
  Lose Their Envelope through Interaction with a Binary Companion: Role of
  Metallicity. \emph{Astronomy and Astrophysics} \textbf{608}, A11 (2017),
  \doi{10.1051/0004-6361/201730472}.

\bibitem{bashi_features_2023}
D.~Bashi, T.~Mazeh, S.~Faigler, Features of {{Gaia DR3}} Spectroscopic Binaries
  {{I}}. {{Tidal}} Circularization of Main-Sequence Stars. \emph{Monthly
  Notices of the Royal Astronomical Society} \textbf{522}, 1184--1195 (2023),
  \doi{10.1093/mnras/stad999}.

\bibitem{finger_reappearance_1996}
M.~H. Finger, R.~B. Wilson, D.~Chakrabarty, Reappearance of the {{X-ray}}
  Binary Pulsar {{2S}} 1417-624. \emph{Astronomy and Astrophysics Supplement
  Series} \textbf{120}, 209--212 (1996).

\bibitem{raichur_apsidal_2010}
H.~Raichur, B.~Paul, Apsidal Motion in {{4U0115}}+63 and Orbital Parameters of
  {{2S1417-624}} and {{V0332}}+53. \emph{Monthly Notices of the Royal
  Astronomical Society} \textbf{406}, 2663--2670 (2010),
  \doi{10.1111/j.1365-2966.2010.16862.x}.

\bibitem{tsygankov_nustar_2016}
S.~S. Tsygankov, \emph{et~al.}, {{NuSTAR}} Discovery of a Cyclotron Absorption
  Line in the Transient {{X-ray}} Pulsar {{2S}} 1553-542. \emph{Monthly Notices
  of the Royal Astronomical Society} \textbf{457}, 258--266 (2016),
  \doi{10.1093/mnras/stv2849}.

\bibitem{wilson_decade_2002}
C.~A. Wilson, M.~H. Finger, M.~J. Coe, S.~Laycock, J.~Fabregat, A {{Decade}} in
  the {{Life}} of {{EXO}} 2030+375: {{A Multiwavelength Study}} of an
  {{Accreting X-Ray Pulsar}}. \emph{The Astrophysical Journal} \textbf{570},
  287--302 (2002), \doi{10.1086/339739}.

\bibitem{fu_timing_2023}
Y.-C. Fu, \emph{et~al.}, Timing Analysis of {{EXO}} 2030+375 during Its 2021
  Giant Outburst Observed with {{Insight-HXMT}}. \emph{Monthly Notices of the
  Royal Astronomical Society} \textbf{521}, 893--901 (2023),
  \doi{10.1093/mnras/stad614}.

\bibitem{stoyanov_orbital_2014}
K.~A. Stoyanov, R.~K. Zamanov, G.~Y. Latev, A.~Y. Abedin, N.~A. Tomov, Orbital
  Parameters of the High-Mass {{X-ray}} Binary {{4U}} 2206+54.
  \emph{Astronomische Nachrichten} \textbf{335}, 1060 (2014),
  \doi{10.1002/asna.201412127}.

\bibitem{blay_multiwavelength_2006}
P.~Blay, \emph{et~al.}, Multiwavelength Monitoring of {{BD}} +53{$^\circ$}2790,
  the Optical Counterpart to {{4U}} 2206+54. \emph{Astronomy and Astrophysics}
  \textbf{446}, 1095--1105 (2006), \doi{10.1051/0004-6361:20053951}.

\bibitem{kaaret_x-ray_2000}
P.~Kaaret, G.~Cusumano, B.~Sacco, X-{{Ray Timing}} of the 34 {{Millisecond
  Binary Pulsar SAX J0635}}+0533. \emph{The Astrophysical Journal}
  \textbf{542}, L41--L43 (2000), \doi{10.1086/312918}.

\bibitem{watson_2s_1981}
M.~G. Watson, R.~S. Warwick, M.~J. Ricketts, {{2S}} 1145-619: An {{X-ray}}
  Pulsar in an Eccentric Binary System? \emph{Monthly Notices of the Royal
  Astronomical Society} \textbf{195}, 197--203 (1981),
  \doi{10.1093/mnras/195.2.197}.

\bibitem{yoneda_sign_2020}
H.~Yoneda, \emph{et~al.}, Sign of {{Hard-X-Ray Pulsation}} from the {$\gamma$}
  -{{Ray Binary System LS}} 5039. \emph{Physical Review Letters} \textbf{125},
  111103 (2020), \doi{10.1103/PhysRevLett.125.111103}.

\bibitem{makishima_further_2023}
K.~Makishima, N.~Uchida, H.~Yoneda, T.~Enoto, T.~Takahashi, Further
  {{Evidence}} for the 9 s {{Pulsation}} in {{LS}} 5039 from {{NuSTAR}} and
  {{ASCA}}. \emph{The Astrophysical Journal} \textbf{959}, 79 (2023),
  \doi{10.3847/1538-4357/ad0bdf}.

\bibitem{mcswain_n_2004}
M.~V. McSwain, \emph{et~al.}, The {{N Enrichment}} and {{Supernova Ejection}}
  of the {{Runaway Microquasar LS}} 5039. \emph{The Astrophysical Journal}
  \textbf{600}, 927--938 (2004), \doi{10.1086/379892}.

\bibitem{casares_binary_2012}
J.~Casares, \emph{et~al.}, On the Binary Nature of the {$\gamma$}-Ray Sources
  {{AGL J2241}}+4454 (= {{MWC}} 656) and {{HESS J0632}}+057 (= {{MWC}} 148).
  \emph{Monthly Notices of the Royal Astronomical Society} \textbf{421},
  1103--1112 (2012), \doi{10.1111/j.1365-2966.2011.20368.x}.

\bibitem{moritani_orbital_2018}
Y.~Moritani, \emph{et~al.}, Orbital Solution Leading to an Acceptable
  Interpretation for the Enigmatic Gamma-Ray Binary {{HESS J0632}}+057.
  \emph{Publications of the Astronomical Society of Japan} \textbf{70}, 61
  (2018), \doi{10.1093/pasj/psy053}.

\bibitem{adams_observation_2021}
C.~B. Adams, \emph{et~al.}, Observation of the {{Gamma-Ray Binary HESS
  J0632}}+057 with the {{H}}.{{E}}.{{S}}.{{S}}., {{MAGIC}}, and {{VERITAS
  Telescopes}}. \emph{The Astrophysical Journal} \textbf{923}, 241 (2021),
  \doi{10.3847/1538-4357/ac29b7}.

\bibitem{matchett_new_2025}
N.~Matchett, B.~{van Soelen}, New Insight into the Orbital Parameters of the
  Gamma-Ray Binary {{HESS J0632}} + 057. \emph{Monthly Notices of the Royal
  Astronomical Society} \textbf{536}, 166--173 (2025),
  \doi{10.1093/mnras/stae2597}.

\bibitem{sidoli_integral_2018}
L.~Sidoli, A.~Paizis, An {{INTEGRAL}} Overview of {{High-Mass X-ray Binaries}}:
  Classes or Transitions? \emph{Monthly Notices of the Royal Astronomical
  Society} \textbf{481}, 2779--2803 (2018), \doi{10.1093/mnras/sty2428}.

\bibitem{harvey_galactic_2022}
M.~Harvey, C.~B. Rulten, P.~M. Chadwick, The {{Galactic}} High Mass {{X-ray}}
  Binary Population with {{Fermi-LAT}}. \emph{Monthly Notices of the Royal
  Astronomical Society} \textbf{512}, 1141--1168 (2022),
  \doi{10.1093/mnras/stac375}.

\bibitem{reig_optical_2020}
P.~Reig, J.~Fabregat, J.~{Alfonso-Garz{\'o}n}, Optical Counterpart to {{Swift
  J0243}}.6+6124. \emph{Astronomy and Astrophysics} \textbf{640}, A35 (2020),
  \doi{10.1051/0004-6361/202038333}.

\bibitem{delgado-marti_orbit_2001}
H.~{Delgado-Mart{\'i}}, A.~M. Levine, E.~Pfahl, S.~A. Rappaport, The {{Orbit}}
  of {{X Persei}} and {{Its Neutron Star Companion}}. \emph{The Astrophysical
  Journal} \textbf{546}, 455--468 (2001), \doi{10.1086/318236}.

\bibitem{zamanov_spectral_2019}
R.~Zamanov, K.~A. Stoyanov, U.~Wolter, D.~Marchev, N.~I. Petrov, Spectral
  Observations of {{X Persei}}: {{Connection}} between {{H$\alpha$}} and
  {{X-ray}} Emission. \emph{Astronomy and Astrophysics} \textbf{622}, A173
  (2019), \doi{10.1051/0004-6361/201834697}.

\bibitem{doroshenko_sgr_2021}
V.~Doroshenko, A.~Santangelo, S.~S. Tsygankov, L.~Ji, {{SGR}} 0755-2933: A New
  High-Mass {{X-ray}} Binary with the Wrong Name. \emph{Astronomy and
  Astrophysics} \textbf{647}, A165 (2021), \doi{10.1051/0004-6361/202039785}.

\bibitem{richardson_high-mass_2023}
N.~D. Richardson, \emph{et~al.}, A High-Mass {{X-ray}} Binary Descended from an
  Ultra-Stripped Supernova. \emph{Nature} \textbf{614}, 45--47 (2023),
  \doi{10.1038/s41586-022-05618-9}.

\bibitem{wilson_sequence_1997}
C.~A. Wilson, \emph{et~al.}, A {{Sequence}} of {{Outbursts}} from the
  {{Transient X-Ray Pulsar GS}} 0834-430. \emph{The Astrophysical Journal}
  \textbf{479}, 388--397 (1997), \doi{10.1086/303841}.

\bibitem{israel_discovery_2000}
G.~L. Israel, \emph{et~al.}, The Discovery of the Optical/{{IR}} Counterpart of
  the 12-s Transient {{X-ray}} Pulsar {{GS}} 0834-43. \emph{Monthly Notices of
  the Royal Astronomical Society} \textbf{314}, 87--91 (2000),
  \doi{10.1046/j.1365-8711.2000.03404.x}.

\bibitem{int_zand_discovery_2001}
J.~J.~M. {in't Zand}, R.~H.~D. Corbet, F.~E. Marshall, Discovery of a 75 {{Day
  Orbit}} in {{XTE J1543-568}}. \emph{The Astrophysical Journal} \textbf{553},
  L165--L168 (2001), \doi{10.1086/320688}.

\bibitem{lutovinov_2s_2016}
A.~A. Lutovinov, D.~A.~H. Buckley, L.~J. Townsend, S.~S. Tsygankov, J.~Kennea,
  {{2S}} 1553-542: A {{Be}}/{{X-ray}} Binary Pulsar on the Far Side of the
  {{Galaxy}}. \emph{Monthly Notices of the Royal Astronomical Society}
  \textbf{462}, 3823--3829 (2016), \doi{10.1093/mnras/stw1889}.

\bibitem{baykal_orbital_2010}
A.~Baykal, E.~G{\"o}{\v g}{\"u}{\c s}, S.~{\c C}a{\v g}da{\c s}~{\.I}nam,
  T.~Belloni, The {{Orbital Period}} of {{Swift J1626}}.6-5156. \emph{The
  Astrophysical Journal} \textbf{711}, 1306--1309 (2010),
  \doi{10.1088/0004-637X/711/2/1306}.

\bibitem{reig_multi-frequency_2011}
P.~Reig, E.~Nespoli, J.~Fabregat, R.~E. Mennickent, Multi-Frequency
  Observations of {{Swift J1626}}.6-5156. \emph{Astronomy and Astrophysics}
  \textbf{533}, A23 (2011), \doi{10.1051/0004-6361/201117301}.

\bibitem{bissinger_observations_2016}
M.~Bissinger, \emph{Observations of {{Be X-ray Binaries}}: {{Spin Period}} and
  {{Spectral Evolution}}}, Ph.D. thesis (2016).

\bibitem{salganik_nature_2022}
A.~Salganik, \emph{et~al.}, On the Nature of the {{X-ray}} Pulsar {{XTE
  J1859}}+083 and Its Broad-Band Properties. \emph{Monthly Notices of the Royal
  Astronomical Society} \textbf{509}, 5955--5963 (2022),
  \doi{10.1093/mnras/stab3362}.

\bibitem{galloway_discovery_2005}
D.~K. Galloway, Z.~Wang, E.~H. Morgan, Discovery of {{Pulsations}} in the
  {{X-Ray Transient 4U}} 1901+03. \emph{The Astrophysical Journal}
  \textbf{635}, 1217--1223 (2005), \doi{10.1086/497573}.

\bibitem{strader_optical_2019}
J.~Strader, L.~Chomiuk, S.~Swihart, E.~Aydi, Optical {{Spectroscopy}} of the
  {{Candidate Be Star Counterpart}} to {{4U1901}}+03. \emph{The Astronomer's
  Telegram} \textbf{12554}, 1 (2019).

\bibitem{verrecchia_identification_2002}
F.~Verrecchia, \emph{et~al.}, The Identification of the Optical/{{IR}}
  Counterpart of the 15.8-s Transient {{X--ray}} Pulsar {{XTE J1946}}+274.
  \emph{Astronomy and Astrophysics} \textbf{393}, 983--989 (2002),
  \doi{10.1051/0004-6361:20021087}.

\bibitem{ozbey_arabaci_detection_2015}
M.~{\"O}zbey~Arabac{\i}, \emph{et~al.}, Detection of a Large {{Be}}
  Circumstellar Disk during {{X-ray}} Quiescence of {{XTE J1946}}+274.
  \emph{Astronomy and Astrophysics} \textbf{582}, A53 (2015),
  \doi{10.1051/0004-6361/201425488}.

\bibitem{negueruela_bex-ray_2003}
I.~Negueruela, G.~L. Israel, A.~Marco, A.~J. Norton, R.~Speziali, The
  {{Be}}/{{X-ray}} Transient {{KS}} 1947+300. \emph{Astronomy and Astrophysics}
  \textbf{397}, 739--745 (2003), \doi{10.1051/0004-6361:20021529}.

\bibitem{galloway_frequency_2004}
D.~K. Galloway, E.~H. Morgan, A.~M. Levine, A {{Frequency Glitch}} in an
  {{Accreting Pulsar}}. \emph{The Astrophysical Journal} \textbf{613},
  1164--1172 (2004), \doi{10.1086/423265}.

\bibitem{negueruela_bex-ray_2001}
I.~Negueruela, A.~T. Okazaki, The {{Be}}/{{X-ray}} Transient {{4U}}
  0115+63/{{V635 Cassiopeiae}}. {{I}}. {{A}} Consistent Model. \emph{Astronomy
  and Astrophysics} \textbf{369}, 108--116 (2001),
  \doi{10.1051/0004-6361:20010146}.

\bibitem{negueruela_bex-ray_1999}
I.~Negueruela, P.~Roche, J.~Fabregat, M.~J. Coe, The {{Be}}/{{X-ray}} Transient
  {{V0332}}+53: Evidence for a Tilt between the Orbit and the Equatorial Plane?
  \emph{Monthly Notices of the Royal Astronomical Society} \textbf{307},
  695--702 (1999), \doi{10.1046/j.1365-8711.1999.02682.x}.

\bibitem{finger_quasi-periodic_1996}
M.~H. Finger, R.~B. Wilson, B.~A. Harmon, Quasi-Periodic {{Oscillations}}
  during a {{Giant Outburst}} of {{A0535}}+262. \emph{The Astrophysical
  Journal} \textbf{459}, 288 (1996), \doi{10.1086/176892}.

\bibitem{wang_constraints_1998}
Z.~X. Wang, D.~R. Gies, Constraints on the {{Radial Velocity Curve}} of {{HDE}}
  245770 = {{A0535}}+26. \emph{Publications of the Astronomical Society of the
  Pacific} \textbf{110}, 1310--1314 (1998), \doi{10.1086/316269}.

\bibitem{coe_discovery_1994}
M.~J. Coe, \emph{et~al.}, Discovery of the Optical Counterpart to the {{CGRO}}
  Transient {{GRO J1008-57}}. \emph{Monthly Notices of the Royal Astronomical
  Society} \textbf{270}, L57--L61 (1994), \doi{10.1093/mnras/270.1.L57}.

\bibitem{coe_now_2007}
M.~J. Coe, \emph{et~al.}, Now You See It, Now You Don't - the Circumstellar
  Disc in the {{GRO J1008-57}} System. \emph{Monthly Notices of the Royal
  Astronomical Society} \textbf{378}, 1427--1433 (2007),
  \doi{10.1111/j.1365-2966.2007.11878.x}.

\bibitem{parkes_shell_1980}
G.~E. Parkes, P.~G. Murdin, K.~O. Mason, The Shell Spectrum of the Optical
  Counterpart of {{GX}} 304-1 ({{4U}} 1258-61). \emph{Monthly Notices of the
  Royal Astronomical Society} \textbf{190}, 537--542 (1980),
  \doi{10.1093/mnras/190.3.537}.

\bibitem{sugizaki_luminosity_2015}
M.~Sugizaki, T.~Yamamoto, T.~Mihara, M.~Nakajima, K.~Makishima, Luminosity and
  Spin-Period Evolution of {{GX}} 304-1 during Outbursts from 2009 to 2013
  Observed with the {{MAXI}}/{{GSC}}, {{RXTE}}/{{PCA}}, and {{Fermi}}/{{GBM}}.
  \emph{Publications of the Astronomical Society of Japan} \textbf{67}, 73
  (2015), \doi{10.1093/pasj/psv039}.

\bibitem{negueruela_astrophysical_2011}
I.~Negueruela, \emph{et~al.}, Astrophysical {{Parameters}} of {{LS}} 2883 and
  {{Implications}} for the {{PSR B1259-63 Gamma-ray Binary}}. \emph{The
  Astrophysical Journal} \textbf{732}, L11 (2011),
  \doi{10.1088/2041-8205/732/1/L11}.

\bibitem{grindlay_optical_1984}
J.~E. Grindlay, L.~D. Petro, J.~E. McClintock, Optical Identification of {{2S}}
  1417-62. \emph{The Astrophysical Journal} \textbf{276}, 621--624 (1984),
  \doi{10.1086/161650}.

\bibitem{scott_discovery_1997}
D.~M. Scott, \emph{et~al.}, Discovery and {{Orbital Determination}} of the
  {{Transient X-Ray Pulsar GRO J1750-27}}. \emph{The Astrophysical Journal}
  \textbf{488}, 831--835 (1997), \doi{10.1086/304740}.

\bibitem{shaw_accretion_2009}
S.~E. Shaw, \emph{et~al.}, The Accretion Powered Spin-up of {{GRO J1750-27}}.
  \emph{Monthly Notices of the Royal Astronomical Society} \textbf{393},
  419--428 (2009), \doi{10.1111/j.1365-2966.2008.14212.x}.

\bibitem{lyne_binary_2015}
A.~G. Lyne, \emph{et~al.}, The Binary Nature of {{PSR J2032}}+4127.
  \emph{Monthly Notices of the Royal Astronomical Society} \textbf{451},
  581--587 (2015), \doi{10.1093/mnras/stv236}.

\bibitem{ho_multiwavelength_2017}
W.~C.~G. Ho, \emph{et~al.}, Multiwavelength Monitoring and {{X-ray}}
  Brightening of {{Be X-ray}} Binary {{PSR J2032}}+4127/{{MT91}} 213 on Its
  Approach to Periastron. \emph{Monthly Notices of the Royal Astronomical
  Society} \textbf{464}, 1211--1219 (2017), \doi{10.1093/mnras/stw2420}.

\bibitem{motch_accretion_1991}
C.~Motch, L.~Stella, E.~{Janot-Pacheco}, M.~Mouchet, Accretion {{Mechanisms}}
  in the {{Be}}/{{X-Ray Transient System A0535}}+26. \emph{The Astrophysical
  Journal} \textbf{369}, 490 (1991), \doi{10.1086/169779}.

\bibitem{baykal_timing_2007}
A.~Baykal, \emph{et~al.}, Timing Studies on {{RXTE}} Observations of {{SAX
  J2103}}.5+4545. \emph{Monthly Notices of the Royal Astronomical Society}
  \textbf{374}, 1108--1114 (2007), \doi{10.1111/j.1365-2966.2006.11231.x}.

\bibitem{reig_discovery_2004}
P.~Reig, \emph{et~al.}, Discovery of the Optical Counterpart to the {{X-ray}}
  Pulsar {{SAX J2103}}.5+4545. \emph{Astronomy and Astrophysics} \textbf{421},
  673--680 (2004), \doi{10.1051/0004-6361:20035786}.

\bibitem{yu_multiclass_2003}
{Yu}, {Shi}, Multiclass Spectral Clustering, in \emph{Proceedings {{Ninth IEEE
  International Conference}} on {{Computer Vision}}} (IEEE, Nice, France)
  (2003), pp. 313--319 vol.1, \doi{10.1109/ICCV.2003.1238361}.

\bibitem{pedregosa_scikit-learn_2011}
F.~Pedregosa, \emph{et~al.}, Scikit-Learn: {{Machine Learning}} in {{Python}}.
  \emph{Journal of Machine Learning Research} \textbf{12}, 2825--2830 (2011),
  \doi{10.48550/arXiv.1201.0490}.

\bibitem{damle_simple_2019}
A.~Damle, V.~Minden, L.~Ying, Simple, Direct and Efficient Multi-Way Spectral
  Clustering. \emph{Information and Inference: A Journal of the IMA}
  \textbf{8}~(1), 181--203 (2019), \doi{10.1093/imaiai/iay008}.

\bibitem{brown_modelling_2019}
R.~O. Brown, M.~J. Coe, W.~C.~G. Ho, A.~T. Okazaki, Modelling Decretion Discs
  in {{Be}}/{{X-ray}} Binaries. \emph{Monthly Notices of the Royal Astronomical
  Society} \textbf{488}, 387--394 (2019), \doi{10.1093/mnras/stz1757}.

\bibitem{granada_populations_2013}
A.~Granada, \emph{et~al.}, Populations of Rotating Stars. {{II}}. {{Rapid}}
  Rotators and Their Link to {{Be-type}} Stars. \emph{Astronomy and
  Astrophysics} \textbf{553}, A25 (2013), \doi{10.1051/0004-6361/201220559}.

\bibitem{vieira_life_2017}
R.~G. Vieira, \emph{et~al.}, The Life Cycles of {{Be}} Viscous Decretion Discs:
  Time-Dependent Modelling of Infrared Continuum Observations. \emph{Monthly
  Notices of the Royal Astronomical Society} \textbf{464}, 3071--3089 (2017),
  \doi{10.1093/mnras/stw2542}.

\bibitem{coe_catalogue_2015}
M.~J. Coe, J.~Kirk, Catalogue of {{Be}}/{{X-ray}} Binary Systems in the {{Small
  Magellanic Cloud}}: {{X-ray}}, Optical and {{IR}} Properties. \emph{Monthly
  Notices of the Royal Astronomical Society} \textbf{452}, 969--977 (2015),
  \doi{10.1093/mnras/stv1283}.

\bibitem{corbet_beneutron_1984}
R.~H.~D. Corbet, Be/Neutron Star Binaries : A Relationship between Orbital
  Period and Neutron Star Spin Period. \emph{Astronomy and Astrophysics}
  \textbf{141}, 91--93 (1984).

\bibitem{hills_effects_1983}
J.~G. Hills, The Effects of Sudden Mass Loss and a Random Kick Velocity
  Produced in a Supernova Explosion on the Dynamics of a Binary Star of
  Arbitrary Orbital Eccentricity. {{Applications}} to {{X-ray}} Binaries and to
  the Binarypulsars. \emph{The Astrophysical Journal} \textbf{267}, 322--333
  (1983), \doi{10.1086/160871}.

\bibitem{brandt_effects_1995}
N.~Brandt, P.~Podsiadlowski, The Effects of High-Velocity Supernova Kicks on
  the Orbital Properties and Sky Distributions of Neutron-Star Binaries.
  \emph{Monthly Notices of the Royal Astronomical Society} \textbf{274},
  461--484 (1995), \doi{10.1093/mnras/274.2.461}.

\bibitem{tauris_runaway_1998}
T.~M. Tauris, R.~J. Takens, Runaway Velocities of Stellar Components
  Originating from Disrupted Binaries via Asymmetric Supernova Explosions.
  \emph{Astronomy and Astrophysics} \textbf{330}, 1047--1059 (1998).

\bibitem{almeida_tarantula_2017}
L.~A. Almeida, \emph{et~al.}, The {{Tarantula Massive Binary Monitoring}}.
  {{I}}. {{Observational}} Campaign and {{OB-type}} Spectroscopic Binaries.
  \emph{Astronomy and Astrophysics} \textbf{598}, A84 (2017),
  \doi{10.1051/0004-6361/201629844}.

\bibitem{mahy_tarantula_2020}
L.~Mahy, \emph{et~al.}, The {{Tarantula Massive Binary Monitoring}}. {{III}}.
  {{Atmosphere}} Analysis of Double-Lined Spectroscopic Systems.
  \emph{Astronomy and Astrophysics} \textbf{634}, A118 (2020),
  \doi{10.1051/0004-6361/201936151}.

\bibitem{shenar_tarantula_2022}
T.~Shenar, \emph{et~al.}, The {{Tarantula Massive Binary Monitoring}}. {{VI}}.
  {{Characterisation}} of Hidden Companions in 51 Single-Lined {{O-type}}
  Binaries: {{A}} Flat Mass-Ratio Distribution and Black-Hole Binary
  Candidates. \emph{Astronomy and Astrophysics} \textbf{665}, A148 (2022),
  \doi{10.1051/0004-6361/202244245}.

\bibitem{mandel_parameter_2010}
I.~Mandel, Parameter Estimation on Gravitational Waves from Multiple Coalescing
  Binaries. \emph{Physical Review D} \textbf{81}, 084029 (2010),
  \doi{10.1103/PhysRevD.81.084029}.

\bibitem{mandel_extracting_2019}
I.~Mandel, W.~M. Farr, J.~R. Gair, Extracting Distribution Parameters from
  Multiple Uncertain Observations with Selection Biases. \emph{Monthly Notices
  of the Royal Astronomical Society} \textbf{486}, 1086--1093 (2019),
  \doi{10.1093/mnras/stz896}.

\bibitem{abbott_population_2021}
R.~Abbott, \emph{et~al.}, Population {{Properties}} of {{Compact Objects}} from
  the {{Second LIGO-Virgo Gravitational-Wave Transient Catalog}}. \emph{The
  Astrophysical Journal} \textbf{913}, L7 (2021),
  \doi{10.3847/2041-8213/abe949}.

\bibitem{scott_multivariate_1992}
D.~W. Scott, \emph{Multivariate {{Density Estimation}}} (Wiley, New York)
  (1992).

\bibitem{goodman_ensemble_2010}
J.~Goodman, J.~Weare, Ensemble Samplers with Affine Invariance. \emph{CAMCoS}
  \textbf{5}~(1), 65--80 (2010), \doi{10.2140/camcos.2010.5.65}.

\bibitem{foreman-mackey_emcee_2013}
D.~{Foreman-Mackey}, D.~W. Hogg, D.~Lang, J.~Goodman, Emcee: {{The MCMC
  Hammer}}. \emph{Publications of the Astronomical Society of the Pacific}
  \textbf{125}, 306 (2013), \doi{10.1086/670067}.

\bibitem{chen_monte_1999}
M.-H. Chen, Q.-M. Shao, Monte {{Carlo Estimation}} of {{Bayesian Credible}} and
  {{HPD Intervals}}. \emph{Journal of Computational and Graphical Statistics}
  \textbf{8}~(1), 69--92 (1999), \doi{10.1080/10618600.1999.10474802}.

\bibitem{dickey_weighted_1971}
J.~M. Dickey, The {{Weighted Likelihood Ratio}}, {{Linear Hypotheses}} on
  {{Normal Location Parameters}}. \emph{The Annals of Mathematical Statistics}
  \textbf{42}~(1), 204--223 (1971), \doi{10.1214/aoms/1177693507}.

\bibitem{fritsch_method_1984}
F.~N. Fritsch, J.~Butland, A {{Method}} for {{Constructing Local Monotone
  Piecewise Cubic Interpolants}}. \emph{SIAM J. Sci. and Stat. Comput.}
  \textbf{5}~(2), 300--304 (1984), \doi{10.1137/0905021}.

\bibitem{gelman_bayesian_2014}
A.~Gelman, \emph{et~al.}, \emph{Bayesian {{Data Analysis}}} (Chapman \& Hall),
  3rd ed. (2014).

\bibitem{peacock_two-dimensional_1983}
J.~A. Peacock, Two-Dimensional Goodness-of-Fit Testing in Astronomy.
  \emph{Monthly Notices of the Royal Astronomical Society} \textbf{202},
  615--627 (1983), \doi{10.1093/mnras/202.3.615}.

\bibitem{fasano_multidimensional_1987}
G.~Fasano, A.~Franceschini, A Multidimensional Version of the
  {{Kolmogorov-Smirnov}} Test. \emph{Monthly Notices of the Royal Astronomical
  Society} \textbf{225}, 155--170 (1987), \doi{10.1093/mnras/225.1.155}.

\bibitem{ndtest}
Z.~Li, github.com/syrte/ndtest, \url{github.com/syrte/ndtest}.

\bibitem{gotberg_spectral_2018}
Y.~G{\"o}tberg, \emph{et~al.}, Spectral Models for Binary Products:
  {{Unifying}} Subdwarfs and {{Wolf-Rayet}} Stars as a Sequence of
  Stripped-Envelope Stars. \emph{Astronomy and Astrophysics} \textbf{615}, A78
  (2018), \doi{10.1051/0004-6361/201732274}.

\bibitem{schootemeijer_clues_2018}
A.~Schootemeijer, Y.~G{\"o}tberg, S.~E. {de Mink}, D.~Gies, E.~Zapartas, Clues
  about the Scarcity of Stripped-Envelope Stars from the Evolutionary State of
  the {{sdO}}+{{Be}} Binary System {$\varphi$} {{Persei}}. \emph{Astronomy and
  Astrophysics} \textbf{615}, A30 (2018), \doi{10.1051/0004-6361/201731194}.

\bibitem{yungelson_elusive_2024}
L.~Yungelson, \emph{et~al.}, Elusive Hot Stripped Helium Stars in the
  {{Galaxy}}. {{I}}. {{Evolutionary}} Stellar Models in the Gap between
  Subdwarfs and {{Wolf-Rayet}} Stars. \emph{Astronomy and Astrophysics}
  \textbf{683}, A37 (2024), \doi{10.1051/0004-6361/202347806}.

\bibitem{gotberg_stellar_2023}
Y.~G{\"o}tberg, \emph{et~al.}, Stellar {{Properties}} of {{Observed Stars
  Stripped}} in {{Binaries}} in the {{Magellanic Clouds}}. \emph{The
  Astrophysical Journal} \textbf{959}, 125 (2023),
  \doi{10.3847/1538-4357/ace5a3}.

\bibitem{ramachandran_partially_2023}
V.~Ramachandran, \emph{et~al.}, A Partially Stripped Massive Star in a {{Be}}
  Binary at Low Metallicity. {{A}} Missing Link towards {{Be X-ray}} Binaries
  and Double Neutron Star Mergers. \emph{Astronomy and Astrophysics}
  \textbf{674}, L12 (2023), \doi{10.1051/0004-6361/202346818}.

\bibitem{villasenor_b-type_2023}
J.~I. Villase{\~n}or, \emph{et~al.}, The {{B-type Binaries Characterisation
  Programme}} - {{II}}. {{VFTS}} 291: A Stripped Star from a Recent Mass
  Transfer Phase. \emph{Monthly Notices of the Royal Astronomical Society}
  \textbf{525}, 5121--5145 (2023), \doi{10.1093/mnras/stad2533}.

\bibitem{ramachandran_x-shooting_2024}
V.~Ramachandran, \emph{et~al.}, X-{{Shooting ULLYSES}}: {{Massive Stars}} at
  Low Metallicity {{VIII}}. {{Stellar}} and Wind Parameters of Newly Revealed
  Stripped Stars in {{Be}} Binaries. \emph{arXiv e-prints}  (2024),
  \doi{10.48550/arXiv.2406.17678}.

\bibitem{rivinius_newborn_2025}
{\relax Th}.~Rivinius, \emph{et~al.}, Newborn {{Be}} Star Systems Observed
  Shortly after Mass Transfer. \emph{Astronomy and Astrophysics} \textbf{694},
  A172 (2025), \doi{10.1051/0004-6361/202347275}.

\bibitem{moe_mind_2017}
M.~Moe, R.~Di~Stefano, Mind {{Your Ps}} and {{Qs}}: {{The Interrelation}}
  between {{Period}} ({{P}}) and {{Mass-ratio}} ({{Q}}) {{Distributions}} of
  {{Binary Stars}}. \emph{The Astrophysical Journal Supplement Series}
  \textbf{230}, 15 (2017), \doi{10.3847/1538-4365/aa6fb6}.

\bibitem{von_zeipel_sur_1910}
H.~{von Zeipel}, Sur l'application Des S{\'e}ries de {{M}}. {{Lindstedt}} {\`a}
  l'{\'e}tude Du Mouvement Des Com{\`e}tes P{\'e}riodiques. \emph{Astronomische
  Nachrichten} \textbf{183}, 345 (1910), \doi{10.1002/asna.19091832202}.

\bibitem{kozai_secular_1962}
Y.~Kozai, Secular Perturbations of Asteroids with High Inclination and
  Eccentricity. \emph{The Astronomical Journal} \textbf{67}, 591--598 (1962),
  \doi{10.1086/108790}.

\bibitem{lidov_evolution_1962}
M.~L. Lidov, The Evolution of Orbits of Artificial Satellites of Planets under
  the Action of Gravitational Perturbations of External Bodies. \emph{Planetary
  and Space Science} \textbf{9}, 719--759 (1962),
  \doi{10.1016/0032-0633(62)90129-0}.

\bibitem{peters_detection_2008}
G.~J. Peters, D.~R. Gies, E.~D. Grundstrom, M.~V. McSwain, Detection of a {{Hot
  Subdwarf Companion}} to the {{Be Star FY Canis Majoris}}. \emph{The
  Astrophysical Journal} \textbf{686}, 1280--1291 (2008), \doi{10.1086/591145}.

\bibitem{peters_far-ultraviolet_2013}
G.~J. Peters, T.~D. Pewett, D.~R. Gies, Y.~N. Touhami, E.~D. Grundstrom,
  Far-Ultraviolet {{Detection}} of the {{Suspected Subdwarf Companion}} to the
  {{Be Star}} 59 {{Cygni}}. \emph{The Astrophysical Journal} \textbf{765}, 2
  (2013), \doi{10.1088/0004-637X/765/1/2}.

\bibitem{mourard_spectral_2015}
D.~Mourard, \emph{et~al.}, Spectral and Spatial Imaging of the {{Be}}+{{sdO}}
  Binary {$\phi$} {{Persei}}. \emph{Astronomy and Astrophysics} \textbf{577},
  A51 (2015), \doi{10.1051/0004-6361/201425141}.

\bibitem{chojnowski_remarkable_2018}
S.~D. Chojnowski, \emph{et~al.}, The {{Remarkable Be}}+{{sdOB Binary HD}}
  55606. {{I}}. {{Orbital}} and {{Stellar Parameters}}. \emph{The Astrophysical
  Journal} \textbf{865}, 76 (2018), \doi{10.3847/1538-4357/aad964}.

\bibitem{shenar_hidden_2020}
T.~Shenar, \emph{et~al.}, The "Hidden" Companion in {{LB-1}} Unveiled by
  Spectral Disentangling. \emph{Astronomy and Astrophysics} \textbf{639}, L6
  (2020), \doi{10.1051/0004-6361/202038275}.

\bibitem{gies_h_2020}
D.~R. Gies, L.~Wang, The {{H$\alpha$ Emission Line Variations}} of {{HR}} 6819.
  \emph{The Astrophysical Journal} \textbf{898}, L44 (2020),
  \doi{10.3847/2041-8213/aba51c}.

\bibitem{harmanec_v1294_2022}
P.~Harmanec, \emph{et~al.}, V1294 {{Aql}} = {{HD}} 184279: {{A}} Bad Boy among
  {{Be}} Stars or an Important Clue to the {{Be}} Phenomenon? \emph{Astronomy
  and Astrophysics} \textbf{666}, A136 (2022),
  \doi{10.1051/0004-6361/202244006}.

\bibitem{wang_orbital_2023}
L.~Wang, D.~R. Gies, G.~J. Peters, Z.~Han, The {{Orbital}} and {{Physical
  Properties}} of {{Five Southern Be}}+{{sdO Binary Systems}}. \emph{The
  Astronomical Journal} \textbf{165}, 203 (2023),
  \doi{10.3847/1538-3881/acc6ca}.

\bibitem{klement_chara_2024}
R.~Klement, \emph{et~al.}, The {{CHARA Array Interferometric Program}} on the
  {{Multiplicity}} of {{Classical Be Stars}}: {{New Detections}} and {{Orbits}}
  of {{Stripped Subdwarf Companions}}. \emph{The Astrophysical Journal}
  \textbf{962}, 70 (2024), \doi{10.3847/1538-4357/ad13ec}.

\bibitem{muller-horn_hip_2025}
J.~{M{\"u}ller-Horn}, \emph{et~al.}, {{HIP}} 15429: A Newborn {{Be}} Star on an
  Eccentric Binary Orbit  (2025), \doi{10.48550/arXiv.2504.06973}.

\bibitem{north_2_2007}
J.~R. North, P.~G. Tuthill, W.~J. Tango, J.~Davis, {$\Gamma$}2 {{Velorum}}:
  Orbital Solution and Fundamental Parameter Determination with {{SUSI}}.
  \emph{Monthly Notices of the Royal Astronomical Society} \textbf{377},
  415--424 (2007), \doi{10.1111/j.1365-2966.2007.11608.x}.

\bibitem{de_la_chevrotiere_spectroscopic_2011}
A.~{de La Chevroti{\`e}re}, A.~F.~J. Moffat, A.~N. Chen{\'e}, Spectroscopic
  Study of the Short-Period {{WN5o}} + {{O8}}.{{5V}} Binary System {{WR127}}
  ({{HD}} 186943). \emph{Monthly Notices of the Royal Astronomical Society}
  \textbf{411}, 635--643 (2011), \doi{10.1111/j.1365-2966.2010.17710.x}.

\bibitem{richardson_first_2021}
N.~D. Richardson, \emph{et~al.}, The {{First Dynamical Mass Determination}} of
  a {{Nitrogen-rich Wolf-Rayet Star Using}} a {{Combined Visual}} and
  {{Spectroscopic Orbit}}. \emph{The Astrophysical Journal} \textbf{908}, L3
  (2021), \doi{10.3847/2041-8213/abd722}.

\bibitem{thomas_orbit_2021}
J.~D. Thomas, \emph{et~al.}, The Orbit and Stellar Masses of the Archetype
  Colliding-Wind Binary {{WR}} 140. \emph{Monthly Notices of the Royal
  Astronomical Society} \textbf{504}, 5221--5230 (2021),
  \doi{10.1093/mnras/stab1181}.

\bibitem{zahn_tidal_1977}
J.~P. Zahn, Tidal Friction in Close Binary Systems. \emph{Astronomy and
  Astrophysics} \textbf{57}, 383--394 (1977).

\bibitem{meibom_robust_2005}
S.~Meibom, R.~D. Mathieu, A {{Robust Measure}} of {{Tidal Circularization}} in
  {{Coeval Binary Populations}}: {{The Solar-Type Spectroscopic Binary
  Population}} in the {{Open Cluster M35}}. \emph{The Astrophysical Journal}
  \textbf{620}, 970--983 (2005), \doi{10.1086/427082}.

\bibitem{zanazzi_tale_2022}
J.~J. Zanazzi, A {{Tale}} of {{Two Circularization Periods}}. \emph{The
  Astrophysical Journal} \textbf{929}, L27 (2022),
  \doi{10.3847/2041-8213/ac6516}.

\bibitem{witte_tidal_1999}
M.~G. Witte, G.~J. Savonije, Tidal Evolution of Eccentric Orbits in Massive
  Binary Systems. {{A}} Study of Resonance Locking. \emph{Astronomy and
  Astrophysics} \textbf{350}, 129--147 (1999).

\bibitem{rimulo_life_2018}
L.~R. R{\'i}mulo, \emph{et~al.}, The Life Cycles of {{Be}} Viscous Decretion
  Discs: Fundamental Disc Parameters of 54 {{SMC Be}} Stars. \emph{Monthly
  Notices of the Royal Astronomical Society} \textbf{476}, 3555--3579 (2018),
  \doi{10.1093/mnras/sty431}.

\bibitem{liu_interaction_2015}
Z.-W. Liu, \emph{et~al.}, The Interaction of Core-Collapse Supernova Ejecta
  with a Companion Star. \emph{Astronomy and Astrophysics} \textbf{584}, A11
  (2015), \doi{10.1051/0004-6361/201526757}.

\bibitem{hirai_possible_2015}
R.~Hirai, S.~Yamada, Possible {{Signatures}} of {{Ejecta-Companion
  Interaction}} in {{iPTF}} 13bvn. \emph{The Astrophysical Journal}
  \textbf{805}, 170 (2015), \doi{10.1088/0004-637X/805/2/170}.

\bibitem{hirai_comprehensive_2018}
R.~Hirai, P.~Podsiadlowski, S.~Yamada, Comprehensive {{Study}} of
  {{Ejecta-companion Interaction}} for {{Core-collapse Supernovae}} in
  {{Massive Binaries}}. \emph{The Astrophysical Journal} \textbf{864}, 119
  (2018), \doi{10.3847/1538-4357/aad6a0}.

\bibitem{ogata_observability_2021}
M.~Ogata, R.~Hirai, K.~Hijikawa, Observability of Inflated Companion Stars
  after Supernovae in Massive Binaries. \emph{Monthly Notices of the Royal
  Astronomical Society} \textbf{505}, 2485--2499 (2021),
  \doi{10.1093/mnras/stab1439}.

\bibitem{hirai_constraining_2023}
R.~Hirai, Constraining Mass Transfer and Common-Envelope Physics with
  Post-Supernova Companion Monitoring. \emph{Monthly Notices of the Royal
  Astronomical Society} \textbf{523}, 6011--6019 (2023),
  \doi{10.1093/mnras/stad1856}.

\bibitem{akira_rocha_mass_2024}
K.~Akira~Rocha, \emph{et~al.}, Mass {{Transfer}} in {{Eccentric Orbits}} with
  {{Self-consistent Stellar Evolution}}. \emph{arXiv e-prints}  (2024),
  \doi{10.48550/arXiv.2411.11840}.

\bibitem{marcussen_banana_2022}
M.~L. Marcussen, S.~H. Albrecht, The {{BANANA Project}}. {{VI}}. {{Close Double
  Stars}} Are {{Well Aligned}} with {{Noticeable Exceptions}}; {{Results}} from
  an {{Ensemble Study Using Apsidal Motion}} and {{Rossiter-McLaughlin
  Measurements}}. \emph{The Astrophysical Journal} \textbf{933}, 227 (2022),
  \doi{10.3847/1538-4357/ac75c2}.

\bibitem{ozel_masses_2016}
F.~{\"O}zel, P.~Freire, Masses, {{Radii}}, and the {{Equation}} of {{State}} of
  {{Neutron Stars}}. \emph{Annual Review of Astronomy and Astrophysics}
  \textbf{54}, 401--440 (2016), \doi{10.1146/annurev-astro-081915-023322}.

\bibitem{miller-jones_geometric_2018}
J.~C.~A. {Miller-Jones}, \emph{et~al.}, The Geometric Distance and Binary Orbit
  of {{PSR B1259-63}}. \emph{Monthly Notices of the Royal Astronomical Society}
  \textbf{479}, 4849--4860 (2018), \doi{10.1093/mnras/sty1775}.

\bibitem{fryer_double_1997}
C.~Fryer, V.~Kalogera, Double {{Neutron Star Systems}} and {{Natal Neutron Star
  Kicks}}. \emph{The Astrophysical Journal} \textbf{489}, 244--253 (1997),
  \doi{10.1086/304772}.

\bibitem{hughes_constraints_1999}
A.~Hughes, M.~Bailes, Constraints on {{Natal Pulsar Kicks}} from {{Eccentric
  Binary Pulsars}}. \emph{The Astrophysical Journal} \textbf{522}, 504--511
  (1999), \doi{10.1086/307605}.

\bibitem{dewi_spin_2005}
J.~D.~M. Dewi, {\relax Ph}.~Podsiadlowski, O.~R. Pols, The Spin
  Period-Eccentricity Relation of Double Neutron Stars: Evidence for Weak
  Supernova Kicks? \emph{Monthly Notices of the Royal Astronomical Society}
  \textbf{363}, L71--L75 (2005), \doi{10.1111/j.1745-3933.2005.00085.x}.

\bibitem{chruslinska_constraints_2017}
M.~Chruslinska, K.~Belczynski, T.~Bulik, W.~Gladysz, Constraints on the
  {{Formation}} of {{Double Neutron Stars}} from the {{Observed
  Eccentricities}} and {{Current Limits}} on {{Merger Rates}}. \emph{Acta
  Astronomica} \textbf{67}, 37--50 (2017), \doi{10.32023/0001-5237/67.1.2}.

\bibitem{andrews_double_2019}
J.~J. Andrews, I.~Mandel, Double {{Neutron Star Populations}} and {{Formation
  Channels}}. \emph{The Astrophysical Journal} \textbf{880}, L8 (2019),
  \doi{10.3847/2041-8213/ab2ed1}.

\bibitem{disberg_kinematic_2024}
P.~Disberg, N.~Gaspari, A.~J. Levan, Kinematic Constraints on the Ages and Kick
  Velocities of {{Galactic}} Neutron Star Binaries. \emph{Astronomy and
  Astrophysics} \textbf{689}, A348 (2024), \doi{10.1051/0004-6361/202450790}.

\bibitem{manchester_australia_2005}
R.~N. Manchester, G.~B. Hobbs, A.~Teoh, M.~Hobbs, The {{Australia Telescope
  National Facility Pulsar Catalogue}}. \emph{The Astronomical Journal}
  \textbf{129}, 1993--2006 (2005), \doi{10.1086/428488}.

\bibitem{peters_gravitational_1964}
P.~C. Peters, Gravitational {{Radiation}} and the {{Motion}} of {{Two Point
  Masses}}. \emph{Physical Review} \textbf{136}~(4B), B1224--B1232 (1964),
  \doi{10.1103/PhysRev.136.B1224}.

\bibitem{vink_mass-loss_2001}
J.~S. Vink, A.~{de Koter}, H.~J. G. L.~M. Lamers, Mass-Loss Predictions for
  {{O}} and {{B}} Stars as a Function of Metallicity. \emph{Astronomy and
  Astrophysics} \textbf{369}~(2), 574--588 (2001),
  \doi{10.1051/0004-6361:20010127}.

\bibitem{spruit_birth_1998}
H.~Spruit, E.~S. Phinney, Birth Kicks as the Origin of Pulsar Rotation.
  \emph{Nature} \textbf{393}, 139--141 (1998), \doi{10.1038/30168}.

\bibitem{heger_presupernova_2005}
A.~Heger, S.~E. Woosley, H.~C. Spruit, Presupernova {{Evolution}} of
  {{Differentially Rotating Massive Stars Including Magnetic Fields}}.
  \emph{The Astrophysical Journal} \textbf{626}, 350--363 (2005),
  \doi{10.1086/429868}.

\bibitem{ott_spin_2006}
C.~D. Ott, A.~Burrows, T.~A. Thompson, E.~Livne, R.~Walder, The {{Spin
  Periods}} and {{Rotational Profiles}} of {{Neutron Stars}} at {{Birth}}.
  \emph{The Astrophysical Journal Supplement Series} \textbf{164}, 130--155
  (2006), \doi{10.1086/500832}.

\bibitem{coleman_kicks_2022}
M.~S.~B. Coleman, A.~Burrows, Kicks and Induced Spins of Neutron Stars at
  Birth. \emph{Monthly Notices of the Royal Astronomical Society} \textbf{517},
  3938--3961 (2022), \doi{10.1093/mnras/stac2573}.

\bibitem{janka_supernova_2022}
H.-T. Janka, A.~Wongwathanarat, M.~Kramer, Supernova {{Fallback}} as {{Origin}}
  of {{Neutron Star Spins}} and {{Spin-kick Alignment}}. \emph{The
  Astrophysical Journal} \textbf{926}, 9 (2022),
  \doi{10.3847/1538-4357/ac403c}.

\bibitem{dodson_vela_2003}
R.~Dodson, D.~Legge, J.~E. Reynolds, P.~M. McCulloch, The {{Vela Pulsar}}'s
  {{Proper Motion}} and {{Parallax Derived}} from {{VLBI Observations}}.
  \emph{The Astrophysical Journal} \textbf{596}, 1137--1141 (2003),
  \doi{10.1086/378089}.

\bibitem{kaplan_precise_2008}
D.~L. Kaplan, S.~Chatterjee, B.~M. Gaensler, J.~Anderson, A {{Precise Proper
  Motion}} for the {{Crab Pulsar}}, and the {{Difficulty}} of {{Testing
  Spin-Kick Alignment}} for {{Young Neutron Stars}}. \emph{The Astrophysical
  Journal} \textbf{677}, 1201--1215 (2008), \doi{10.1086/529026}.

\bibitem{yao_evidence_2021}
J.~Yao, \emph{et~al.}, Evidence for Three-Dimensional Spin-Velocity Alignment
  in a Pulsar. \emph{Nature Astronomy} \textbf{5}, 788--795 (2021),
  \doi{10.1038/s41550-021-01360-w}.

\bibitem{johnston_evidence_2005}
S.~Johnston, \emph{et~al.}, Evidence for Alignment of the Rotation and Velocity
  Vectors in Pulsars. \emph{Monthly Notices of the Royal Astronomical Society}
  \textbf{364}, 1397--1412 (2005), \doi{10.1111/j.1365-2966.2005.09669.x}.

\bibitem{johnston_evidence_2007}
S.~Johnston, \emph{et~al.}, Evidence for Alignment of the Rotation and Velocity
  Vectors in Pulsars - {{II}}. {{Further}} Data and Emission Heights.
  \emph{Monthly Notices of the Royal Astronomical Society} \textbf{381},
  1625--1637 (2007), \doi{10.1111/j.1365-2966.2007.12352.x}.

\bibitem{wang_neutron_2006}
C.~Wang, D.~Lai, J.~L. Han, Neutron {{Star Kicks}} in {{Isolated}} and {{Binary
  Pulsars}}: {{Observational Constraints}} and {{Implications}} for {{Kick
  Mechanisms}}. \emph{The Astrophysical Journal} \textbf{639}, 1007--1017
  (2006), \doi{10.1086/499397}.

\bibitem{rankin_further_2007}
J.~M. Rankin, Further {{Evidence}} for {{Alignment}} of the {{Rotation}} and
  {{Velocity Vectors}} in {{Pulsars}}. \emph{The Astrophysical Journal}
  \textbf{664}, 443--447 (2007), \doi{10.1086/519018}.

\bibitem{noutsos_pulsar_2012}
A.~Noutsos, M.~Kramer, P.~Carr, S.~Johnston, Pulsar Spin-Velocity Alignment:
  Further Results and Discussion. \emph{Monthly Notices of the Royal
  Astronomical Society} \textbf{423}, 2736--2752 (2012),
  \doi{10.1111/j.1365-2966.2012.21083.x}.

\bibitem{noutsos_pulsar_2013}
A.~Noutsos, D.~H. F.~M. Schnitzeler, E.~F. Keane, M.~Kramer, S.~Johnston,
  Pulsar Spin-Velocity Alignment: Kinematic Ages, Birth Periods and Braking
  Indices. \emph{Monthly Notices of the Royal Astronomical Society}
  \textbf{430}, 2281--2301 (2013), \doi{10.1093/mnras/stt047}.

\bibitem{rankin_toward_2015}
J.~M. Rankin, Toward an {{Empirical Theory}} of {{Pulsar Emission}}. {{XI}}.
  {{Understanding}} the {{Orientations}} of {{Pulsar Radiation}} and
  {{Supernova}} ``{{Kicks}}''. \emph{The Astrophysical Journal} \textbf{804},
  112 (2015), \doi{10.1088/0004-637X/804/2/112}.

\bibitem{biryukov_evidence_2024}
A.~Biryukov, G.~Beskin, Evidence for the Spin-Kick Alignment of Pulsars from
  the Statistics of Their Magnetic Inclinations. \emph{arXiv e-prints}  (2024),
  \doi{10.48550/arXiv.2412.12017}.

\bibitem{berger_search_2001}
D.~H. Berger, D.~R. Gies, A {{Search}} for {{High-Velocity Be Stars}}.
  \emph{The Astrophysical Journal} \textbf{555}, 364--367 (2001),
  \doi{10.1086/321461}.

\bibitem{boubert_kinematics_2018}
D.~Boubert, N.~W. Evans, On the Kinematics of a Runaway {{Be}} Star Population.
  \emph{Monthly Notices of the Royal Astronomical Society} \textbf{477},
  5261--5278 (2018), \doi{10.1093/mnras/sty980}.

\bibitem{stairs_psr_2001}
I.~H. Stairs, \emph{et~al.}, {{PSR J1740-3052}}: A Pulsar with a Massive
  Companion. \emph{Monthly Notices of the Royal Astronomical Society}
  \textbf{325}, 979--988 (2001), \doi{10.1046/j.1365-8711.2001.04447.x}.

\bibitem{madsen_timing_2012}
E.~C. Madsen, \emph{et~al.}, Timing the Main-Sequence-Star Binary Pulsar
  {{J1740-3052}}. \emph{Monthly Notices of the Royal Astronomical Society}
  \textbf{425}, 2378--2385 (2012), \doi{10.1111/j.1365-2966.2012.21691.x}.

\bibitem{andersen_chime_2023}
B.~C. Andersen, \emph{et~al.}, {{CHIME Discovery}} of a {{Binary Pulsar}} with
  a {{Massive Nondegenerate Companion}}. \emph{The Astrophysical Journal}
  \textbf{943}, 57 (2023), \doi{10.3847/1538-4357/aca485}.

\bibitem{parent_study_2022}
E.~Parent, \emph{et~al.}, Study of 72 {{Pulsars Discovered}} in the {{PALFA
  Survey}}: {{Timing Analysis}}, {{Glitch Activity}}, {{Emission Variability}},
  and a {{Pulsar}} in an {{Eccentric Binary}}. \emph{The Astrophysical Journal}
  \textbf{924}, 135 (2022), \doi{10.3847/1538-4357/ac375d}.

\bibitem{kaspi_massive_1994}
V.~M. Kaspi, \emph{et~al.}, A {{Massive Radio Pulsar Binary}} in the {{Small
  Magellanic Cloud}}. \emph{The Astrophysical Journal} \textbf{423}, L43
  (1994), \doi{10.1086/187231}.

\bibitem{kaspi_evidence_1996}
V.~M. Kaspi, M.~Bailes, R.~N. Manchester, B.~W. Stappers, J.~F. Bell, Evidence
  from a Precessing Pulsar Orbit for a Neutron-Star Birth Kick. \emph{Nature}
  \textbf{381}, 584--586 (1996), \doi{10.1038/381584a0}.

\bibitem{kumar_orbital_1998}
P.~Kumar, E.~J. Quataert, On the {{Orbital Decay}} of the {{PSR J0045-7319
  Binary}}. \emph{The Astrophysical Journal} \textbf{493}, 412--425 (1998),
  \doi{10.1086/305091}.

\bibitem{su_dynamical_2022}
Y.~Su, D.~Lai, Dynamical Tides in Eccentric Binaries Containing Massive
  Main-Sequence Stars: Analytical Expressions. \emph{Monthly Notices of the
  Royal Astronomical Society} \textbf{510}~(4), 4943--4951 (2022),
  \doi{10.1093/mnras/stab3698}.

\bibitem{trimble_motions_1968}
V.~Trimble, Motions and {{Structure}} of the {{Filamentary Envelope}} of the
  {{Crab Nebula}}. \emph{The Astronomical Journal} \textbf{73}, 535 (1968),
  \doi{10.1086/110658}.

\bibitem{gunn_nature_1970}
J.~E. Gunn, J.~P. Ostriker, On the {{Nature}} of {{Pulsars}}. {{III}}.
  {{Analysis}} of {{Observations}}. \emph{The Astrophysical Journal}
  \textbf{160}, 979 (1970), \doi{10.1086/150487}.

\bibitem{cordes_neutron_1998}
J.~M. Cordes, D.~F. Chernoff, Neutron {{Star Population Dynamics}}. {{II}}.
  {{Three-dimensional Space Velocities}} of {{Young Pulsars}}. \emph{The
  Astrophysical Journal} \textbf{505}, 315--338 (1998), \doi{10.1086/306138}.

\bibitem{arzoumanian_velocity_2002}
Z.~Arzoumanian, D.~F. Chernoff, J.~M. Cordes, The {{Velocity Distribution}} of
  {{Isolated Radio Pulsars}}. \emph{The Astrophysical Journal} \textbf{568},
  289--301 (2002), \doi{10.1086/338805}.

\bibitem{brisken_proper-motion_2003}
W.~F. Brisken, A.~S. Fruchter, W.~M. Goss, R.~M. Herrnstein, S.~E. Thorsett,
  Proper-{{Motion Measurements}} with the {{VLA}}. {{II}}. {{Observations}} of
  28 {{Pulsars}}. \emph{The Astronomical Journal} \textbf{126}, 3090--3098
  (2003), \doi{10.1086/379559}.

\bibitem{faucher-giguere_birth_2006}
C.-A. {Faucher-Gigu{\`e}re}, V.~M. Kaspi, Birth and {{Evolution}} of {{Isolated
  Radio Pulsars}}. \emph{The Astrophysical Journal} \textbf{643}, 332--355
  (2006), \doi{10.1086/501516}.

\bibitem{verbunt_observed_2017}
F.~Verbunt, A.~Igoshev, E.~Cator, The Observed Velocity Distribution of Young
  Pulsars. \emph{Astronomy and Astrophysics} \textbf{608}, A57 (2017),
  \doi{10.1051/0004-6361/201731518}.

\bibitem{stockinger_three-dimensional_2020}
G.~Stockinger, \emph{et~al.}, Three-Dimensional Models of Core-Collapse
  Supernovae from Low-Mass Progenitors with Implications for {{Crab}}.
  \emph{Monthly Notices of the Royal Astronomical Society} \textbf{496},
  2039--2084 (2020), \doi{10.1093/mnras/staa1691}.

\bibitem{van_den_heuvel_x-ray_2004}
E.~P.~J. {van den Heuvel}, X-{{Ray Binaries}} and {{Their Descendants}}:
  {{Binary Radio Pulsars}}; {{Evidence}} for {{Three Classes}} of {{Neutron
  Stars}}?, in \emph{Proceedings of the {5th INTEGRAL Workshop} on the
  {INTEGRAL Universe}} (ESA SP-552, Munich, Germany), vol. 552 (2004), p. 185,
  \doi{10.48550/arXiv.astro-ph/0407451}.

\bibitem{van_den_heuvel_double_2007}
E.~P.~J. {van den Heuvel}, Double {{Neutron Stars}}: {{Evidence For Two
  Different Neutron-Star Formation Mechanisms}}, in \emph{AIP Conference
  Proceedings} (AIP, Cefalu), vol. 924 (2007), pp. 598--606,
  \doi{10.1063/1.2774916}.

\bibitem{knigge_two_2011}
C.~Knigge, M.~J. Coe, P.~Podsiadlowski, Two Populations of {{X-ray}} Pulsars
  Produced by Two Types of Supernova. \emph{Nature} \textbf{479}, 372--375
  (2011), \doi{10.1038/nature10529}.

\bibitem{waters_relation_1989}
L.~B. F.~M. Waters, M.~H. {van Kerkwijk}, The Relation between Orbital and Spin
  Periods in Massive {{X-ray}} Binaries. \emph{Astronomy and Astrophysics}
  \textbf{223}, 196--206 (1989).

\bibitem{cheng_spin_2014}
Z.~Q. Cheng, Y.~Shao, X.~D. Li, On the {{Spin Period Distribution}} in
  {{Be}}/{{X-Ray Binaries}}. \emph{The Astrophysical Journal} \textbf{786}, 128
  (2014), \doi{10.1088/0004-637X/786/2/128}.

\bibitem{xu_bimodal_2019}
X.-T. Xu, X.-D. Li, On the {{Bimodal Spin-period Distribution}} of
  {{Be}}/{{X-Ray Pulsars}}. \emph{The Astrophysical Journal} \textbf{872}, 102
  (2019), \doi{10.3847/1538-4357/aafee0}.

\bibitem{schwab_further_2010}
J.~Schwab, {\relax Ph}.~Podsiadlowski, S.~Rappaport, Further {{Evidence}} for
  the {{Bimodal Distribution}} of {{Neutron-star Masses}}. \emph{The
  Astrophysical Journal} \textbf{719}, 722--727 (2010),
  \doi{10.1088/0004-637X/719/1/722}.

\bibitem{bray_neutron_2016}
J.~C. Bray, J.~J. Eldridge, Neutron Star Kicks and Their Relationship to
  Supernovae Ejecta Mass. \emph{Monthly Notices of the Royal Astronomical
  Society} \textbf{461}, 3747--3759 (2016), \doi{10.1093/mnras/stw1275}.

\bibitem{bray_neutron_2018}
J.~C. Bray, J.~J. Eldridge, Neutron Star Kicks - {{II}}. {{Revision}} and
  Further Testing of the Conservation of Momentum `kick' Model. \emph{Monthly
  Notices of the Royal Astronomical Society} \textbf{480}, 5657--5672 (2018),
  \doi{10.1093/mnras/sty2230}.

\bibitem{giacobbo_revising_2020}
N.~Giacobbo, M.~Mapelli, Revising {{Natal Kick Prescriptions}} in {{Population
  Synthesis Simulations}}. \emph{The Astrophysical Journal} \textbf{891}~(2),
  141 (2020), \doi{10.3847/1538-4357/ab7335}.

\bibitem{willcox_constraints_2021}
R.~Willcox, \emph{et~al.}, Constraints on {{Weak Supernova Kicks}} from
  {{Observed Pulsar Velocities}}. \emph{The Astrophysical Journal}
  \textbf{920}, L37 (2021), \doi{10.3847/2041-8213/ac2cc8}.

\bibitem{kapil_calibration_2023}
V.~Kapil, I.~Mandel, E.~Berti, B.~M{\"u}ller, Calibration of Neutron Star Natal
  Kick Velocities to Isolated Pulsar Observations. \emph{Monthly Notices of the
  Royal Astronomical Society} \textbf{519}, 5893--5901 (2023),
  \doi{10.1093/mnras/stad019}.

\bibitem{andrews_evolutionary_2015}
J.~J. Andrews, W.~M. Farr, V.~Kalogera, B.~Willems, Evolutionary {{Channels}}
  for the {{Formation}} of {{Double Neutron Stars}}. \emph{The Astrophysical
  Journal} \textbf{801}, 32 (2015), \doi{10.1088/0004-637X/801/1/32}.

\bibitem{kruckow_progenitors_2018}
M.~U. Kruckow, T.~M. Tauris, N.~Langer, M.~Kramer, R.~G. Izzard, Progenitors of
  Gravitational Wave Mergers: Binary Evolution with the Stellar Grid-Based Code
  {{ComBinE}}. \emph{Monthly Notices of the Royal Astronomical Society}
  \textbf{481}~(2), 1908--1949 (2018), \doi{10.1093/mnras/sty2190}.

\bibitem{shao_role_2018}
Y.~Shao, X.-D. Li, On the {{Role}} of {{Supernova Kicks}} in the {{Formation}}
  of {{Galactic Double Neutron Star Systems}}. \emph{The Astrophysical Journal}
  \textbf{867}, 124 (2018), \doi{10.3847/1538-4357/aae648}.

\bibitem{wang_neutron_2025}
X.~I. Wang, X.-D. Li, On {{Neutron Star Natal Kicks}} in {{High-Mass X-Ray
  Binaries}}: {{Insights}} from {{Population Synthesis}} ~(arXiv:2504.19672)
  (2025), \doi{10.48550/arXiv.2504.19672}.

\bibitem{bodaghee_clustering_2012}
A.~Bodaghee, J.~A. Tomsick, J.~Rodriguez, J.~B. James, Clustering between
  {{High-mass X-Ray Binaries}} and {{OB Associations}} in the {{Milky Way}}.
  \emph{The Astrophysical Journal} \textbf{744}, 108 (2012),
  \doi{10.1088/0004-637X/744/2/108}.

\bibitem{zuo_displacement_2015}
Z.-Y. Zuo, Displacement of {{X-ray}} Binaries: Constraints on the Natal Kicks.
  \emph{Astronomy and Astrophysics} \textbf{573}, A58 (2015),
  \doi{10.1051/0004-6361/201424604}.

\bibitem{repetto_galactic_2017}
S.~Repetto, A.~P. Igoshev, G.~Nelemans, The {{Galactic}} Distribution of
  {{X-ray}} Binaries and Its Implications for Compact Object Formation and
  Natal Kicks. \emph{Monthly Notices of the Royal Astronomical Society}
  \textbf{467}, 298--310 (2017), \doi{10.1093/mnras/stx027}.

\bibitem{bodaghee_evidence_2021}
A.~Bodaghee, \emph{et~al.}, Evidence for {{Low Kick Velocities}} among
  {{High-mass X-Ray Binaries}} in the {{Small Magellanic Cloud}} from the
  {{Spatial Correlation Function}}. \emph{The Astrophysical Journal}
  \textbf{919}, 81 (2021), \doi{10.3847/1538-4357/ac11f4}.

\bibitem{prisegen_kinematic_2020}
M.~Pri{\v s}egen, Kinematic Distinction of the Two Subpopulations of {{X-ray}}
  Pulsars. \emph{Astronomy and Astrophysics} \textbf{640}, A86 (2020),
  \doi{10.1051/0004-6361/201935642}.

\bibitem{odoherty_observationally_2023}
T.~N. O'Doherty, \emph{et~al.}, An Observationally Derived Kick Distribution
  for Neutron Stars in Binary Systems. \emph{Monthly Notices of the Royal
  Astronomical Society} \textbf{521}, 2504--2524 (2023),
  \doi{10.1093/mnras/stad680}.

\bibitem{zhao_evidence_2023}
Y.~Zhao, \emph{et~al.}, Evidence for Mass-Dependent Peculiar Velocities in
  Compact Object Binaries: Towards Better Constraints on Natal Kicks.
  \emph{Monthly Notices of the Royal Astronomical Society} \textbf{525},
  1498--1519 (2023), \doi{10.1093/mnras/stad2226}.

\bibitem{martin_supernova_2009}
R.~G. Martin, C.~A. Tout, J.~E. Pringle, Supernova Kicks and Misaligned {{Be}}
  Star Binaries. \emph{Monthly Notices of the Royal Astronomical Society}
  \textbf{397}, 1563--1576 (2009), \doi{10.1111/j.1365-2966.2009.15031.x}.

\bibitem{hummel_spectacular_1998}
W.~Hummel, On the Spectacular Variations of {{Be}} Stars. {{Evidence}} for a
  Temporarily Tilted Circumstellar Disk. \emph{Astronomy and Astrophysics}
  \textbf{330}, 243--252 (1998).

\bibitem{wex_timing_1998}
N.~Wex, \emph{et~al.}, Timing Models for the Long Orbital Period Binary Pulsar
  {{PSR B1259-63}}. \emph{Monthly Notices of the Royal Astronomical Society}
  \textbf{298}, 997--1004 (1998), \doi{10.1046/j.1365-8711.1998.01700.x}.

\bibitem{Hunter:2007}
J.~D. Hunter, Matplotlib: A 2D graphics environment. \emph{Computing in Science
  \& Engineering} \textbf{9}~(3), 90--95 (2007), \doi{10.1109/MCSE.2007.55}.

\bibitem{numpy}
C.~R. Harris, \emph{et~al.}, Array programming with {NumPy}. \emph{Nature}
  \textbf{585}~(7825), 357--362 (2020), \doi{10.1038/s41586-020-2649-2},
  \url{https://doi.org/10.1038/s41586-020-2649-2}.

\bibitem{pandas_13819579}
{The pandas development team}, pandas-dev/pandas: Pandas (2024),
  \doi{10.5281/zenodo.13819579}, \url{https://doi.org/10.5281/zenodo.13819579}.

\bibitem{mckinney-proc-scipy-2010}
{W}es {M}c{K}inney, {D}ata {S}tructures for {S}tatistical {C}omputing in
  {P}ython, in \emph{{P}roceedings of the 9th {P}ython in {S}cience
  {C}onference}, {S}t\'efan van~der {W}alt, {J}arrod {M}illman, Eds. (2010),
  pp. 56 -- 61, \doi{10.25080/Majora-92bf1922-00a}.

\bibitem{python}
G.~Van~Rossum, F.~L. Drake, \emph{Python 3 Reference Manual} (CreateSpace,
  Scotts Valley, CA) (2009).

\bibitem{chatGPT}
openAI, chatGPT: 3.5 (2025), \url{chatgpt.com}.

\bibitem{corner-Foreman-Mackey-2016}
D.~{Foreman-Mackey}, {corner.py: Scatterplot matrices in Python}. \emph{The
  Journal of Open Source Software} \textbf{1}, 24 (2016),
  \doi{10.21105/joss.00024}.

\bibitem{corner.py_4592454}
D.~Foreman-Mackey, \emph{et~al.}, dfm/corner.py: corner.py v.2.2.1 (2021),
  \doi{10.5281/zenodo.4592454}, \url{https://doi.org/10.5281/zenodo.4592454}.

\bibitem{cython:2011}
S.~Behnel, \emph{et~al.}, Cython: The Best of Both Worlds. \emph{Computing in
  Science Engineering} \textbf{13}~(2), 31--39 (2011),
  \doi{10.1109/MCSE.2010.118}.

\bibitem{emcee_10996751}
D.~Foreman-Mackey, \emph{et~al.}, dfm/emcee: v3.1.6 (2024),
  \doi{10.5281/zenodo.10996751}, \url{https://doi.org/10.5281/zenodo.10996751}.

\bibitem{collette_python_hdf5_2014}
A.~Collette, \emph{Python and HDF5} (O'Reilly) (2013).

\bibitem{h5py_7560547}
A.~Collette, \emph{et~al.}, h5py/h5py: 3.8.0 (2023),
  \doi{10.5281/zenodo.7560547}, \url{https://doi.org/10.5281/zenodo.7560547}.

\end{thebibliography}
\bibliographystyle{sciencemag}
}

\end{document}